\documentclass[amsthm]{elsart}

\usepackage{yjsco}
\usepackage{natbib}

\usepackage{amssymb, amsbsy, amsfonts,color}
\usepackage{amsmath, amsthm,epsfig}
\usepackage{latexsym,float}

\input{xy}
\xyoption{all}

\allowdisplaybreaks[4]
\relax

\newcommand{\dd}{\mathfrak{d}}
\newcommand{\depth}[1]{\delta(#1)}
\newcommand{\depthF}[2]{\delta_{#2}(#1)}

\newcommand{\PFDE}{{\sf PFDE}}
\newcommand{\DOT}{{\sf DOT}}
\newcommand{\DOTX}{{\sf DOT$^*$}}
\newcommand{\SR}{{\sf SR}}
\newcommand{\PT}{{\sf PT}}
\newcommand{\CP}{{\sf CP}}
\newcommand{\PP}{{\sf PP}}
\newcommand{\RP}{{\sf RP}}
\newcommand{\CE}{{\sf CE}}

\newcommand{\SigmaP}{{\sf Sigma}}
\newcommand{\shiftS}{\hspace*{0.4cm}}
\newcommand{\MyFrame}[1]{
\fcolorbox[rgb]{0,0,0}{1,1,1}{
\begin{minipage}{12.8cm}#1\end{minipage}}
}

\newcommand{\SolSpaceT}[2]{{\rm V}(\vect{#1},{#2})}
\newcommand{\SolSpace}[3]{{\rm V}({\vect{#1}},\vect{#2},{#3})}
\newcommand{\SolSpaceaV}[3]{{\rm V}({\vecT{#1}},\vect{#2},{#3})}
\newcommand{\coeff}[3]{\text{coeff}({#1},{#3})}

\newcommand{\SA}{\mathbb{S}}
\newcommand{\VV}{\mathbb{V}}
\newcommand{\NN}{\mathbb{N}}
\newcommand{\ZZ}{\mathbb{Z}}
\newcommand{\QQ}{\mathbb{Q}}
\newcommand{\KK}{\mathbb{K}}
\newcommand{\FF}{\mathbb{F}}
\newcommand{\EE}{\mathbb{E}}
\newcommand{\GG}{\mathbb{G}}
\newcommand{\HH}{\mathbb{H}}
\newcommand{\DD}{\mathbb{D}}
\newcommand{\sigmaSE}{$\Sigma^*$}
\newcommand{\piE}{$\Pi$}
\newcommand{\pisiSE}{$\Pi\Sigma^*$}
\newcommand{\pisiDE}{$\Pi\Sigma^{\delta}$}
\newcommand{\sigmaDE}{$\Sigma^{\delta}$}
\newcommand{\fracF}[1]{{#1}_{\text{\tiny(r)}}}
\newcommand{\vect}[1]{\boldsymbol{#1}}
\newcommand{\vecT}[1]{({#1})}

\newcommand{\vecTSS}[1]{(\begin{smallmatrix}{#1}\end{smallmatrix})}
\newcommand{\notion}[1]{{\it #1}}
\newcommand{\dfield}[2]{({#1},{#2})}
\newcommand{\const}[2]{{\rm const}_{#2}{#1}}
\newcommand{\set}[2]{{\{#1\,|\,#2\}}}
\newcommand{\setsmall}[2]{{\{#1\,|\,#2\}}}
\newcommand{\fct}[3]{{#1:#2 \to #3}}

\newtheorem{resA}{Result}

\newcommand{\res}[2]{\noindent\fbox{\hspace*{-0.05cm}
\begin{minipage}[t]{13.25cm}
\begin{resA}#1
#2
\end{resA}
\end{minipage}\hspace*{-0.05cm}}}

\newcommand{\FrameD}[1]{\noindent\fbox{\hspace*{-0.05cm}
\parbox{13.25cm}{#1}\hspace*{-0.05cm}}}

\floatstyle{ruled}
\newfloat{myalg}{ht}{lop}
\floatname{myalg}{Algorithm}
\newcounter{linectr}
\newenvironment{algor}[5]
{\footnotesize\begin{myalg}{\caption{#3}#1}\begin{minipage}{13.2cm}
\noindent{\sf
In: }{\small#4}\\[-0.0cm]
\noindent{\sf Out: }{\small#5}
\begin{list}{{\tiny\arabic{linectr}}}{\usecounter{linectr} \labelwidth3ex\itemsep0ex\labelsep0.7ex\leftmargin3.5ex\parskip-0.2cm\parsep0.05cm\listparindent0ex}\small}
{\end{list}\end{minipage}\end{myalg}\normalsize}

\begin{document}

\begin{frontmatter}

\title{A Refined Difference Field Theory for Symbolic Summation}

\thanks{Supported by the SFB-grant F1305 and the project P20347-N18 of the Austrian FWF}

\author{Carsten Schneider}
\address{Research Institute for Symbolic Computation\\
J. Kepler University Linz\\
A-4040 Linz, Austria}
\ead{Carsten.Schneider@risc.uni-linz.ac.at}

\begin{abstract}
In this article we present a refined summation theory based on Karr's difference field approach. The resulting algorithms find sum representations with optimal nested depth. For instance, the algorithms have been applied successively to evaluate Feynman integrals from Perturbative Quantum Field Theory.
\end{abstract}


\end{frontmatter}

\vspace*{-0.7cm}

\section{Introduction}

Over the past few years rapid progress has been made in the field of symbolic summation. The beginning was made by Gosper's telescoping algorithm~\cite{Gosper:78} for hypergeometric terms and Zeilberger's extension of it to creative telescoping~\cite{Zeilberger:91}. An algebraic clarification of
Gosper's setting has been carried out by Paule~\cite{Paule:95}. Meanwhile various important
variations or generalizations have been developed, like for
$q$--hypergeometric terms~\cite{PauleRiese:97}, the mixed
case~\cite{Bauer:99}, or the $\partial$-finite case~\cite{Chyzak:00}.

In particular, Karr's telescoping algorithm~\cite{Karr:81,Karr:85} based on his theory of
difference fields provides a fundamental general framework for symbolic summation. His algorithm is, in a sense, the summation counterpart to
Risch's algorithm \cite{Risch:69,Risch:70} for indefinite
integration. Karr introduced the so-called $\Pi\Sigma$-extensions,
in which parameterized first order linear difference equations can be solved in full generality; see below. As a consequence, Karr's algorithm cannot only deal with telescoping and creative telescoping over \hbox{($q$-)}hypergeometric terms, but also over rational terms consisting of arbitrarily nested sums and products. More generally, it turned out that also  parameterized linear difference equations can be solved in such difference fields~\cite{Schneider:05a}. This enables to solve recurrence relations with coefficients in terms of indefinite nested sums and products; it also gives rise to algorithms for rather general classes, for instance, holonomic sequences~\cite{Schneider:05d}.

An important general aspect of using difference field methods is the following: In order to exploit the full power of the algorithmic machinery, it is necessary to find for a summand, given in terms of indefinite nested sums and products, a ``good'' representation in a suitable \pisiSE-field extension $\EE$ of $\FF$; note that some similar considerations for indefinite integration appeared in~\cite{Singer:85}. Based on the results of~\cite{Karr:81} Karr comes to the following  somehow misleading conclusion~\cite[p.~349]{Karr:81}:

\smallskip

\begin{center}
\begin{minipage}{12.7cm}
\it Loosely speaking, if $f$ is summable in $\EE$, then part of it
is summable in $\FF$, and the rest consists of pieces whose formal
sums have been adjoined to $\FF$ in the construction of $\EE$. This
makes the construction of extension fields in which $f$ is summable
somewhat uninteresting and justifies the tendency to look for sums
of $f\in\FF$ only in $\FF$.
\end{minipage}
\end{center}

\smallskip

\noindent In other words, following Karr's point of view, one either succeeds to express a given sum of $f$ in $\FF$, or, if one fails, one adjoins the sum formally to $\FF$ which leads to a bigger field $\EE$. But, it turns out that Karr's theory of difference field extensions can be refined. Namely, as shown below, his strategy in general produces sum representations that are not optimal with respect to simplification; see, e.g., Examples~\ref{Exp:TruncatedNaive} and~\ref{Exp:DepthIncreaseExp}.

As a measure of simplification we introduce the notion of nested depth. And the main part of this article deals with the problem of finding sum representations which are optimal with respect to this property. Based on results
of~\cite{Schneider:04a,Schneider:05f,Schneider:07d} we develop a refined version of Karr's summation theory, which leads to the definition of the so called depth-optimal \pisiSE-extensions. Various important properties hold in such extensions which are relevant in symbolic summation. Moreover, an efficient telescoping algorithm which computes sum representations with optimal nested depth is presented. Throughout this article all these ideas will be illustrated by one guiding example, namely the identity
\begin{equation}\label{Equ:TruncatedSumId}
\text{\small$\displaystyle\sum_{k=1}^K\frac{\displaystyle\sum_{i=1}^kx^{i-1}\tbinom{m+i-1}{m}}{k+m}
=\left(\sum_{k=1}^K\frac{1}{k+m}\right)\left(\sum_{k=1}^Kx^{k-1}\tbinom{m+k-1}{m}\right)
-\sum_{k=1}^K\tbinom{m-k-1}{m}x^{k-1}\sum_{i=1}^{k-1}\frac{1}{m+i}$},
\end{equation}
\noindent which was needed in~\cite{Schneider:07g} to generalize identities from statistics.

We stress that our algorithms are of particular importance to simplify
d'Alembertian solutions~\cite{Abramov:94,Schneider:T01}, a subclass of Liouvillian
solutions~\cite{Singer:99}, of a given recurrence; for applications see, e.g.,~\cite{Schneider:04d,Schneider:06c,Schneider:07a,Schneider:07c}.
Furthermore, we obtain a refined version of creative telescoping which can find recurrences with smaller order; for applications see, e.g.,~\cite{Schneider:03,Schneider:T05b,Schneider:06e,Schneider:08c}.
In addition, we show how our algorithms can be used to compute efficiently algebraic relations of nested sums, like harmonic sums~\cite{Bluemlein:99,Vermaseren:99}

\vspace*{-0.2cm}

\begin{equation}\label{Equ:HSums}
S_{m_1,\dots,m_r}(n)=
\sum_{i_1=1}^n\frac{{\rm sign}(m_1)^{i_1}}{i_1^{|m_1|}}
\sum_{i_2=1}^{i_1}\frac{{\rm sign}(m_2)^{i_2}}{i_2^{|m_2|}}\dots
\sum_{i_r=1}^{i_{r-1}}\frac{{\rm sign}(m_r)^{i_r}}{i_r^{|m_r|}},
\end{equation}
$m_1,\dots,m_r\in\ZZ\setminus\{0\}$.
We illustrate by concrete examples~\cite{Schneider:07h,Schneider:07i} from Perturbative Quantum Field Theory how our algorithms can evaluate efficiently Feynman diagrams.

\smallskip

The general structure of this article is as follows. In Section~\ref{Sec:SymbSum} we
introduce the basic summation problems in difference fields. In
Section~\ref{Sec:MainResults} we present in summarized form our refined summation
theory of depth-optimal \pisiSE-extensions in which the central results are
supplemented by concrete examples. Some
first properties of depth-optimal \pisiSE-extensions are proven then in Section~\ref{Sec:MainResPart}. After considering a variation of Karr's reduction technique in Section~\ref{Sec:Reduction} we are ready to design algorithms to construct depth-optimal \pisiSE-extensions in Section~\ref{Sec:ConstructDeltaExt}. As a consequence we can prove the main results, stated in Section~\ref{Sec:MainResults}, in Section~\ref{Sec:MainProofs}. Finally, we present
applications from particle physics in Section~\ref{Sec:Application}.

\section{Refined Telescoping in
\pisiSE-extensions}\label{Sec:SymbSum}

Let $\FF$ be a \notion{difference field} with field automorphism $\fct{\sigma}{\FF}{\FF}$. Note that
$$\const{\FF}{\sigma}:=\set{c\in\FF}{\sigma(c)=c}$$
forms a subfield of $\FF$; we call
$\KK:=\const{\FF}{\sigma}$ the \notion{constant
field}\footnote{All fields have
characteristic zero. As a consequence, the constant field contains the rational numbers.} of the difference field $\dfield{\FF}{\sigma}$.
Subsequently, we consider the following two problems:
\begin{enumerate}
\item[\SR]({\it Sequence Representation}): {\bf Given} sequences $f_1(k),\dots,f_n(k)\in\KK^{\NN}$; try to {\bf construct} an appropriate difference field
$\dfield{\FF}{\sigma}$ with elements $f_1,\dots,f_n\in\FF$ where the shift-behavior $f_i(k+1)$ for $1\leq i\leq n$ is reflected by $\sigma(f_i)$.

\item[\PT] ({\it Parameterized Telescoping}): {\bf Given} $\dfield{\FF}{\sigma}$ with $\KK=\const{\FF}{\sigma}$
and $f_1,\dots,f_{n}\in\FF$; {\bf
find} all $c_1,\dots,c_n\in\KK$ and $g\in\FF$ such that
\begin{equation}\label{Equ:ParaTele}
\sigma(g)-g=c_1f_1+\dots+c_n f_n.
\end{equation}
\end{enumerate}

\noindent Then reinterpreting such a solution $g\in\FF$ with
$c_1,\dots,c_n\in\KK$ in terms of a sequence $g(k)$ gives
$$g(k+1)-g(k)=c_1 f_1(k)+\dots+c_n f_n(k)$$
which then holds in a certain range $a\leq k\leq b$. Hence, summing this
equation over $k$ gives
$$g(b+1)-g(a)=c_1\sum_{k=a}^bf_1(k)+\dots+c_n\sum_{k=a}^bf_n(k).$$
If we restrict to $n=1$ in~\eqref{Equ:ParaTele} and search for a
solution with $c_1=1$, we solve the telescoping problem: \textbf{Given} $f\in\FF$; \textbf{find}
$g\in\FF$ such that
\begin{equation}\label{Equ:Telef}
\sigma(g)-g=f.
\end{equation}
Moreover, Zeilberger's creative telescoping~\cite{Zeilberger:91} can
be formulated by translating $f(N+i-1,k)$ into $f_i\in\FF$ for a parameter $N$ which occurs in the constant field $\KK$.

\medskip

Karr's summation theory~\cite{Karr:81,Karr:85}
treats these problems in the so-called $\Pi\Sigma$-difference fields. In our work we restrict to
\pisiSE-extensions~\cite{Schneider:T01} being slightly less general but covering all sums and products treated explicitly in Karr's work.
Those fields are introduced by difference field extensions. A
difference field $\dfield{\EE}{\sigma}$ is a \notion{difference
field extension} of a difference field $\dfield{\FF}{\sigma'}$ if
$\FF$ is a subfield of $\FF$ and $\sigma(f)=\sigma'(f)$ for all
$f\in\FF$; since $\sigma$ and $\sigma'$ agree on $\FF$, we usually
do not distinguish between them anymore.

\begin{defn}
A difference field extension $\dfield{\FF(t)}{\sigma}$ of
$\dfield{\FF}{\sigma}$ is a \notion{\sigmaSE-extension}
(resp.~\notion{\piE-extension}), if $t$ is transcendental over
$\FF$, $\sigma(t)=t+a$ (resp. $\sigma(t)=a\,t$) for some $a\in\FF^*$
and $\const{\FF(t)}{\sigma}=\const{\FF}{\sigma}$; if it is clear
from the context, we say that $t$ is a \sigmaSE-extension
(resp.~a \piE-extension). A \notion{\pisiSE-extension} is either a \piE-extension or a \sigmaSE-extension.

\noindent A \notion{
\piE-extension (resp.~\sigmaSE-extension/\pisiSE-extension)}
$\dfield{\FF(t_1)\dots(t_e)}{\sigma}$ of $\dfield{\FF}{\sigma}$ is a
tower of such \piE-extensions (resp.\
\sigmaSE-extensions/\pisiSE-extensions). Such an extension if defined over $\HH$ if $\HH$ is a subfield of $\FF$
and for all $1\leq i\leq e$, $\sigma(t_i)/t_i$ or
$\sigma(t_i)-t_i$ is in $\HH(t_1,\dots,t_{i-1})$; note: $\HH(t_1,\dots,t_{i-1})=\HH$, if $i=1$.

\noindent A \pisiSE-extension $\dfield{\FF(t_1)\dots(t_e)}{\sigma}$ of
$\dfield{\FF}{\sigma}$ is called \notion{generalized d'Alembertian}, or in short \notion{polynomial}, if
$\sigma(t_i)-t_i\in\FF[t_1,\dots,t_{i-1}]$ or $\sigma(t_i)/t_i\in\FF$ for all $1\leq i\leq e$.

\noindent A \notion{\pisiSE-field $\dfield{\FF}{\sigma}$ over $\KK$}
is a \pisiSE-extension $\dfield{\FF}{\sigma}$ of
$\dfield{\KK}{\sigma}$ with $\const{\KK}{\sigma}=\KK$.
\end{defn}

\subsection{A solution of problem~\PT.}

Karr derived an
algorithm~\cite{Karr:81} that solves the following more general
problem which under the specialization $a_1=1$ and $a_2=-1$
gives~\eqref{Equ:ParaTele}.

\smallskip

\noindent\FrameD{\PFDE\ ({\it Parameterized First order Difference Equations}): {\bf Given} $\vect{0}\neq\vecT{a_1,a_2}\in\FF^2$ and
$\vect{f}=\vecT{f_1,\dots,f_n}\in\FF^n$; {\bf find} all
$\vecT{c_1,\dots,c_n,g}\in\KK^n\times\FF$ such that
\begin{equation}\label{Equ:PLDE}
a_1\,\sigma(g)+a_2\,g=c_1\,f_1+\dots+c_n\,f_n.
\end{equation}
}

\begin{rem}\label{Rem:SigmaSit}
Karr's algorithm or our simplified version~\cite{Schneider:05a} can be applied if $\dfield{\FF}{\sigma}$ is a \pisiSE-extension of $\dfield{\GG}{\sigma}$ where $\dfield{\GG}{\sigma}$ satisfies certain properties; see~\cite{Schneider:06d}. As a consequence, we obtain algorithms for problem~\PFDE\ if $\dfield{\GG}{\sigma}$ is given as follows:

\begin{description}
\item[(1)] $\KK=\const{\GG}{\sigma}$: As worked out
in~\cite[Thm.~3.2,Thm.~3.5]{Schneider:05c}, we obtain a complete algorithm, if $\KK$ is as a rational function field over an algebraic number field.

\item[(2)] The free difference field
$\dfield{\GG}{\sigma}$ with $\GG=\KK(\dots,x_{-1},x_0,x_1,\dots)$,
$\sigma(x_i)=x_{i+1}$ for $i\in\ZZ$, and $\KK$ is as in~\textbf{(1)}. In this setting
generic sequences can be treated; see~\cite{Schneider:06d,Schneider:06e}.

\item[(3)] The radical difference field
$\dfield{\KK(k)(\dots,x_{-1},x_0,x_1,\dots)}{\sigma}$ with
$\sigma(k)=k+1$ and $\sigma(x_i)=x_{i+1}$ where $x_i^2=k$;
$\KK$ is given as in~\textbf{(1)}. This allows to handle $\sqrt{k}$;  see~\cite{Schneider:07f}.

\item[(4)] $\dfield{\GG}{\sigma}$ is a \pisiSE-extension of one of the difference fields described in {\bf(1)}--{\bf(3)}.
\end{description}
\end{rem}

\subsection{A naive approach for problem~\SR.}

\noindent Sum-product expressions can be represented in \pisiSE-fields with the following result~\cite{Karr:81}.

\begin{thm}\label{Thm:PiSigma}
Let $\dfield{\FF(t)}{\sigma}$ be a difference field extension of
$\dfield{\FF}{\sigma}$ with~$\sigma(t)=at+f$.
\begin{description}
\item[(1)] $t$ a \sigmaSE-extension iff $a=1$ and there is no
$g\in\FF$ s.t.~\eqref{Equ:Telef}.
\item[(2)] $t$
is a \piE-extension iff $t\neq0$, $f=0$ and there are no $g\in\FF^*, m>0$
s.t.\ $\sigma(g)=a^m g$.
\end{description}
\end{thm}

Consequently,
we are allowed to adjoin a sum formally by a \sigmaSE-extension if and only if there does not exist a solution of the telescoping problem. The product case works similarly; for further information and problematic cases we refer
to~\cite{Schneider:05c}.

\begin{exmp}\label{Exp:TruncatedNaive}
We try to simplify the left hand side of~\eqref{Equ:TruncatedSumId} by telescoping, or equivalently, by representing~\eqref{Equ:TruncatedSumId} in a \pisiSE-field. For simplicity of representation, it will be convenient to rewrite this expression as

\vspace*{-0.7cm}
\begin{equation}\label{Equ:TruncatedSum}
\sum_{k=1}^K\frac{1}{k+m}\overbrace{\sum_{i=1}^k\frac{i}{x(m+i)}x^i\binom{m+i}{m}}^{=s(k)}.
\end{equation}
\normalsize
\begin{description}
 \item[(1)] We start with the difference field $\dfield{\QQ(x,m)}{\sigma}$ with $\sigma(c)=c$ for all $c\in\QQ(x,m)$, i.e., $\KK=\QQ(m,x)$ is the constant field. Since there is no $g\in\KK$ with $\sigma(g)-g=1$, we can define the \sigmaSE-extension $\dfield{\KK(k)}{\sigma}$ of $\dfield{\KK}{\sigma}$ with $\sigma(k)=k+1$.
\item[(2)]Since there are no $n>0$ and $g\in\KK(k)^*$ with $\sigma(g)=x^ng$ (for algorithms see~\cite{Karr:81}), we can define the
\piE-extension $\dfield{\KK(k)(q)}{\sigma}$ of $\dfield{\KK(k)}{\sigma}$ with $\sigma(q)=x q$. Similarly, we introduce the \piE-extension
$\dfield{\KK(k)(q)(b)}{\sigma}$ of $\dfield{\KK(k)(q)}{\sigma}$ with $\sigma(b)=\frac{1+m+k}{k+1}b$. By construction, $\sigma$ reflects the shift in $k$ with $S_k x^k=xx^k$ and $S_k\binom{m+k}{m}=\frac{1+m+k}{k+1}\binom{m+k}{m}$.
\item[(3)]Next, we try to simplify $s(k)$ by telescoping. Since we fail, i.e., there is no $g\in\KK(k)(q)(b)$ with $\sigma(g)-g=qb=\sigma(\frac{kqb}{(m+k)x})$, we add the \sigmaSE-extension $\dfield{\KK(k)(q)(b)(s)}{\sigma}$ of $\dfield{\KK(k)(q)(b)}{\sigma}$ with $\sigma(s)=s+qb$; note that $S_k s(k)=s(k+1)=s(k)+x^k\binom{m+k}{m}$.
\item[(4)] Finally, we look for a $g\in\KK(k)(q)(b)(s)$ such that

\vspace*{-0.3cm}

\begin{equation}\label{Equ:TruncatedTele}
\sigma(g)-g=\frac{s+q\,b}{1+k+m}.
\end{equation}
Since there is none, see Example~\ref{Exp:TruncatedNaiveRat}, we adjoin the sum~\eqref{Equ:TruncatedSum} in form of the \sigmaSE-extension $\dfield{\KK(k)(q)(b)(s)(S)}{\sigma}$ of $\dfield{\KK(k)(q)(b)(s)}{\sigma}$ with $\sigma(S)=S+\frac{s+q\,b}{1+k+m}$.
\end{description}
Summarizing, using this straight-forward approach the sum~\eqref{Equ:TruncatedSum} could not be simplified: the two nested sum-quantifier is reflected by the nested definition of $\dfield{\KK(k)(q)(b)(s)(S)}{\sigma}$.
\end{exmp}

\subsection{A refined approach for problem~\SR: The depth of nested sums and products}

Let $\dfield{\FF}{\sigma}$ be a \pisiSE-extension of
$\dfield{\GG}{\sigma}$ with $\FF:=\GG(t_1)\dots(t_e)$ where $\sigma(t_i)=a_i\,t_i$ or
$\sigma(t_i)=t_i+a_i$ for $1\leq i\leq e$. The
\notion{depth function for elements of $\FF$ over $\GG$},
$\fct{\delta_{\GG}}{\FF}{\NN_0}$, is defined as follows.
\begin{enumerate}
\item For any $g\in\GG$, $\depth{g}:=0$.

\item If $\delta_{\GG}$ is defined for
$\dfield{\GG(t_1)\dots(t_{i-1})}{\sigma}$ with $i>1$, we define
$\depthF{t_{i}}{\GG}:=\depthF{a_i}{\GG}+1$; for
$g=\frac{g_1}{g_2}\in\GG(t_1)\dots(t_i)$, with $g_1,g_2\in\GG[t_1,\dots,t_{i}]$ coprime, we define
$$\depthF{g}{\GG}:=\max(\setsmall{\depthF{t_i}{\GG}}{t_i\text{ occurs in
}g_1\text{ or }g_2}\cup\{0\}).$$
\end{enumerate}
\noindent For $\vect{f}=\vecT{f_1,\dots,f_n}\in\FF^n$, $\depthF{\vect{f}}{\GG}:=\max_{1\leq i\leq
n}\depthF{f_i}{\GG}$.
The \notion{depth of $\dfield{\FF}{\sigma}$}, in short
$\depthF{\FF}{\GG}$, is given by
$\depthF{\vecT{0,t_1,\dots,t_e}}{\GG}$.
Similarly, the \notion{extension depth} of a \pisiSE-extension
$\dfield{\FF(x_1)\dots(x_r)}{\sigma}$ of $\dfield{\FF}{\sigma}$ is $\depthF{\vecT{0,x_1,\dots,x_r}}{\GG}$. This extension is \textit{ordered} if $\depthF{x_1}{\GG}\leq\dots\leq\depthF{x_r}{\GG}$; if $\const{\FF}{\sigma}=\FF$, we call $\dfield{\FF(x_1)\dots(x_r)}{\sigma}$ an \notion{ordered} \pisiSE-field.

\begin{exmp}
In the \pisiSE-extension $\dfield{\KK(k)(q)(b)(s)(S)}{\sigma}$ of $\dfield{\KK}{\sigma}$ from Example~\ref{Exp:TruncatedNaive} the depth (function) is given by
$\depthF{k}{\KK}=1, \depthF{q}{\KK}=1, \depthF{b}{\KK}=2, \depthF{s}{\KK}=3$, and $\depthF{S}{\KK}=4$.
\end{exmp}

Throughout this article the depth is defined over the ground field
$\dfield{\GG}{\sigma}$; we set $\delta:=\delta_{\GG}$. We might use
the depth function without mentioning $\GG$. Then we assume that the corresponding difference fields are \pisiSE-extensions of
$\dfield{\GG}{\sigma}$. Moreover, note that the definition of $\delta$ depends on the particular way the extension field $\FF$ is build from $\GG$.

\begin{exmp}
We consider the sum~\eqref{Equ:TruncatedSum} and take the \pisiSE-field $\dfield{\KK(k)(q)(b)(s)}{\sigma}$ with

\vspace*{-0.3cm}

\begin{align}\label{Equ:TrunDF}
\sigma(k)=k+1,&&\sigma(q)=x\,q,&&\sigma(b)=\frac{1+m+k}{k+1}b,&&\sigma(s)=s+q\,b,
\end{align}

\vspace*{-0.15cm}

\noindent which we introduced in Example~\ref{Exp:TruncatedNaive}. Now we proceed differently: We compute the \pisiSE-extension $\dfield{\KK(k)(q)(s)(h)(H)}{\sigma}$ of $\dfield{\KK(k)(q)(s)}{\sigma}$ with

\vspace*{-0.3cm}

\begin{align}\label{Eqn:TruncatedDoubleExt}
\sigma(h)=h+\frac{1}{1+k+m},&&\sigma(H)=H-bqh
\end{align}

\vspace*{-0.15cm}

\noindent in which we find the solution $g=sh+H$ of~\eqref{Equ:TruncatedTele}; for details see Example~\ref{Exp:ConstructDepthOptExt}.
Note that $\depth{h}=2$ and $\depth{H}=3$, in particular, $\depth{g}=3$.
Reinterpreting $g$ as a sequence and checking initial values produces~\eqref{Equ:TruncatedSumId}. We emphasize that this way we have reduced the depth in~\eqref{Equ:TruncatedSumId} since $g$ and the summand $f=\frac{s+q\,b}{1+k+m}$ in~\eqref{Equ:TruncatedTele} have the same depth   $\depth{g}=\depth{f}$.
\end{exmp}

This example motivates us to consider the following refined
telescoping problem.

\smallskip

\FrameD{\noindent\DOT\ ({\it Depth Optimal Telescoping}): \textbf{Given} a
\pisiSE-extension $\dfield{\FF}{\sigma}$ of $\dfield{\GG}{\sigma}$
and $f\in\FF$; \textbf{find}, if possible, a \sigmaSE-extension
$\dfield{\EE}{\sigma}$ of $\dfield{\FF}{\sigma}$ with $g\in\EE$ s.t.~\eqref{Equ:Telef} and\footnotemark
$\depth{g}=\depth{f}$.}

\footnotetext{Note that for any \pisiSE-extension $\dfield{\EE}{\sigma}$ of $\dfield{\FF}{\sigma}$ and any $g\in\EE$ as in~\eqref{Equ:Telef} we have $\depth{g}\geq\depth{f}$. Note also that it suffices to restrict to extensions with $\depth{\EE}=\depth{f}$.}

\begin{exmp}\label{Exp:S12S21}
Our goal is to encode the harmonic sums $S_{4,2}(k)$ and $S_{2,4}(k)$ in a \pisiSE-field.
\begin{description}
\item[(1)] First we express $S_{4,2}(k)=\sum_{i=1}^k\frac{S_2(i)}{i^4}$ with the inner sum $S_2(k)=\sum_{j=1}^k\frac{1}{j^2}$ in the \pisiSE-field $\dfield{\QQ(k)(s_2)(s_{4,2})}{\sigma}$ with $\sigma(k)=k+1$, $\sigma(s_2)=s_2+\frac{1}{(k+1)^2}$ and $\sigma(s_{4,2})=s_{4,2}+\frac{\sigma(s_2)}{(k+1)^4}$, i.e., $s_2$ and $s_{4,2}$ represent $S_2(k)$ and $S_{4,2}(k)$, respectively; note that we failed to express $S_{4,2}(k)$ in an extension with depth $<\depth{s_{4,2}}=3$.
\item[(2)] Next, we consider $S_{2,4}(k)=\sum_{i=1}^k\frac{S_4(i)}{i^2}$.
We start with $S_4(k)$ and construct the \sigmaSE-extension $\dfield{\QQ(s_2)(s_{4,2})(s_4)}{\sigma}$ of $\dfield{\QQ(s_2)(s_{4,2})}{\sigma}$ with $\sigma(s_4)=s_4+\frac{1}{(k+1)^4}$. Finally, we treat the sum $S_{2,4}(k)$ and look for a $g$ such that
$\sigma(g)-g=\frac{\sigma(s_4)}{(k+1)^2}$.

\item[\textsf{The naive approach:}]
Since there is no $g\in\QQ(k)(s_2)(s_{4,2})(s_4)$, we take the \sigmaSE-extension $\dfield{\QQ(k)(s_2)(s_{4,2})(s_4)(s_{2,4})}{\sigma}$ of $\dfield{\QQ(s_2)(s_{4,2})(s_4)}{\sigma}$ with $\sigma(s_{2,4})=s_{2,4}+\frac{\sigma(s_4)}{(k+1)^2}$.

\item[\textsf{The refined approach:}] We can compute the \sigmaSE-extension $\dfield{\QQ(k)(s_2)(s_{4,2})(s_4)(s_6)}{\sigma}$ of  $\dfield{\QQ(k)(s_2)(s_{4,2})(s_4)}{\sigma}$ with $\sigma(s_6)=s_6+\frac{1}{(k+1)^6}$ in which we find the solution $g=s_6+s_2\,s_4-s_{4,2}$. Note that this alternative solution has the same depth, namely $\depth{s_{4,2}}=\depth{g}=3$, but the underlying \pisiSE-field is simpler. As result, we obtain
\begin{equation}\label{Equ:S24S42}
S_{2,4}(N)=S_6(N)+S_2(N)S_4(N)-S_{4,2}(N).
\end{equation}
\end{description}
\end{exmp}

\noindent To sum up, the following version of telescoping is relevant.

\medskip

\noindent\FrameD{\DOTX: \textbf{Given} a
\pisiSE-extension $\dfield{\FF}{\sigma}$ of $\dfield{\GG}{\sigma}$
and $f\in\FF$; \textbf{find} a \sigmaSE-extension
$\dfield{\EE}{\sigma}$ of $\dfield{\FF}{\sigma}$ with \textit{minimal} extension depth such that~\eqref{Equ:Telef} for some $g\in\EE$.}

\medskip

\noindent We shall refine Karr's theory such that we can find a common solution to~\DOT\ and~\DOTX.

\section{A refined summation theory: Depth-optimal
\pisiSE-extensions}\label{Sec:MainResults}

\begin{defn}
A difference field extension $\dfield{\FF(s)}{\sigma}$ of
$\dfield{\FF}{\sigma}$ with $\sigma(s)=s+f$ is called
\notion{depth-optimal} \sigmaSE-extension, in short
\sigmaDE-extension, if there is no \sigmaSE-extension $\dfield{\EE}{\sigma}$ of $\dfield{\FF}{\sigma}$ with extension depth $\leq\depth{f}$ and $g\in\EE$ such that~\eqref{Equ:Telef}. A \pisiSE-extension
$\dfield{\FF(t_1)\dots(t_e)}{\sigma}$ of $\dfield{\FF}{\sigma}$ is depth-optimal, in short a \pisiDE-extension, if all \sigmaSE-exten\-sions\footnote{Note that \sigmaDE-extensions are \sigmaSE-extensions by Theorem~\ref{Thm:PiSigma}.1. Note also that \piE-extensions are not refined here; this gives room for further investigations; see, e.g.,~\cite{Schneider:05c}.} are
depth-optimal. A \pisiDE-field consists of \piE- and \sigmaDE-extensions.
\end{defn}

Our main result is that problems~\SR, \DOT\ and~\DOTX\ can be solved algorithmically in \pisiDE-extensions. Moreover, we will derive various properties that are of general relevance to the field of symbolic summation and that do not hold for \pisiSE-extensions in general.

\smallskip

\noindent\textit{In all our Results~1--9, stated below and proved in Section~\ref{Sec:MainProofs}, we suppose that $\dfield{\FF}{\sigma}$ is a \pisiDE-extension of $\dfield{\GG}{\sigma}$ and $\delta=\delta_{\GG}$. From an algorithmic point of view we assume that  $\dfield{\GG}{\sigma}$ is $\sigma$-computable:}

\begin{defn}
A difference field $\dfield{\GG}{\sigma}$ is \notion{$\sigma$-computable},
if one can execute the usual polynomial arithmetic of multivariate
polynomials over $\GG$ (including factorization), and if one can
solve problem~\PFDE\ algorithmically in any \pisiSE-extension $\dfield{\FF}{\sigma}$ of $\dfield{\GG}{\sigma}$.
\end{defn}

\noindent For instance, $\dfield{\GG}{\sigma}$ can be any of the fields given in Remark~\ref{Rem:SigmaSit}. In our examples we restrict to the case $\const{\GG}{\sigma}=\GG$, i.e., $\dfield{\FF}{\sigma}$ is a \pisiSE-field over $\GG$.

\subsection{Main Results}\label{Sec:Result19}

\noindent{\bf1.~Construction.} Problem~\SR\ can be handled algorithmically in \pisiDE-extensions.

\res{\label{Res:DepthConstruct}}
{For any $f\in\FF$ there is a
\sigmaDE-extension $\dfield{\EE}{\sigma}$ of $\dfield{\FF}{\sigma}$
with $g\in\EE$ s.t.~\eqref{Equ:Telef}; if
$\dfield{\FF}{\sigma}$ is a polynomial extension of
$\dfield{\GG}{\sigma}$ and $f$ is a polynomial with coefficients from $\GG$, $\dfield{\EE}{\sigma}$ can be constructed as a polynomial extension of $\dfield{\GG}{\sigma}$ and $g$ is a polynomial with coefficients from $\GG$. If $\dfield{\GG}{\sigma}$ is
$\sigma$-computable, $\dfield{\EE}{\sigma}$ and $g$ can be given explicitly.
}

\begin{exmp}\label{Exp:ConstructDepthOptExt}
The \pisiSE-field $\dfield{\KK(k)(q)(b)(s)}{\sigma}$ with~\eqref{Equ:TrunDF} is depth-optimal since \DOT\ with $\FF=\KK(k)(q)(b)$ and $f=qb$ has no solution. Moreover, $\dfield{\KK(k)(q)(b)(s)(h)(H)}{\sigma}$ is a \sigmaDE-extension of $\dfield{\KK(k)(q)(b)(s)}{\sigma}$ with~\eqref{Eqn:TruncatedDoubleExt}. With the solution $g=sh+H$ of~\eqref{Equ:TruncatedTele} we represent the sum~\eqref{Equ:TruncatedSum} in a \pisiDE-field; for algorithmic details see Example~\ref{Exp:TruncatedFullRat}.
\end{exmp}

\noindent\textbf{2.~Reordering.} Let $\dfield{\GG(t_1)\dots(t_e)}{\sigma}$ be a \pisiSE-extension of
$\dfield{\GG}{\sigma}$ with $\sigma(t_i)=a_i\,t_i$ or
$\sigma(t_i)=t_i+a_i$ for $1\leq i\leq e$. If there is a permutation $\tau\in S_e$ with
$a_{\tau(i)}\in\GG(t_{\tau(1)})\dots(t_{\tau(i-1)})$ for all $1\leq i\leq
e$, $\dfield{\GG(t_{\tau(1)})\dots(t_{\tau(e)})}{\sigma}$ is again a \pisiSE-extension of $\dfield{\GG}{\sigma}$ and $\GG(t_1)\dots(t_e)$ is isomorphic to $\GG(t_{\tau(1)})\dots(t_{\tau(e)})$ as fields. In short, we say that
$\dfield{\GG(t_1)\dots(t_e)}{\sigma}$ can be
\notion{reordered to} $\dfield{\GG(t_{\tau(1)})\dots(t_{\tau(e)})}{\sigma}$; on the field level we identify such fields. Clearly, by definition of nested depth there is always a reordering that brings a given field to its ordered form, i.e., $\depth{t_{i-1}}\leq\depth{t_i}$ for $1\leq i\leq e$.

Note that reordering of \pisiDE-extensions without destroying depth-optimality is not so obvious: Putting \sigmaSE-extensions in front or removing them, might change the situation of problem~\DOTX. But one of our main results says that reordering indeed does not matter.

\res{\label{Res:Reordering}}
{Any possible reordering of $\dfield{\FF}{\sigma}$ is again a
\pisiDE-extension of $\dfield{\GG}{\sigma}$.}

\vspace*{-0.2cm}

\begin{exmp}\label{Exp:S12Reordered}
Let $\dfield{\QQ(k)(s_2)(s_{4,2})(s_4)(s_6)}{\sigma}$ be the \pisiDE-field from Example~\ref{Exp:S12S21}. Then, e.g., the ordered \pisiSE-field $\dfield{\QQ(k)(s_2)(s_4)(s_6)(s_{4,2})}{\sigma}$ is depth-optimal.
\end{exmp}

\noindent\textbf{3.~Depth--stability.} The following example illustrates the importance of Result~\ref{Res:DepthStable}.

\vspace*{-0.2cm}

\begin{exmp}\label{Exp:DepthIncreaseExp}
Let $\dfield{\QQ(k)(s_2)(s_{4,2})(s_4)(s_{2,4})}{\sigma}$ be the \pisiSE-field from Example~\ref{Exp:S12S21} which is not depth-optimal. We find the solution $g=s_{2,4}+s_{4,2}-s_2\,s_4$ of~\eqref{Equ:Telef} with $f:=\frac{1}{(k+1)^6}$. Hence, $3=\depth{g}>\depth{f}+1=2$. In other words, we obtained the identity $S_6(k)=S_{2,4}(k)+S_{4,2}(k)-S_2(k)S_4(k)$ where the depth is increased by telescoping; compare~\eqref{Equ:S24S42}.
\end{exmp}

\vspace*{-0.3cm}

\res{\label{Res:DepthStable}}
{For any $f,g\in\FF$ as in~\eqref{Equ:Telef}
we have
\begin{equation}\label{Equ:DepthStable}
\delta(f)\leq\delta(g)\leq\delta(f)+1.
\end{equation}
}

\smallskip

\noindent We remark that Result~\ref{Res:DepthStable} can be exploited algorithmically: In order to find all solutions of~\eqref{Equ:Telef}, one only has to take into account those extensions with depth $\leq\depth{f}+1$.

\medskip

\noindent\textbf{4.~Extension--stability.} The most crucial property is the following: Suppose we are given a \sigmaDE-extension  $\dfield{\SA}{\sigma}$ of $\dfield{\FF}{\sigma}$.
Then we can embed any \sigmaSE-extension $\dfield{\HH}{\sigma}$ of $\dfield{\FF}{\sigma}$ in a \sigmaDE-extension $\dfield{\EE}{\sigma}$ of
$\dfield{\SA}{\sigma}$ without increasing the depth.

\begin{exmp}
The \pisiSE-field $\dfield{\FF}{\sigma}$ from Example~\ref{Exp:S12Reordered} with $\FF=\QQ(k)(s_2)(s_{4,2})$ is depth-optimal, and $\dfield{\SA}{\sigma}$ with $\SA=\FF(s_6)$ is a \sigmaDE-extension of $\dfield{\FF}{\sigma}$. Now consider in addition the \pisiSE-extension $\dfield{\HH}{\sigma}$ of $\dfield{\FF}{\sigma}$ with
$\HH=\FF(s_4)(s_{2,4})$; see Example~\ref{Exp:S12S21}. Then we can take the \sigmaDE-extension $\dfield{\EE}{\sigma}$ of $\dfield{\SA}{\sigma}$ with $\EE=\SA(s_4)$, see Example~\ref{Exp:S12Reordered}, and we can define the field homomorphism $\fct{\tau}{\HH}{\EE}$ with $\tau(f)=f$ for all $f\in\FF$,
$\tau(s_4)=s_4$ and $\tau(s_{2,4})=s_6+s_2\,s_4-s_{4,2}$.
By construction, $\tau$ is injective and $\sigma_{\EE}(\tau(h))=\tau(\sigma_{\HH}(h))$ for all $h\in\HH$. In other words, we have embedded $\dfield{\HH}{\sigma}$ in $\dfield{\EE}{\sigma}$ with \eqref{Equ:DepthIneq} for all $a\in\HH$.
\end{exmp}

\noindent More precisely, $\fct{\tau}{\FF}{\FF'}$ is called a
{\it$\sigma$-mono\-morphism/$\sigma$-iso\-morphism} for $\dfield{\FF}{\sigma}$ and $\dfield{\FF'}{\sigma'}$ if $\tau$ is a field
mono\-morphism/iso\-mor\-phism  with
$\sigma'(\tau(a))=\tau(\sigma(a))$ for all $a\in\FF$.\\ Let $\dfield{\FF}{\sigma}$ and $\dfield{\FF'}{\sigma}$ be
difference field extensions of $\dfield{\HH}{\sigma}$. An
\notion{$\HH$-mono\-morphism}/\notion{$\HH$-iso\-morphism}
$\fct{\tau}{\FF}{\FF'}$ is a
$\sigma$-mo\-no\-mor\-phism/$\sigma$-iso\-mor\-phism with
$\tau(a)=a$ for all $a\in\HH$.

\smallskip

\res{\label{Res:ConstructSigmaExtOnTop}}
{Let $\dfield{\SA}{\sigma}$ be a
\sigmaDE-extension of $\dfield{\FF}{\sigma}$. Then for any
\sigmaSE-extension $\dfield{\HH}{\sigma}$ of $\dfield{\FF}{\sigma}$
with extension depth $d$ there is a \sigmaDE-extension
$\dfield{\EE}{\sigma}$ of $\dfield{\SA}{\sigma}$ with extension
depth $\leq d$ and an $\FF$-monomorphism $\fct{\tau}{\HH}{\EE}$
such that
\begin{equation}\label{Equ:DepthIneq}
\depth{\tau(a)}\leq\depth{a}
\end{equation}
for all $a\in\HH$. Such $\dfield{\SA}{\sigma}$ and $\tau$ can be constructed explicitly if $\dfield{\GG}{\sigma}$ is $\sigma$-computable.}

\medskip

\noindent\textbf{5.~Depth--optimal transformation.} By Result~\ref{Res:ExtensionDepthStable} any \pisiSE-extension of $\dfield{\FF}{\sigma}$ can be transformed to a \pisiDE-extension with the same or an improved depth-behavior. Hence the refinement to \pisiDE-extensions does not restrict the range of applications; on the contrary, the refinement to \pisiDE-extensions can lead only to better depth behavior.

\smallskip

\res{\label{Res:ExtensionDepthStable}}
{Let $\dfield{\EE}{\sigma}$ be a
\pisiSE-extension (resp.~\sigmaSE-extension) of
$\dfield{\FF}{\sigma}$. Then there is a \pisiDE-extension
(resp.~\sigmaDE-extension) $\dfield{\DD}{\sigma}$ of
$\dfield{\FF}{\sigma}$ and an $\FF$-monomorphism
$\fct{\tau}{\EE}{\DD}$ s.t.~\eqref{Equ:DepthIneq} for
all $a\in\EE$. $\dfield{\DD}{\sigma}$ and $\tau$ can be given explicitly if $\dfield{\GG}{\sigma}$ is $\sigma$-computable.
}

\medskip

\noindent\textbf{6.~Product--freeness.} \piE-extensions are irrelevant for problem~\DOT.\\
\smallskip
\res{\label{Res:ProdFree}}
{Let $f\in\FF$ and let $\dfield{\EE}{\sigma}$ be
a \pisiSE-extension of $\dfield{\FF}{\sigma}$ with $g\in\EE$ s.t.~\eqref{Equ:Telef}. Then there is a
\sigmaDE-extension $\dfield{\SA}{\sigma}$ of $\dfield{\FF}{\sigma}$ with a solution
$g'\in\SA$ of~\eqref{Equ:Telef} s.t.~$\delta(g')\leq\delta(g)$.
}

\medskip

\noindent\textbf{7.~Alternative definition.} Thus we obtain the following equivalent definition.\\
\smallskip
\res{\label{Res:ProdSumDef}}{
A \sigmaSE-extension
$\dfield{\FF(s)}{\sigma}$ of $\dfield{\FF}{\sigma}$ with
$\sigma(s)=s+f$ is depth-optimal iff there is no \pisiSE-extension $\dfield{\EE}{\sigma}$ of $\dfield{\FF}{\sigma}$ with extension depth $\leq\depth{f}$ and $g\in\EE$ s.t.~\eqref{Equ:Telef}.
}

\medskip

\noindent\textbf{8.~A common solution to~\DOT\ and~\DOTX} can be found  by Result~\ref{Res:DepthConstruct} and\\
\smallskip
\res{\label{Res:DOT}}
{Let $f\in\FF$ and let $\dfield{\EE}{\sigma}$ with $\EE=\FF(s_1)\dots(s_e)$ be
a \sigmaDE-extension of $\dfield{\FF}{\sigma}$ with extension depth $\dd$ and with $g\in\EE$ such
that~\eqref{Equ:Telef}. Then the following holds.
\begin{description}
\item[(1)] For any \pisiSE-extension
$\dfield{\HH}{\sigma}$ of $\dfield{\FF}{\sigma}$ and any solution $g'\in\HH$ of~\eqref{Equ:Telef},  $\delta(g)\leq\delta(g')$.
\item[(2)] Suppose that $\depth{s_e}=\dd$ and that $g\in\EE\setminus\FF(s_1)\dots(s_{e-1})$.  Then for any \pisiSE-extension
$\dfield{\HH}{\sigma}$ of $\dfield{\FF}{\sigma}$ with a solution $g'\in\HH$ of~\eqref{Equ:Telef} the extension depth is  $\geq\dd$.
\end{description}
}

\medskip

\noindent\textbf{9.~Refined parameterized telescoping.}\\
\smallskip
\res{\label{Res:ParaTele}}
{Let $\vect{f}\in\FF^n$ with
$\dd:=\delta(\vect{f})$. Then there is a \sigmaDE-extension
$\dfield{\EE}{\sigma}$ of $\dfield{\FF}{\sigma}$ with extension depth
$\leq\dd$ such that: For any \pisiSE-extension
$\dfield{\HH}{\sigma}$ of $\dfield{\FF}{\sigma}$ with extension depth
$\leq\dd$ and any $g\in\HH$, $\vect{c}\in\KK^n$
s.t.~\eqref{Equ:ParaTele} there is a $g'\in\EE$ s.t.\
$\sigma(g')-g'=\vect{c}\,\vect{f}$ and $\delta(g')\leq\delta(g)$.
If $\dfield{\GG}{\sigma}$ is $\sigma$-computable,
$\dfield{\EE}{\sigma}$ can be constructed explicitly.
}

\begin{exmp}\label{Equ:CreaSol}
For $F_m(k)=\binom{m}{k}\,S_1(k)$ we take
the \pisiDE-field $\dfield{\QQ(m)(k)(b)(s_1)}{\sigma}$ over $\QQ(m)$ with $\sigma(k)=k+1$, $\sigma(b)\frac{m-k}{k+1}b$ and
$\sigma(s_1)=s_1+\frac{1}{k+1}$; then $F_m(k)$ and $F_{m+1}(k)=\frac{m+1}{m-k+1}F_m(k)$ can be represented by $\vect{f}=\vecT{f_1,f_2}=\vecT{bs_1,\frac{m+1}{m-k+1}bs_1}$, respectively.
We set $\GG:=\QQ(m)(k)(b)$ and $\delta:=\delta_{\GG}$, and get $\depth{\vect{f}}=1$. Then we find the \sigmaDE-extension $\dfield{\EE}{\sigma}$ of $\dfield{\GG(s_1)}{\sigma}$ with $\EE:=\GG(s_1)(h)$, $\sigma(h)=h+\frac{b}{k+1}$ and  $\depth{h}=1$ which fulfills the properties from Result~\ref{Res:ParaTele}. In particular, we get $(c_1,c_2)=(2,-1)$ and $g=\frac{kbs_1}{m-k+1}-h$ s.t.~\eqref{Equ:ParaTele}.
Hence, for $g(k)=\frac{k}{m-k+1}\binom{m}{k}S_1(k) - \sum_{i=1}^k\frac{1}{m-i+1}\binom{m}{i}$, we get
$g(k+1)-g(k)=2 F_m(k)-F_{m+1}(k)$. Summation over $k$ from $0$ to $m$ gives
$2S(m)-S(m+1)=-\sum_{i=1}^m\frac{1}{m-i+1}\binom{m}{i}$
for

\vspace*{-0.2cm}

$$S(m)=\sum_{k=0}^mS_1(k)\binom{m}{k}.$$

\vspace*{-0.0cm}

\noindent Together with $\sum_{i=1}^m\frac{1}{m-i+1}\binom{m}{i}=\frac{-1+2^{m+1}}{m+1}$, we arrive at the recurrence relation

\vspace*{-0.2cm}

$$S(m+1)-2S(m)=-\frac{-1+2^{m+1}}{m+1}$$

\vspace*{-0.0cm}

\noindent and can read off the closed form $S(m)=2^{m}\big(S_1(m) - \sum_{i=1}^m\frac{1}{i2^i}\big).$ Note: only a recurrence of order two is produced for $S(m)$ by using standard creative telescoping; see~\cite[Sec.~2.4]{Schneider:07a}.
\end{exmp}

\noindent{\bf How to proceed.} We will prove Results~1--9 as follows. In Section~\ref{Sec:MainResPart}
we first show weaker versions of Results~\ref{Res:Reordering}--\ref{Res:ConstructSigmaExtOnTop};
there we impose that all \pisiDE-extensions are ordered.
After general preparation in Section~\ref{Sec:Reduction}, these results allow us to produce
Theorem~\ref{Thm:ExtendForTeleOrdered}
(cf.~Result~\ref{Res:DepthConstruct}) in
Section~\ref{Sec:ConstructDeltaExt}.  Given all these
properties, we will show our
Results~1--9 in full generality in Section~\ref{Sec:MainProofs}.

\subsection{Recalling basic properties of \pisiSE-extensions}

Let $\dfield{\FF}{\sigma}$ be a difference field with
$\KK=\const{\FF}{\sigma}$, $\vect{f}=\vecT{f_1,\dots,f_n}\in\FF^n$
and $p\in\FF$. We write
$\sigma(\vect{f}):=\vecT{\sigma(f_1),\dots,\sigma(f_n)}$ and
$\vect{f}\,p:=\vecT{f_1\,p,\dots,f_n\,p}$.
Let $\VV$ be a subspace of $\FF$ over $\KK$. The \notion{solution space} for
$\vect{0}\neq\vect{a}=\vecT{a_1,a_2}\in\FF^2$ and $\vect{f}$ in $\VV$ is defined by
$$\SolSpace{a}{f}{\VV}:=\set{\vecT{c_1,\dots,c_n,g}\in\KK^n\times
\VV}{\text{\eqref{Equ:PLDE} holds}}.$$ It forms a subspace of the
$\KK$-vector space $\KK^{n}\times\FF$. In particular, note that
the dimension is at most $n+1$; see~\cite{Karr:81}. If
$\vect{a}=\vecT{1,-1}$, we write in short
$$\SolSpaceT{f}{\VV}:=\SolSpaceaV{1,-1}{f}{\VV}.$$
Thus finding bases of $\SolSpaceT{f}{\VV}$ or
$\SolSpace{a}{f}{\VV}$ solves problem~\PT\ or~\PFDE, respectively.

Let $\FF(t)$ be a rational function field. For a polynomial
$p\in\FF[t]$ the degree is denoted by $\deg(p)$; we set
$\deg(0)=-1$. We define
\begin{align*}
\fracF{\FF(t)}&:=\set{\tfrac{p}{q}}{p,q\in\FF[t], \deg(p)<\deg(q)};&
\FF[t]_k&:=\set{p\in\FF[t]}{\deg(p)\leq k}
\end{align*}
for $k\geq-1$.
Let $\vect{f}=\vecT{f_1,\dots,f_n}\in\FF[t]^n$. Then
$\coeff{f_i}{t}{r}$ gives the $r$-th coefficient of $f_i\in\FF[t]$.
Moreover, we define
$\coeff{\vect{f}}{t}{r}=\vecT{\coeff{f_1}{t}{r},\dots,\coeff{f_n}{t}{r}}.$

\medskip

\noindent\textbf{Extensions and reordering.}\label{Page:Ordering}
We will exploit the following fact frequently: If
$\dfield{\GG(t)}{\sigma}$ is a \pisiSE-extension of
$\dfield{\GG}{\sigma}$ and $\dfield{\GG(t)(t_1)\dots(t_e)}{\sigma}$
is a \pisiSE-extension of $\dfield{\GG(t)}{\sigma}$ over $\GG$, we can reorder $\dfield{\GG(t)(t_1)\dots(t_e)}{\sigma}$ to the
\pisiSE-extension $\dfield{\GG(t_1)\dots(t_e)(t)}{\sigma}$ of
$\dfield{\GG}{\sigma}$.\\
We call a difference field $\dfield{\GG'(t_1)\dots(t_e)}{\sigma}$ a
\pisiSE-ex\-tension (resp.~\sigmaSE-extension/\piE-ex\-tension) of
$\dfield{\GG(t_1)\dots(t_e)}{\sigma}$ if there is a
\pisiSE-extension (resp.~\sigmaSE-extension/\piE-extension)
$\dfield{\GG(t_1)\dots(t_e)(x_1)\dots(x_r)}{\sigma}$ of
$\dfield{\GG(t_1)\dots(t_e)}{\sigma}$ over $\GG$ such that we get the
difference field $\dfield{\GG'(t_1)\dots(t_e)}{\sigma}$ by
reordering the difference field
$\dfield{\GG(t_1)\dots(t_e)(x_1)\dots(x_r)}{\sigma}$. Note that here
$\GG'$ could be $\GG(x_1)\dots(x_r)$, but additional reordering
might be possible. In a nutshell, we enlarge the ground field $\GG$
by additional \pisiSE-extensions to the field $\GG'$ where
$\dfield{\GG'(t_1)\dots(t_e)}{\sigma}$ is still a \pisiSE-extension
of $\dfield{\GG(t_1)\dots(t_e)}{\sigma}$.

\medskip

\noindent\textbf{Basic properties.}
The next properties follow from Karr's
theory~\cite{Karr:81,Karr:85}.

\begin{lem}[{\cite[Lemmas~4.1,4.2]{Karr:85}}]\label{Lemma:Karr}
Let $\dfield{\HH(x)}{\sigma}$ with $\sigma(x)=\alpha x+\beta$ be a
\pisiSE-extension of $\dfield{\HH}{\sigma}$. Let $a,f\in\HH$ and
suppose there is a solution $g\in\HH(x)$ with $\sigma(g)-a g=f$, but no solution in $\HH$. If $x$ is a \piE-extension, then $f=0$ and $a=\frac{\sigma(h)}{h}\alpha^m$ for some $h\in\HH^*$ and $m\neq0$; if $x$
is a \sigmaSE-extension, then $f\neq0$ and $a=1$.
\end{lem}

\begin{cor}\label{Cor:ExtendPiOverSigma}
Let $\dfield{\SA}{\sigma}$ be a \sigmaSE-extension of
$\dfield{\FF}{\sigma}$. If $\dfield{\FF(t)}{\sigma}$ is a
\piE-extension of $\dfield{\FF}{\sigma}$ with $\sigma(t)=a\,t$, then
$\dfield{\SA(t)}{\sigma}$ is a \piE-extension of
$\dfield{\SA}{\sigma}$ with $\sigma(t)=a\,t$.
\end{cor}
\begin{proof}
Write $\SA=\FF(s_1)\dots(s_e)$ with the \sigmaSE-extensions $s_i$. If $e=0$, nothing has to be shown. Suppose that
$\dfield{\SA(t)}{\sigma}$ is not a \piE-extension of
$\dfield{\SA}{\sigma}$. Then we find a $g\in\FF(s_1)\dots(s_e)$ with $\sigma(g)/g=a^m$ for some $m>0$ by Theorem~\ref{Thm:PiSigma}.2. By Lemma~\ref{Lemma:Karr} it follows that $g\in\FF$. Hence $\dfield{\FF(t)}{\sigma}$ is not a \piE-extension of $\dfield{\FF}{\sigma}$ by Theorem~\ref{Thm:PiSigma}.2.
\end{proof}

\begin{prop}\label{Prop:PropertiesInPiSi}
Let $\dfield{\EE}{\sigma}$ be a \pisiSE-extension of
$\dfield{\FF}{\sigma}$ with $\KK:=\const{\FF}{\sigma}$ and
$f\in\FF$.
\begin{description}
\item[(1)] If there is a $g\in\EE\setminus\FF$ such that~\eqref{Equ:Telef},
then there is no $g\in\FF$ such that~\eqref{Equ:Telef}.
\item[(2)] Let $\EE=\FF(t_1)\dots(t_e)$ with $\sigma(t_i)-t_i\in\FF$ or
$\frac{\sigma(t_i)}{t_i}\in\FF$ for $1\leq i\leq e$. If $g\in\EE$ s.t.~\eqref{Equ:Telef},
then $g=\sum_{i=1}^e c_i\,t_i+w$ where $c_i\in\KK$ and $w\in\FF$;
moreover, $c_i=0$, if $\sigma(t_i)/t_i\in\FF$.
\end{description}
\end{prop}

\begin{proof} {\bf(1)}~Assume there are such $g'\in\FF$ and
$g\in\EE\setminus\FF$. Then $\sigma(g-g')=(g-g')$, and thus
$\const{\EE}{\sigma}\neq\const{\FF}{\sigma}$, a contradiction that
$\dfield{\EE}{\sigma}$ is a \pisiSE-extension of
$\dfield{\FF}{\sigma}$. {\bf(2)}~is a special
case of~\cite[Result, page~314]{Karr:85}; see also~\cite[Thm.~4.2.1]{Schneider:T01}.
\end{proof}

\begin{prop}\label{Prop:HomProperties}
Let $\dfield{\FF}{\sigma}$, $\dfield{\FF'}{\sigma'}$ be difference
fields with a $\sigma$-isomorphism $\fct{\tau}{\FF}{\FF'}$.
\begin{description}
\item[(1)] Let $\dfield{\FF(t)}{\sigma}$ and $\dfield{\FF'(t')}{\sigma'}$
be \sigmaSE-extensions of $\dfield{\FF}{\sigma}$ and
$\dfield{\FF'}{\sigma'}$, respectively, with
$g\in\FF'(t')\setminus\FF'$ such that $\sigma'(g)-g=\tau(\sigma(t)-t)$.
Then there is a $\sigma$-isomorphism $\fct{\tau'}{\FF(t)}{\FF'(t')}$
with $\tau'(t)=g$ and $\tau'(f)=\tau(f)$ for all $f\in\FF$.

\item[(2)] Let $\dfield{\FF(t)}{\sigma}$ be a \pisiSE-extension of
$\dfield{\FF}{\sigma}$ with $\sigma(t)=\alpha\,t+\beta$. Then there is a \pisiSE-extension $\dfield{\FF'(t')}{\sigma}$ of
$\dfield{\FF'}{\sigma}$ with
$\sigma(t')=\tau(\alpha)t'+\tau(\beta)$. Moreover, there is the
$\sigma$-iso\-morphism $\fct{\tau'}{\EE}{\EE'}$ where $\tau'(t)=t'$
and $\tau'(a)=\tau(a)$ for all $a\in\FF$.

\item[(3)] Let $\dfield{\EE}{\sigma}$ be a \pisiSE-extension of
$\dfield{\FF}{\sigma}$  Then there is a \pisiSE-extension
$\dfield{\EE'}{\sigma}$ of $\dfield{\FF'}{\sigma}$ with a
$\sigma$-iso\-morphism $\fct{\tau'}{\EE}{\EE'}$ where
$\tau'(a)=\tau(a)$ for all $a\in\FF$.
\end{description}
\end{prop}

\begin{proof} {\bf(1)}~Let $\beta:=\sigma(t)-t\in\FF$, and let
$g\in\FF'(t')\setminus\FF'$ such that $\sigma'(g)-g=\tau(\beta)$. By
Proposition~\ref{Prop:PropertiesInPiSi}.2 there are a $0\neq
c\in\const{\FF}{\sigma}$ and a $w\in\FF'$ such that $g=c\,t'+w$.
Since $t'$ is transcendental over $\FF'$, also $g$ is transcendental
over $\FF'$. Therefore we can define the field isomorphism
$\fct{\tau'}{\FF(t)}{\FF'(g)}$ with $\tau'(t)=g$ and
$\tau'(f)=\tau(f)$ for all $f\in\FF$. We have
$\tau'(\sigma(t))=\tau'(t+\beta)=
g+\tau(\beta)=\sigma'(g)=\sigma'(\tau'(t))$ and thus $\tau'$ is a
$\sigma$-isomorphism. Since $\FF'(g)=\FF(t')$, the first part is proven. {\bf(2)}~Let $\dfield{\FF(t)}{\sigma}$ be a \pisiSE-extension of
$\dfield{\FF}{\sigma}$ with $\sigma(t)=\alpha\,t+\beta$. Since
$\tau$ is a $\sigma$-isomorphism, there is a \pisiSE-extension
$\dfield{\FF'(t')}{\sigma}$ of $\dfield{\FF'}{\sigma'}$ with
$\sigma(t')=\tau(\alpha)t'+\tau(\beta)$ by
Theorem~\ref{Thm:PiSigma}.1. We can construct the field isomorphism
$\fct{\tau'}{\FF(t)}{\FF'(t')}$ with $\tau'(t)=t'$ and
$\tau'(a)=\tau(a)$ for all $a\in\FF$. Since
$\sigma(\tau'(t))=\tau'(\sigma(t))$, $\tau'$ is a
$\sigma$-isomorphism. Iterative application of~{\bf(2)} shows~{\bf(3)}.
\end{proof}

\section{Preparing the stage I: Properties of ordered \pisiDE-extensions}\label{Sec:MainResPart}

We will show the first properties of depth-optimal
\pisiSE-extensions; some of the following results and proofs are simplified
and streamlined versions of~\cite{Schneider:05f}

\begin{prop}\label{Prop:TwoNestedExt}
A \pisiSE-extension $\dfield{\GG(t_1)\dots(t_e)}{\sigma}$ of
$\dfield{\GG}{\sigma}$ with $\depth{t_i}\leq2$, $\sigma(t_1)=t_1+1$
and $\const{\GG}{\sigma}=\GG$ is depth-optimal.
\end{prop}

\begin{proof} $t_1$ is depth-optimal. Suppose that $t_k$ is not depth-optimal with $2\leq k\leq e$. Set $f:=\sigma(t_k)-t_k\in\FF:=\GG(t_1)\dots(t_{k-1})$. Then there is a \pisiSE-extension
$\dfield{\FF(x_1)\dots(x_r)}{\sigma}$ of $\dfield{\FF}{\sigma}$ with
$\depth{x_i}=1$ and $g\in\FF(x_1)\dots(x_r)\setminus\FF$ such that~\eqref{Equ:Telef}. By Prop.~\ref{Prop:PropertiesInPiSi}.2,
$q_j:=\sigma(x_j)-x_j\in\GG$ for some $x_j$. Then
$\sigma(q_jt_1)-q_jt_1=q_j$;
a contradiction to Theorem~\ref{Thm:PiSigma}.1.
\end{proof}

\begin{lem}\label{Lemma:DepthStableSummation}
Let $\dfield{\EE}{\sigma}$ with $\EE=\FF(t_1)\dots(t_e)$ be an
ordered \pisiDE-extension of $\dfield{\FF}{\sigma}$ with
$\dd:=\depth{\FF}$ and $\depth{t_1}>\dd$. Let $f\in\FF$ with
$\depth{f}<\dd$. Then for any $g\in\EE$ with~\eqref{Equ:Telef}, $\depth{g}\leq\dd$.
\end{lem}
\begin{proof}
Suppose we have~\eqref{Equ:Telef} with $g\in\EE$ and
$m:=\depth{g}>\dd$. Hence $g$ depends on one of the $t_k$, i.e., let $k\geq1$ and $\HH:=\FF(t_1)\dots(t_{k-1})$ such that $g\in\HH(t_k)\setminus\HH$.
By Proposition~\ref{Prop:PropertiesInPiSi}.2, $g=c\,t_k+h$ where
$h\in\HH$, $0\neq c\in\const{\FF}{\sigma}$ and
$\beta:=\sigma(t_k)-t_k\in\HH$. Since the extension is ordered,
$\depth{t_k}=m$. By Proposition~\ref{Prop:PropertiesInPiSi}.1 there
is no $g'\in\HH$ with $\sigma(g')-g'=f$. Therefore by
Theorem~\ref{Thm:PiSigma}.1 one can construct a \sigmaSE-extension
$\dfield{\HH(s)}{\sigma}$ of $\dfield{\HH}{\sigma}$ with
$\sigma(s)=s+f$ where $\depth{s}=\depth{f}+1\leq\dd<m$. Note that
$\sigma(g')-(g')=\beta$ with $g':=(s-h)/c\in\HH(s)$. Hence $t_k$ is
not depth-optimal, a contradiction. Therefore $\depth{g}\leq\dd$.
\end{proof}

\begin{thm}[cf. Result~\ref{Res:DepthStable}]\label{Thm:DepthStableOrdered}
Let $\dfield{\FF}{\sigma}$ be an ordered \pisiDE-extension
of $\dfield{\GG}{\sigma}$ and $f\in\FF^*$. If $g\in\FF$ as in~\eqref{Equ:Telef}, then~\eqref{Equ:DepthStable}.
\end{thm}
\begin{proof}
Since $\depth{\sigma(g)}=\depth{g}\geq\depth{\sigma(g)-g}$, $\depth{g}\geq\depth{f}$.
If $\depth{\FF}=\depth{f}$, then $\depth{g}=\depth{f}$. Otherwise,
since $\dfield{\FF}{\sigma}$ is ordered, we can split $\FF$ into the
\pisiDE-extension $\dfield{\FF}{\sigma}$ of $\dfield{\HH}{\sigma}$
with $\FF=\HH(t_1)\dots(t_e)$ ($e\geq0$) and the \pisiDE-extension
$\dfield{\HH}{\sigma}$ of $\dfield{\GG}{\sigma}$ where
$\depth{t_i}>\depth{f}+1$ for each $1\leq i\leq e$ and
$\depth{\HH}=\depth{f}+1$. By
Lemma~\ref{Lemma:DepthStableSummation}, $\depth{g}\leq\depth{f}+1$.
\end{proof}

\begin{lem}\label{Lemma:ShiftSumToLeft}
Let $\dfield{\FF(t_1)\dots(t_e)}{\sigma}$ be a \pisiDE-extension of
$\dfield{\FF}{\sigma}$ and $\dfield{\FF(t_1)\dots(t_e)(x)}{\sigma}$
be a \sigmaSE-extension of $\dfield{\FF(t_1)\dots(t_e)}{\sigma}$
with $\depth{x}<\depth{t_i}$ for all $1\leq i\leq e$. By reordering,
$\dfield{\FF(x)(t_1)\dots(t_e)}{\sigma}$ is a \pisiDE-extension of
$\dfield{\FF(x)}{\sigma}$.
\end{lem}
\begin{proof}
We show the lemma by induction. If $e=0$, nothing has to be shown.
Consider $\dfield{\FF(t_1)\dots(t_{e})(x)}{\sigma}$ as claimed above
with $e>0$. Then by the induction assumption
$\dfield{\FF(t_1)(x)(t_2)\dots(t_{e})}{\sigma}$ is a
\pisiDE-extension of $\dfield{\FF(t_1)(x)}{\sigma}$. Note that
$\dfield{\FF(x)(t_1)}{\sigma}$ is a \pisiSE-extension of
$\dfield{\FF}{\sigma}$. If $t_1$ is a \piE-extension, we are done.
Otherwise, suppose that $t_1$ is a \sigmaSE-extension with
$f:=\sigma(t_1)-t_1\in\FF$ which is not depth-optimal. Then
there is a \sigmaSE-extension $\dfield{\HH}{\sigma}$ of
$\dfield{\FF(x)}{\sigma}$ with extension depth $<\depth{t_1}$
and $g\in\HH$ such that~\eqref{Equ:Telef}. Since
$\depth{x}<\depth{t_1}$, $\dfield{\HH}{\sigma}$ is a
\sigmaSE-extension of $\dfield{\FF}{\sigma}$ with extension
depth $<\depth{t_1}$. A contradiction that
$\dfield{\FF(t_1)}{\sigma}$ is a \sigmaDE-extension of
$\dfield{\FF}{\sigma}$.
\end{proof}

The following two propositions will be heavily used in Section~\ref{Sec:ConstructDeltaExt}.

\begin{prop}\label{Prop:ExtensionOnTop}
Let $\dfield{\EE}{\sigma}$ with $\EE=\FF(t_1)\dots(t_e)$ be an ordered
\pisiDE-extension of $\dfield{\FF}{\sigma}$ where $\dd:=\depth{\FF}$
and $\depth{t_i}>\dd$. Suppose that
$\dfield{\FF(x_1)\dots(x_r)}{\sigma}$ is a \sigmaSE-extension of
$\dfield{\FF}{\sigma}$ with $\beta_i:=\sigma(x_i)-x_i$ and
$\depth{x_i}\leq\dd$ for $1\leq i\leq r$. Then the following holds.
\begin{description}
\item[(1)] There is the \sigmaSE-extension
$\dfield{\EE(x_1)\dots(x_r)}{\sigma}$ of $\dfield{\EE}{\sigma}$ with
$\sigma(x_i)=x_i+\beta_i$ for all $i$ with $1\leq i\leq r$.
\item[(2)] In particular, by
reordering, we get the \pisiSE-extension
$\dfield{\FF(x_1)\dots(x_r)(t_1)\dots(t_e)}{\sigma}$ of
$\dfield{\FF}{\sigma}$;
$\dfield{\FF(x_1)\dots(x_r)(t_1)\dots(t_e)}{\sigma}$ is a
\pisiDE-extension of $\dfield{\FF(x_1)\dots(x_r)}{\sigma}$.
\end{description}
\end{prop}
\begin{proof}
{\bf(1)}~Let $i\geq1$ be minimal s.t.\
$\dfield{\EE(x_1)\dots(x_i)}{\sigma}$ is not a \sigmaSE-extension of
$\dfield{\EE}{\sigma}$. Take $g\in\EE(x_1)\dots(x_{i-1})$
s.t.~$\sigma(g)-g=\beta_{i}$. Hence $\depth{g}\leq\dd$, i.e.,
$g\in\FF(x_1)\dots(x_{i-1})$ by
Lemma~\ref{Lemma:DepthStableSummation}; this contradicts
Thm.~\ref{Thm:PiSigma}.1. Iterative
application of Lemma~\ref{Lemma:ShiftSumToLeft} proves~{\bf(2)}.
\end{proof}

\begin{prop}\label{Prop:ReorderEqualDepth}
Let $\dfield{\FF(t_1)\dots(t_e)}{\sigma}$ be a \pisiDE-extension of $\dfield{\FF}{\sigma}$ with $\depth{t_i}=\dd$ for $1\leq i\leq e$ and $\depth{\FF}\leq\dd$. For $\tau\in S_e$, $\dfield{\FF(t_{\tau(1)})\dots(t_{\tau(e)})}{\sigma}$ is a \pisiDE-extension of $\dfield{\FF}{\sigma}$.
\end{prop}

\begin{proof}
Let $e\geq1$ ($e=0$ is trivial); take $u$ with $1\leq u\leq e$ such that $\dfield{\FF(t_{\tau(1)})\dots(t_{\tau(u)})}{\sigma}$ is not a \sigmaDE-extension of $\dfield{\FF(t_{\tau(1)})\dots(t_{\tau(u-1)})}{\sigma}$ with $f:=\sigma(t_{\tau(u)})-t_{\tau(u)}\in\FF$. Choose a \sigmaSE-extension $\dfield{\FF(t_{\tau(1)})\dots(t_{\tau(u-1)})(s_1)\dots(s_r)}{\sigma}$ of $\dfield{\FF(t_{\tau(1)})\dots(t_{\tau(u-1)})}{\sigma}$ with $\depth{s_i}<\dd$ and $g\in\FF(t_{\tau(1)})\dots(t_{\tau(u-1)})(s_1)\dots(s_r)$ such that~\eqref{Equ:Telef}. Note that $\dfield{\FF(s_1)\dots(s_r)}{\sigma}$ is a \sigmaSE-extension of $\dfield{\FF}{\sigma}$ by reordering; hence by Prop.~\ref{Prop:ExtensionOnTop}.2, $\dfield{\FF(t_1)\dots(t_e)(s_1)\dots(s_r)}{\sigma}$ is a \sigmaSE-extension of $\dfield{\FF(t_1)\dots(t_e)}{\sigma}$.
Let $S=\{\tau(i)|1\leq i<u\wedge\sigma(t_{\tau(i)})-t_{\tau(i)}\in\FF\}$.
By Prop.~\ref{Prop:PropertiesInPiSi}.2, $g=\sum_{i\in S}c_it_i+h$ where $c_i\in\const{\FF}{\sigma}$ and $h\in\FF(s_1)\dots(s_r)$. Let $v\in S$ be maximal such $c_v\neq0$; if $c_i=0$ for all $i\in S$, set $v=0$.
If $v<\tau(u)$, $g\in\FF(x_1)\dots(x_r)(t_1)\dots(t_w)$ for $w:=\tau(u)-1$. Otherwise, $\sigma(g')-g'=f_v$ with $g'=\frac{1}{c_v}(t_{\tau(u)}-h-\sum_{i\in S\setminus\{v\}}c_it_i)\in\FF(x_1)\dots(x_r)(t_1)\dots(t_{w})$ for $w:=v-1$. Summarizing,
$\dfield{\FF(t_1)\dots(t_{w+1})}{\sigma}$ is not a \sigmaDE-extension of $\dfield{\FF(t_1)\dots(t_{w})}{\sigma}$, a contradiction.
\end{proof}

\begin{thm}[cf.~Res.~\ref{Res:ConstructSigmaExtOnTop}]\label{Thm:ConstructSigmaExtOnTopOrdered}
Let $\dfield{\FF}{\sigma}$ be a \pisiSE-extension of
$\dfield{\GG}{\sigma}$ and let $\dfield{\SA}{\sigma}$ be a
\sigmaSE-extension of $\dfield{\FF}{\sigma}$ which can be brought to an ordered
\pisiDE-extension of $\dfield{\GG}{\sigma}$. Then for
any \sigmaSE-extension $\dfield{\HH}{\sigma}$ of
$\dfield{\FF}{\sigma}$ with extension depth $d$ there is a
\sigmaSE-extension $\dfield{\EE}{\sigma}$ of $\dfield{\SA}{\sigma}$
with extension depth $\leq d$ and an $\FF$-monomorphism
$\fct{\tau}{\HH}{\EE}$ s.t.~\eqref{Equ:DepthIneq} for
$a\in\HH$.
\end{thm}
\begin{proof}
Let $\dfield{\DD}{\sigma}$ be an ordered \pisiDE-extension of
$\dfield{\FF}{\sigma}$ that we get by reordering the
\sigmaSE-extension $\dfield{\SA}{\sigma}$ of $\dfield{\FF}{\sigma}$.
Moreover, let $\dfield{\HH}{\sigma}$ be a \sigmaSE-extension of
$\dfield{\FF}{\sigma}$ with extension depth $d$, i.e.,
$\HH:=\FF(t_1)\dots(t_e)$. Suppose that
$\depth{t_i}\leq\depth{t_{i+1}}$, otherwise we can reorder it
without loosing any generality. We show that there is a
\sigmaSE-extension $\dfield{\EE}{\sigma}$ of $\dfield{\DD}{\sigma}$
with extension depth $\leq d$ and an $\FF$-monomorphism
$\fct{\tau}{\HH}{\EE}$ with $\depth{\tau(a)}\leq\depth{a}$ for
$a\in\HH$. Then reordering of $\dfield{\DD}{\sigma}$ proves the
statement for the extension $\dfield{\SA}{\sigma}$ of $\dfield{\FF}{\sigma}$.\\
If $e=0$, i.e., $\HH=\FF$, the statement is
proven by taking $\dfield{\EE}{\sigma}:=\dfield{\DD}{\sigma}$ with
the $\FF$-monomorphism $\fct{\tau}{\FF}{\DD}$ where $\tau(a)=a$ for
all $a\in\FF$.

Otherwise, suppose that $\HH:=\HH'(t)$ with
$\HH':=\FF(t_1)\dots(t_{e-1})$ and $\beta:=\sigma(t)-t\in\HH'$ where
$d':=\delta(t_{e-1})$ and $d:=\delta(t)\geq d'$. Moreover, assume
that there is a \sigmaSE-extension $\dfield{\EE'}{\sigma}$ of
$\dfield{\DD}{\sigma}$ with extension depth $\leq d'$ together with an
$\FF$-monomorphism $\fct{\tau}{\HH'}{\EE'}$ such that~\eqref{Equ:DepthIneq} for all $a\in\HH'$. Now define
$f:=\tau(\beta)\in\EE$. By assumption
\begin{equation}\label{Thm:SummandIsSmaller}
\depth{f}\leq\depth{\beta}<d.
\end{equation}
\textbf{Case~1:} Suppose that there is no $g\in\EE'$
as in~\eqref{Equ:Telef}. Then we can construct the \sigmaSE-extension
$\dfield{\EE'(y)}{\sigma}$ of $\dfield{\EE'}{\sigma}$ with
$\sigma(y)=y+f$ by Theorem~\ref{Thm:PiSigma}.1 and can define the
$\FF$-monomorphism $\fct{\tau'}{\HH(t)}{\EE'(y)}$ such that
$\tau'(a)=\tau(a)$ for all $a\in\HH'$ and $\tau'(t)=y$.
With~\eqref{Thm:SummandIsSmaller} we get $\depth{y}=\depth{f}+1\leq
d$. By the induction assumption, $\depth{\tau'(a)}\leq\depth{a}$ for
all $a\in\HH'(t)$. Clearly, the \sigmaSE-extension
$\dfield{\EE'(y)}{\sigma}$ of
$\dfield{\DD}{\sigma}$ has extension depth~$\leq d$.\\
\textbf{Case~2:} Suppose there is a $y\in\EE'$ with $\sigma(y)-y=f$.
Since $\dfield{\EE'}{\sigma}$ is a \sigmaSE-extension of
$\dfield{\DD}{\sigma}$ with extension depth $\leq d'$ ($d'\leq d$)
we can apply Lemma~\ref{Lemma:ShiftSumToLeft} and obtain by
reordering of $\dfield{\EE'}{\sigma}$ an ordered \pisiDE-extension
$\dfield{\GG(z_1)\dots(z_l)(x_1)\dots(x_u)}{\sigma}$ of
$\dfield{\GG(z_1)\dots(z_l)}{\sigma}$ where
$\depth{\GG(z_1)\dots(z_l)}\leq d$ and $\depth{x_j}>d$ for all
$1\leq j\leq u$. Hence with~\eqref{Thm:SummandIsSmaller} we can
apply Lemma~\ref{Lemma:DepthStableSummation} and it follows that
$\depth{y}\leq d$, i.e.,
\begin{equation}\label{Equ:ySolDepth}
\depth{y}\leq\depth{t}.
\end{equation}
Since $\tau$ is a monomorphism, there is no $g$ in the image
$\tau(\HH')$ such that~\eqref{Equ:Telef}. Since
$\dfield{\tau(\HH')(y)}{\sigma}$ is a difference field (it is a
sub-difference field of $\dfield{\EE'}{\sigma}$), $y$ is
transcendental over $\tau(\HH')$ by Theorem~\ref{Thm:PiSigma}.1. In
particular, we get the $\FF$-monomorphism
$\fct{\tau'}{\HH'(t)}{\EE'}$ with $\tau'(a)=\tau(a)$ for all
$a\in\HH'$ and $\tau'(t)=y$. With~\eqref{Equ:ySolDepth} and our
induction assumption it follows that $\depth{\tau'(a)}\leq\depth{a}$
for all $a\in\HH'(t)$. This completes the induction step.
\end{proof}

\section{Preparing the stage II: A variation of Karr's
reduction}\label{Sec:Reduction}

We modify Karr's reduction for problem~\PT: Given a \pisiSE-extension
$\dfield{\HH(t)}{\sigma}$ of $\dfield{\HH}{\sigma}$ and
$\vect{f}\in\HH(t)$; find a basis $B$ of
$\VV:=\SolSpaceT{f}{\HH(t)}$, as follows: First
split $\vect{f}\in\HH(t)^n$ by polynomial division in the form
$\vect{f}=\vect{h}+\vect{p}$ such that $\vect{h}\in\fracF{\HH(t)}^n$ and
$\vect{p}\in\HH[t]^n$; in short we write
\begin{equation}\label{Equ:Splitf}
\vect{f}=\vect{h}+\vect{p}\in\fracF{\HH(t)}^n\oplus\HH[t]^n.
\end{equation}
Then the following lemma, a direct consequence
of~\cite[Lemma~3.1]{Schneider:07d}, is crucial.

\begin{lem}\label{Lemma:SplitProblem}
Let $\dfield{\HH(t)}{\sigma}$ be a \pisiSE-extension of
$\dfield{\HH}{\sigma}$ with $\sigma(t)=\alpha\,t+\beta$
and~\eqref{Equ:Splitf}. Then $\vect{c}\in\KK^n$,
$g=\rho+g'\in\fracF{\HH(t)}\oplus\HH[t]$ solve~\eqref{Equ:ParaTele}
iff $\sigma(\rho)-\rho=\vect{c}\vect{h}$ and
$\sigma(g')-g'=\vect{c}\vect{p}$.
\end{lem}

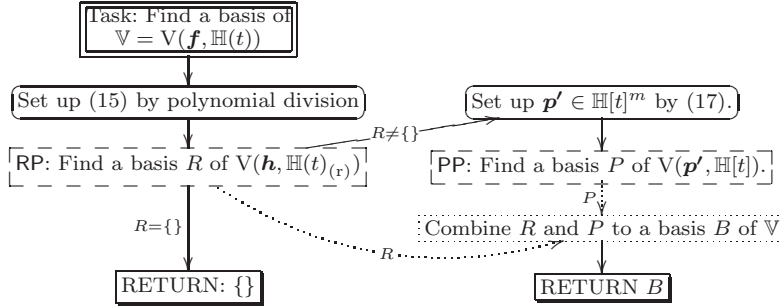
\begin{figure}[ht]
\scriptsize $$\xymatrix@C=0.7cm@R=0.4cm{
*+[F=]\txt{Task: Find a basis of\\
$\VV=\SolSpaceT{f}{\HH(t)}$}\ar[d]\\
*+[F-:<3pt>]{\txt{Set up~\eqref{Equ:Splitf} by polynomial division}}\ar[d]&*+[F-:<3pt>]{\txt{Set up $\vect{p'}\in\HH[t]^m$ by~\eqref{Equ:PolyF}.}}\ar[d]\\
*+[F--]{\txt{\RP:
Find a basis $R$ of
$\SolSpaceT{h}{\fracF{\HH(t)}}$}}\ar@{.>}|{R}@/_1.8pc/[dr]\ar[dd]_{R=\{\}}\ar[ru]|<(.3){R\neq\{\}}&*+[F--]{\txt{\PP:
Find a basis $P$ of $\SolSpaceT{p'}{\HH[t]}$.
}}\ar@{.>}_{P}[d]\\
&*+[F.]{\txt{Combine $R$ and $P$ to a basis $B$ of
$\VV$}}\ar[d]\\
*+[F-,]\txt{RETURN: $\{\}$}&*+[F-,]\txt{RETURN $B$} }$$ \normalsize
\caption{The rational reduction.}\label{Fig:RatRed}
\end{figure}

Note that we get a first strategy: Find bases for $\SolSpaceT{h}{\fracF{\HH(t)}}$ and $\SolSpaceT{p}{\HH[t]}$, and afterwards combine the solutions accordingly to get a basis of $\SolSpaceT{f}{\HH(t)}$.
As it will turn out, the following version, presented in Figure~\ref{Fig:RatRed}, is more appropriate: First solve the rational problem (\RP);
if there is no solution, there is no solution for the
original problem. Otherwise plug in the rational solutions into our
ansatz~\eqref{Equ:ParaTele} and continue to find the polynomial solutions (problem~\PP); for details see
Remark~\ref{Rem:RatDetails}.

\vspace*{0.2cm}

\begin{rem}\label{Rem:RatDetails}
Find a basis $R=\set{\vecT{d_{i1},\dots,d_{in},\rho_i}}{1\leq
i\leq m}$ of $\SolSpaceT{h}{\fracF{\HH(t)}}$; note that $m\leq n+1$. If $R=\{\}$, then
$\SolSpaceT{f}{\HH(t)}=\{\vect{0}\}$ by
Lemma~\ref{Lemma:SplitProblem}. Otherwise, define
$\vect{D}=\vecT{d_{ij}}\in\KK^{m\times n}$ and
$\vect{\rho}=\vecT{\rho_1,\dots,\rho_m}$. Then we look for all
$\vect{e}\in\KK^m$ and $g'\in\HH[t]$ such that
\begin{equation}\label{Equ:SpitProbPolyPart}
\sigma(\vect{e}\vect{\rho}+g')-(\vect{e}\vect{\rho}+g')=\vect{e}\vect{D}\vect{f}.
\end{equation}
Since $\vect{D}\vect{f}=\vect{D}\vect{h}+\vect{D}\vect{p}$ and $\sigma(\vect{e}\vect{\rho})-(\vect{e}\vect{\rho})=\vect{e}\vect{D}\vect{h}$
by construction, problem~\eqref{Equ:SpitProbPolyPart} is equivalent to looking for all
$\vect{e}\in\KK^m$ and $g'\in\HH[t]$ s.t.\
$\sigma(g')-g'=\vect{e}\vect{D}\vect{p}.$ Hence, set up
\begin{equation}\label{Equ:PolyF}
\vect{p'}:=\vect{D}\,\vect{p}
\end{equation}
where $\vect{p'}\in\HH[t]^m$ and find a basis $P=\set{\vecT{e_{i1},\dots,e_{im},g'_i}}{1\leq
i\leq l}$ of $\VV'=\SolSpaceT{p'}{\HH[t]}$.
Note that $P\neq\{\}$, since there is the trivial solution
$\sigma(1)-1=0$; define $\vect{E}=\vecT{e_{ij}}$ and
$\vect{g'}=\vecT{g'_1,\dots,g'_l}$. Then
with~\eqref{Equ:SpitProbPolyPart} it follows that
$\sigma(\vect{E}\vect{\rho}+\vect{p'})-(\vect{E}\vect{\rho}+\vect{g'})=\vect{E}\vect{D}\vect{f}.$
Thus, if we define
$\vecT{g_1,\dots,g_l}:=\vect{E}\vect{\rho}+\vect{g'}$ and
$\vecT{c_{i,j}}:=\vect{E}\vect{D}\in\KK^{l\times n}$, we get a set
of generators $B=\set{\vecT{c_{i1},\dots,c_{in},g_i}}{1\leq i\leq
l}$ that spans a subspace of $\VV:=\SolSpaceT{f}{\HH(t)}$. By simple linear algebra arguments it follows that $B$ is a basis of $\VV$.
\end{rem}

\begin{exmp}\label{Exp:TruncatedNaiveRat}
Consider the ordered \pisiDE-field $\dfield{\KK(k)(q)(b)(s)}{\sigma}$ over $\KK=\QQ(x,m)$ with~\eqref{Equ:TrunDF}
and $\vect{f}=\vecT{\frac{bq+s}{1+k+m}}$. By~\eqref{Equ:Splitf} we have $\vect{h}=\vecT{0}$ and $\vect{p}=\vect{f}$. Following the reduction of Figure~\ref{Fig:RatRed}, we need a basis $R$ of $\SolSpaceT{h}{\fracF{\KK(k)(q)(b)(s)}}$; obviously $R=\{\vecT{1,0}\}$. Thus $\vect{p'}=\vect{p}=\vect{f}$ by~\eqref{Equ:PolyF}. In Example~\ref{Exp:TruncatedNaivePoly} we will show that $P=\{\vecT{0,1}\}$ is a basis of $\SolSpaceT{f}{\KK(k)(q)(b)[s])}$. By Remark~\ref{Rem:RatDetails} a basis of $\SolSpaceT{f}{\KK(k)(q)(b)(s)}$ is $\{\vecT{0,1}\}$.
\end{exmp}

\noindent{\it Remark.} In~\cite{Karr:81} the reversed strategy was proposed: First consider the polynomial and afterwards
the rational problem. Related to problems~\DOT\ and \DOTX, the following remark
is in place. In Lemma~\ref{Lemma:RatExt} we will show that
the solutions of~\RP\ are independent of the type of extension that are needed to solve problems~\DOT,\DOTX. Thus, we will consider problem~\RP\ first. If there is no solution, we can stop; see
Corollary~\ref{Cor:PolyRatExt}. Otherwise, we will attack problems~\DOT,\DOT\ on $\vect{p'}$ which is usually simpler ($m\leq n$) than $\vect{p}$.

\medskip

\noindent The next lemma will be needed in
Section~\ref{Sec:DepthSpeedUp} to solve problems~\DOT,\DOTX\ efficiently.

\begin{lem}\label{Lemma:EntriesInPolCase}
Let $\dfield{\HH(t)}{\sigma}$ be a \pisiSE-extension of
$\dfield{\HH}{\sigma}$ and $\vect{f}\in\HH(t)^n$. If $a\in\HH$
occurs in $\vect{f}$, then the reduction in Figure~\ref{Fig:RatRed}
can be applied so that $a$ occurs in $\vect{p'}$.
\end{lem}
\begin{proof}
Let $a\in\HH$ occur in the $i$th position of $\vect{f}$. Write
$\vect{f}=\vect{h}+\vect{p}\in\fracF{\HH(t)}^n\oplus\HH[t]^n$. Then the
$i$th entry in $\vect{h}$ is zero. Hence, we can take a basis $R$
for $\SolSpaceT{h}{\fracF{\HH(t)}}$ where the $i$th unit-vector
is in $R$. Applying~\eqref{Equ:PolyF} it follows that $a$
occurs in~$\vect{p'}$.
\end{proof}

\subsection{The rational problem~\RP}\label{Sec:RatCase}

Subproblem~\RP\ has been solved in~\cite[Sections~3.4,
3.5]{Karr:81}. Alternatively, this task can be accomplished by
computing a basis $R'$ of $\VV':=\SolSpaceT{h}{\HH(t)}$ by using, e.g.,
algorithm~\cite{Schneider:05a} which is based on results from~\cite{Bron:00}. Namely,
by~\cite[Cor.~1,2]{Karr:81} it follows that
$\VV'=\SolSpaceT{h}{\fracF{\HH(t)}}\oplus(\{0\}^n\times\KK).$
Hence a basis $R$ for $\SolSpaceT{h}{\fracF{\HH(t)}}$ can be derived
by simple manipulation of the basis $R'$.\\
We remark that both approaches can be solved algorithmically if
$\dfield{\HH}{\sigma}$ is $\sigma$-computable.

\subsection{The polynomial problem~\PP}\label{Sec:PolyCase}

As in Karr's reduction~\cite{Karr:81} we bound the degree of the polynomial solutions.

\begin{lem}[\cite{Karr:81},Cor.~1,2]\label{Lemma:DegBound}
Let $\dfield{\HH(t)}{\sigma}$ be a \pisiSE-extension of
$\dfield{\HH}{\sigma}$ and let
$\vect{p'}=\vecT{p_1,\dots,p_m}\in\HH[t]^m$. Then $\SolSpaceT{p'}{\HH[t]_b}=\SolSpaceT{p'}{\HH[t]}$ for
\begin{equation}
b:=\begin{cases}\label{Equ:DegBound}
\max(\deg(p_1),\dots,\deg(p_m),1)&\text{ if $t$ is a \piE-extension}\\
\max(\deg(p_1),\dots,\deg(p_m),1)+1&\text{ if $t$ is a \sigmaSE-extension.}
\end{cases}
\end{equation}
\end{lem}

Thus we set up $r:=b\geq0$ and
$\vect{f_r}:=\vect{p'}\in\HH[t]^m_{r}$ and look for a basis $B_r$ of
$\VV_r:=\SolSpaceT{f_r}{\HH[t]_r}$.
We will accomplish this task by solving instances of problem~\PFDE\
in $\dfield{\HH}{\sigma}$; see also~\cite[Thm.~12]{Karr:81}
or~\cite[Sec.~3.3]{Schneider:05a}. Note that this is possible if $\dfield{\HH}{\sigma}$ is $\sigma$-computable.

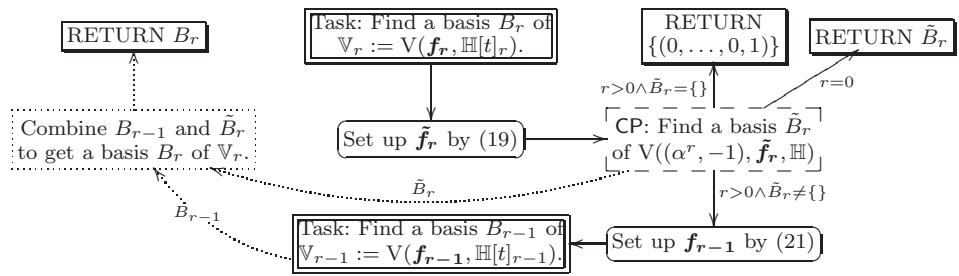
\begin{figure}[th]
\scriptsize
$$\xymatrix@R=0.6cm@C=0.5cm{
*+[F-,]\txt{RETURN $B_r$}&*+[F=]{\txt{Task: Find a basis $B_r$ of\\
$\VV_r:=\SolSpaceT{f_r}{\HH[t]_r}$.}}\ar[d]&*+[F-,]\txt{RETURN\\
$\{\vecT{0,\dots,0,1}\}$}&*+[F-,]+<-1.3cm,0cm>\txt{RETURN
$\tilde{B}_r$}\\
*+[F.]\txt{Combine $B_{r-1}$ and $\tilde{B}_r$\\
to get a basis $B_r$ of $\VV_{r}$.}\ar@{.>}[u]&*+[F-:<3pt>]\txt{Set
up $\vect{\tilde{f}_r}$
by~\eqref{Equ:Defineftilde}}\ar[r]&*+[F--]\txt{\CP:
Find a basis $\tilde{B}_r$\\of
$\SolSpaceaV{\alpha^r,-1}{\tilde{f}_r}{\HH}$}
\ar[ur]_<(0.5){r=0}\ar@{.>}@/^2pc/[ll]_{\tilde{B}_r}\ar[u]^{r>0\wedge\tilde{B}_r=\{\}}\ar[d]^{r>0\wedge\tilde{B}_r\neq\{\}}\\
&*+[F=]{\txt{Task: Find a basis $B_{r-1}$ of\\
$\VV_{r-1}:=\SolSpaceT{f_{r-1}}{\HH[t]_{r-1}}$.}}\ar@{.>}@/^2pc/[lu]|<(.55){B_{r-1}}
&*+[F-:<3pt>]\txt{Set up $\vect{f_{r-1}}$ by~\eqref{Equ:DefinefdeltaM}}\ar[l]\\
}$$ \caption{The polynomial
reduction.}\label{Fig:PolyRed}
\end{figure}

If $r=0$, we are already in the base case. Otherwise, let
$r>0$. Then we try to get all $g=\sum_{i=0}^{r}
g_it^i\in\HH[t]_{r}$ and $\vect{c}\in\KK^m$ such that
$\sigma(g)-g=\vect{c}\vect{f_{r}}$ as follows. First, we derive the
possible leading coefficients $g_{r}$ in $\dfield{\HH}{\sigma}$,
then we plug in the resulting solutions into
$\sigma(g)-g=\vect{c}\vect{f_{r}}$ and look for the remaining
$\sum_{i=0}^{r-1} g_it^i$ by recursion. The technical details are
given in Remark~\ref{Rem:PolyTechnical}, and a graphical illustration is
presented in Figure~\ref{Fig:PolyRed}: Here the task of finding a
basis of $\VV_r$ is reduced to finding a basis of the ``leading
coefficients'' (problem~\CP) and to finding a basis of the ``remaining coefficients'' $\VV_{r-1}$.

\vspace*{0.2cm}

\vspace*{-0.1cm}

\begin{rem}\label{Rem:PolyTechnical}
The main task is to find a basis $B_r$ of $\SolSpaceT{f_r}{\HH[t]_r}$. First,
define
\begin{equation}\label{Equ:Defineftilde}
\vect{\tilde{f}_{r}}:=\coeff{\vect{f_r}}{t}{r}
\end{equation}
where $\vect{\tilde{f}_{r}}\in\HH^m$. Then find a basis
$\tilde{B_r}=\{\vecT{c_{i1},\dots,c_{im},w_i}\}_{1\leq
i\leq\lambda}$ of $\SolSpaceaV{\alpha^r,-1}{\tilde{f}_{r}}{\HH}$. If
$\tilde{B}_r=\{\}$, then $\vect{c}=\vect{0}$ and $g\in\HH[t]_{r-1}$
are the only choices, i.e., $B_r=\{\vecT{0,\dots,0,1}\}$.
Otherwise, if $\tilde{B}_r\neq\{\}$, define
\begin{equation}\label{Equ:gRedVec}
\vect{C}:=\vecT{c_{ij}}\in\KK^{\lambda\times m},\quad\vect{g}:=\vecT{w_1\,t^{r},\dots,w_{\lambda}\,t^{r}}\in
t^{r}\,\HH^{\lambda},
\end{equation}
\begin{equation}\label{Equ:DefinefdeltaM}
\vect{f_{r-1}}:=\vect{C}\,\vect{f_{r}}-(\sigma(\vect{g})-\vect{g}).
\end{equation}
By construction, $\vect{f_{r-1}}\in\HH[t]_{r-1}^{\lambda}$. Now we
proceed as follows. Find all $h\in\HH[t]_{r-1}$ and
$\vect{d}\in\KK^{\lambda}$ such that
\begin{equation}\label{Equ:SubProblemProp}
\sigma(h+\vect{d}\,\vect{g})-(h+\vect{d}\,\vect{g})=\vect{d}\,\vect{C}\,\vect{f_{r}}
\end{equation}
which is equivalent to $\sigma(h)-h=\vect{d}\,\vect{f_{r-1}}.$ Therefore the subtask is to find a basis $B_{r-1}$ of $\VV_{r-1}:=\SolSpaceT{f_{r-1}}{\HH[t]_{r-1}}$;
note that $B_{r-1}\neq\{\}$, since $\sigma(1)-1=0$.

Given $B_{r-1}=\{\vecT{d_{i1},\dots,d_{i\lambda},h_i}\}_{1\leq i\leq l}$ and $\tilde{B}_r$, a basis for
$\SolSpaceT{f_r}{\HH[t]_{r}}$ can be constructed as follows. Define $\vect{D}:=\vecT{d_{ij}}\in\KK^{l\times \lambda}$,
$\vect{h}:=\vecT{h_1,\dots,h_{l}}\in\HH[t]^{l}_{r-1}$,
$\vect{E}=\vecT{e_{ij}}=\vect{C}\vect{D}\in\KK^{l\times m}$, and
$\vecT{g_1,\dots,g_l}:=\vecT{p_1,\dots,p_l}+\vect{D}\vect{q}\in\HH[t]_r^l$.
Then by~\eqref{Equ:SubProblemProp} the set
$B_r=\set{\vecT{e_{i1},\dots,e_{im},g_i}}{1\leq i\leq l}$ spans a
subspace of $\VV_r=\SolSpaceT{p}{\HH[t]}$.
By simple linear algebra arguments it follows that $B_r$ is a basis of $\VV_r$.
\end{rem}

\begin{exmp}\label{Exp:TruncatedNaivePoly}
Given the \pisiSE-field $\dfield{\HH(s)}{\sigma}$ over $\KK$ with $\HH=\KK(k)(q)(b)$ and $\vect{p'}=\vecT{\frac{bq+s}{1+k+m}}$ from Example~\ref{Exp:TruncatedNaiveRat}, we compute a basis of $\SolSpaceT{p'}{\HH[s]}$ as follows. We start the reduction of Figure~\ref{Fig:PolyRed} with $r:=2$, see~\eqref{Equ:DegBound}, and $\vect{f_2}:=\vect{p'}=\vecT{\frac{bq+s}{1+k+m}}$.
\begin{description}
\item[$r=2$:]
By~\eqref{Equ:Defineftilde} we get
$\vect{\tilde{f}_2}=\vecT{0}$; a basis of $\SolSpaceT{\tilde{f}_2}{\HH}$ is $\tilde{B}_2=\{\vecT{1,0},\vecT{0,1}\}$.

\item[$r=1$:]
We get $\vect{f_{1}}=\vecT{\tfrac{bq+s}{1+k+m},-b^2q^2-2bqs}$ by~\eqref{Equ:DefinefdeltaM} and
$\vect{\tilde{f}_1}=\vecT{\frac{1}{1+k+m},-2bq}$ by~\eqref{Equ:Defineftilde}. By another reduction in $\HH$ we compute the basis $\tilde{B}_1=\{\vecT{0,0,1}\}$ of $\SolSpaceT{\tilde{f}_1}{\HH}$.

\item[$r=0$:] $\vect{f_0}=\vecT{0}$ by~\eqref{Equ:DefinefdeltaM}. Clearly, $B_0=\{\vecT{1,0},\vecT{0,1}\}$ is a basis of $\SolSpaceT{f_0}{\HH}$.
\end{description}
Finally, we combine $\tilde{B}_2$, $\tilde{B}_1$ and $B_0$ and get the basis $B_1=\{\vecT{0,0,1}\}$, $B_2=\{\vecT{0,1}\}$ of $\SolSpaceT{f_1}{\HH[s]_1}$, $\SolSpaceT{f_2}{\HH[s]_2}$, respectively. Thus $B_2$ is a basis of $\SolSpaceT{p'}{\KK(k)(q)(b)[s]}$.\\
We \textrm{emphasize} that the summand $\frac{1}{1+k+m}$ of $h$ given in~\eqref{Eqn:TruncatedDoubleExt} occurs in $\vect{\tilde{f}_1}$. This observation is crucial for our refined summation algorithm; see Example~\ref{Exp:TruncatedFullPoly}.
\end{exmp}

\begin{rem}\label{Rem:RatPolyRed}
If $r=0$, or if $r>0$ and $t$ is a
\sigmaSE-extension ($\alpha=1$), problem~\CP\ is nothing else than
problem~\PT\ in the ground field $\dfield{\HH}{\sigma}$. Hence, we can apply again the
reductions presented in the Figures~\ref{Fig:RatRed}
and~\ref{Fig:PolyRed} to the subfield $\HH$. More precisely, if $\dfield{\GG(t_1)\dots(t_e)}{\sigma}$ is a \pisiSE-extension of a $\sigma$-computable $\dfield{\GG}{\sigma}$, we reduce problem~\PT\ to \PT\ in the fields below whenever possible, and change to the more general situation~\PFDE\ only when it is necessary. This strategy will be the basis to construct \sigmaDE-extensions in Section~\ref{Sec:ConstructDeltaExt}.
\end{rem}

\noindent The following lemmata are needed for our refined
algorithms. Lemma~\ref{Lemma:SingeExtPLDE} is immediate by construction; it is used to prove Corollary~\ref{Cor:IncremRedRef}. Lemma~\ref{Lemma:EntriesInGroundField} is crucial in
Section~\ref{Sec:DepthSpeedUp}.

\begin{lem}\label{Lemma:SingeExtPLDE}
Let $\dfield{\FF(x_1)\dots(x_e)(t)}{\sigma}$ be a \pisiSE-extension
of $\dfield{\FF}{\sigma}$ with $\sigma(t)=\alpha\,t+\beta$ where
$\alpha,\beta\in\FF$. Set $\HH:=\FF(x_1)\dots(x_e)$, let $r>0$,
$\vect{f_r}\in\FF[t]^m_r$
and $\vect{\tilde{f}_{r}}\in\FF^m$ 
with~\eqref{Equ:Defineftilde}. If the coefficients with the monomial
$t^r$ in $\VV:=\SolSpaceT{f_r}{\HH[t]_r}$ are free of the $x_i$, it
suffices to take a basis of
$\SolSpaceaV{\alpha^r,-1}{\tilde{f}_r}{\FF}$ to get a basis of $\VV$
following the reduction in Figure~\ref{Fig:PolyRed}.
\end{lem}

\begin{lem}\label{Lemma:EntriesInGroundField}
Let $\dfield{\HH(t)}{\sigma}$ be a \pisiSE-extension of
$\dfield{\HH}{\sigma}$ with $\sigma(t)=\alpha t+\beta$ and
$\vect{p'}\in\HH[t]^m$. Suppose we succeed in reducing the problem to
the base case $r=0$ with $\vect{f_0}\in\FF^{\lambda}$.
\begin{description}
\item[(1)] The reduction can be applied s.t.~all entries
of $\vect{p'}$ which are in $\FF$ also occur in $\vect{f_0}$.
\item[(2)] Moreover, if $\alpha=1$, we can guarantee that $\beta$ occurs in $\vect{f_0}$.
\end{description}
\end{lem}
\begin{proof}
{\bf(1)}~Suppose that $a\in\FF$ occurs in the $i$th position of
$\vect{p'}$. Define $b\geq0$ by~\eqref{Equ:DegBound} and set $r:=b$.
If $r=0$, nothing has to be shown. Otherwise, suppose that $r>0$.
Since $a\in\HH$, the $i$th entry in
$\vect{\tilde{f}_r}$, defined by~\eqref{Equ:Defineftilde}, is zero.
Hence we can choose a basis $\tilde{B}_r$ of
$\SolSpaceaV{\alpha^r,-1}{\tilde{f}_r}{\HH}$ where the $i$th
unit-vector is in $\tilde{B}_r$ and all other elements of
$\tilde{B}_r$ have zero in the $i$th position. With the corresponding $\vect{C}$ and~\eqref{Equ:gRedVec} we get
$\vect{f_{r-1}}$ where $a$ pops up. This construction can be done for all such entries of $\vect{p'}$ which are in $\FF$. If we continue with this refined reduction to the case $\vect{f_0}$, all $\FF$-entries of $\vect{p'}$ occur in $\vect{f_0}$.\\
{\bf(2)}~Assume that $t$ is a \sigmaSE-extension. Hence $r:=b>0$ by~\eqref{Equ:DegBound}. Suppose we are in the reduction for $r=1$. Since $\SolSpaceT{\tilde{f}_1}{\HH}$ contains $\vect{b}:=\vecT{0,\dots,0,-1}$, we can choose a basis $\tilde{B}_1$ with
$\vect{b}\in\tilde{B}_1$. By construction of $\vect{f_0}$ it follows that $\beta=\sigma(t)-t$ occurs in $\vect{f_0}$.
\end{proof}

\subsection{Key properties for refined algorithms}\label{Sec:StrategyExt}

We focus on the problem when extensions do not
contribute to solutions of~\PFDE.

\begin{lem}\label{Lemma:RatExt}
Let $\dfield{\HH(t)}{\sigma}$ be a \pisiSE-extension of
$\dfield{\HH}{\sigma}$ and
$\dfield{\HH(t)(x_1)\dots(x_e)}{\sigma}$ be a \pisiSE-ext.\ of
$\dfield{\HH(t)}{\sigma}$ over $\HH$. For
$\vect{h}\in\fracF{\HH(t)}^n$,
$\SolSpaceT{h}{\fracF{\HH(x_1)\dots(x_e)(t)}}=\SolSpaceT{h}{\fracF{\HH(t)}}.$
\end{lem}
\begin{proof}
Suppose we find an additional solution in a \pisiSE-extension
$\dfield{\HH(t)(x_1)\dots(x_e)}{\sigma}$ of
$\dfield{\HH(t)}{\sigma}$ over $\HH$, i.e., there is a
$g\in\fracF{\HH(x_1)\dots(x_{e})(t)}$ such that $g$ depends on $x_e$
and $\sigma(g)-g=\vect{c}\vect{h}$ for some
$\vect{c}\in\KK^n$. Take such a solution and define
$f:=\vect{c}\vect{h}\in\fracF{\HH(t)}$. Now reorder the extension to the
\pisiSE-extension $\dfield{\HH(t)(x_1)\dots(x_{e})}{\sigma}$ of
$\dfield{\HH}{\sigma}$. With
Proposition~\ref{Prop:PropertiesInPiSi}.2 it follows that $g=d
x_e+w$ for some $d\in\KK^*$ and $w\in\HH(t)(x_1)\dots(x_{e-1})$.
Write $w=\frac{w_1}{w_2}$ with $w_1,w_2\in\HH(x_1)\dots(x_{e})[t]$; $w_2\neq0$.
Then $g=\frac{d x_e w_2+w_1}{w_2}$. Since $w_1$ and $w_2$ are free
of $x_e$, $\deg_t(d x_e w_2+w_1)\geq\deg_t(w_2)$, i.e.,
$g\notin\fracF{\FF(x_1)\dots(x_e)(t)}$.
\end{proof}

\begin{cor}\label{Cor:PolyRatExt}
Let $\dfield{\HH(t)}{\sigma}$ be a \pisiSE-extension of
$\dfield{\HH}{\sigma}$ and let
$\dfield{\HH(t)(x_1)\dots(x_e)}{\sigma}$ be a \pisiSE-extension of
$\dfield{\HH(t)}{\sigma}$ over $\HH$. Let $\vect{f}\in\FF(t)^n$ and
take $\VV:=\SolSpaceT{f}{\HH(t)}$ and
$\VV':=\SolSpaceT{f}{\HH(t)(x_1)\dots(x_e)}$. Write
$\vect{f}=\vect{h}+\vect{p}$ as in~\eqref{Equ:Splitf}.
\begin{description}
\item[(1)] If $\SolSpaceT{h}{\fracF{\HH(t)}}=\{\}$, then
$\VV=\VV'=\{0\}^n\times\KK$.
\item[(2)] Otherwise, define $\vect{p'}\in\HH[t]^m$ and $b$
by~\eqref{Equ:PolyF} and~\eqref{Equ:DegBound}.
Then $\VV=\VV'$ iff $\SolSpaceT{p'}{\HH[t]_b}=\SolSpaceT{p'}{\HH(x_1)\dots(x_r)[t]_b}.$
If $R$, $P$ are bases of $\SolSpaceT{f}{\fracF{\HH(t)}}$ and $\SolSpaceT{f}{\HH(x_1)\dots(x_r)[t]_b}$, respectively, we get a basis of
$\SolSpaceT{f}{\HH(x_1)\dots(x_r)(t)}$ as given in
Remark~\ref{Rem:RatDetails}.
\end{description}
\end{cor}
\begin{proof}
By Lemma~\ref{Lemma:RatExt} we have
$\SolSpaceT{h}{\fracF{\HH(t)}}=\SolSpaceT{h}{\fracF{\HH(x_1)\dots(x_e)(t)}}$, i.e., $R$ is also a basis of
$\SolSpaceT{h}{\HH(x_1)\dots(x_e)(t)}$. Thus, if $R=\{\}$, then $\VV=\VV'=\{0\}^n\times\KK$; see Fig.~\ref{Fig:RatRed}. This proves~\textbf{(1)}. Otherwise, let $\vect{p'}$ and $b$ be as assumed. Note that $b$ bounds the polynomial solutions in $\HH[t]$ {\it and} $\HH(x_1)\dots(x_e)[t]$ by Lemma~\ref{Lemma:DegBound}. Hence, if
$\SolSpaceT{p'}{\HH[t]_b}=\SolSpaceT{p'}{\HH(x_1)\dots(x_e)[t]_b}$,
then by Remark~\ref{Rem:RatDetails} we get $\VV=\VV'$.
Conversely, if
$\SolSpaceT{p'}{\HH[t]_b}\subsetneq\SolSpaceT{p'}{\HH(x_1)\dots(x_e)[t]_b}$,
then $\VV\subsetneq\VV'$ by construction. This proves~\textbf{(2)}.
\end{proof}

\noindent Consequently, if one wants to find an extension with additional solutions, one has to focus on problem~\CP; see
Fig.~\ref{Fig:PolyRed}. With
Lemma~\ref{Lemma:NoExtForPi} we can refine this observation in
Corollary~\ref{Cor:IncremRedRef}.

\begin{lem}\label{Lemma:NoExtForPi}
Let $\dfield{\HH(t)}{\sigma}$ be a \piE-extension of
$\dfield{\HH}{\sigma}$ and let
$\dfield{\HH(t)(x_1)\dots(x_e)}{\sigma}$ be a \pisiSE-extension of
$\dfield{\HH(t)}{\sigma}$ over $\HH$. Set
$\alpha:=\sigma(t)/t\in\HH$. Then for $\vect{f}\in\HH$ and $r>0$,
$\SolSpaceaV{\alpha^r,-1}{f}{\HH(x_1)\dots(x_e)}=\SolSpaceaV{\alpha^r,-1}{f}{\HH}.$
\end{lem}

\begin{proof}
Suppose that
$\SolSpaceaV{\alpha^r,-1}{f}{\HH(x_1)\dots(x_e)}\supsetneq\SolSpaceaV{\alpha^r,-1}{f}{\HH}$.
Then there are $\vect{c}\in\KK^n$ and
$g\in\HH(x_1)\dots(x_j)\setminus\HH(x_1)\dots(x_{j-1})$ for some $j\geq1$ such that
$\alpha^r\sigma(g)-g=\vect{c}\vect{f}=:f$. Note: there is no
$g_0\in\HH(t)(x_1)\dots(x_{j-1})$ with $\alpha^r\sigma(g_0)-g_0=f$. (Otherwise,
$\alpha^r\sigma(g-g_0)-(g-g_0)$. Since $g\neq g_0$, we get
$\sigma(g')/g'=\alpha^{r}$ for $g'=1/(g-g_0)$. But, by reordering we
get the \piE-extension $\dfield{\HH(x_1)\dots(x_{j-1})(t)}{\sigma}$
of $\dfield{\HH(x_1)\dots(x_{j-1})}{\sigma}$, a contradiction by
Theorem~\ref{Thm:PiSigma}.2.)  Thus
we can apply Lemma~\ref{Lemma:Karr}: If $x_j$ is a
\piE-extension with $\sigma(x_j)=a\,x_j$ for some
$a\in\HH(x_1)\dots(x_{j-1})$, $\alpha^r=\frac{\sigma(h)}{h}a^m$
for some $m\neq0$ and $h\in\HH(x_1)\dots(x_{j-1})$. Hence,
$\alpha^r=\sigma(g)/g$ with $g=h\,x_j^m$. Otherwise, if $x_j$ is a
\sigmaSE-extension, $\alpha^r=1$. Summarizing,
$\sigma(g)/g=\alpha^r$ for some $g\in\HH(x_1)\dots(x_{j})$. Since $\dfield{\HH(x_1)\dots(x_{j})(t)}{\sigma}$ is a \piE-extension of
$\dfield{\HH(x_1)\dots(x_{j})}{\sigma}$, this contradicts
Theorem~\ref{Thm:PiSigma}.2.
\end{proof}

\begin{cor}\label{Cor:IncremRedRef}
Let $\dfield{\HH(t)}{\sigma}$ be a \pisiSE-extension of
$\dfield{\HH}{\sigma}$ and take a \pisiSE-extension
$\dfield{\HH(t)(x_1)\dots(x_l)(y_1)\dots(y_k)}{\sigma}$ of
$\dfield{\HH(t)}{\sigma}$ over $\HH$ where for $1\leq i\leq k$ we have $\tfrac{\sigma(y_i)}{y_i}$ or $\sigma(y_i)-y_i\in\HH(x_1)\dots(x_l)$.
Let $r>0$, $\vect{f_r}\in\HH[t]^m_r$, and set
$\VV:=\SolSpaceT{f_{r}}{\HH[t]_r}$ and $\VV':=\SolSpaceT{f_{r}}{\HH(x_1)\dots(x_l)(y_1)\dots(y_k)[t]_r}$. Define
$\vect{\tilde{f}_r}$ and $\vect{f_{r-1}}$ as
in~\eqref{Equ:Defineftilde} and~\eqref{Equ:DefinefdeltaM}.
\begin{description}
\item[(1) $\sigma(t)-t\in\FF$:]
If
$\SolSpaceT{\tilde{f}_r}{\HH(x_1)\dots(x_l)}=\SolSpaceT{\tilde{f}_r}{\HH}$
and\\
$\SolSpaceT{f_{r-1}}{\HH(x_1)\dots(x_l)(y_1)\dots(y_k)[t]_{r-1}}=\SolSpaceT{f_{r-1}}{\HH[t]_{r-1}}$, then $\VV'=\VV$.\\
Given the basis $B_{r-1}$ of
$\SolSpaceT{f_{r-1}}{\HH(x_1)\dots(x_l)(y_1)\dots(y_k)[t]_{r-1}}$ and the basis $\tilde{B}_{r}$
of $\SolSpaceT{\tilde{f}_r}{\HH(x_1)\dots(x_l)}$, one gets a basis of
$\VV'$ as stated in Remark~\ref{Rem:PolyTechnical}.

\item[(2) $\alpha:=\frac{\sigma(t)}{t}\in\FF$:]
If
$\SolSpaceT{f_{r-1}}{\HH(x_1)\dots(x_l)(y_1)\dots(y_k)[t]_{r-1}}=\SolSpaceT{f_{r-1}}{\HH[t]_{r-1}}$, then $\VV'=\VV$.
Given the basis $B_{r-1}$ of
$\SolSpaceT{f_{r-1}}{\HH(x_1)\dots(x_l)(y_1)\dots(y_k)[t]_{r-1}}$ and the basis $\tilde{B}_{r}$
of $\SolSpaceaV{\alpha^r,-1}{\tilde{f}_r}{\HH}$, one gets a basis of
$\VV'$ as stated in Remark~\ref{Rem:PolyTechnical}.
\end{description}
\end{cor}

\begin{proof}
{\bf(1)}~$t$ is a \sigmaSE-extension: By
Prop.~\ref{Prop:PropertiesInPiSi}.2 the leading coefficients with degree $r$ in the solutions of $\VV'$ are free of $y_1,\dots,y_k$. Hence, by
Lemma~\ref{Lemma:SingeExtPLDE} it suffices to take a basis of
$\SolSpaceT{\tilde{f}_r}{\HH(x_1)\dots(x_l)}$ to get a
basis of $\VV'$ following Remark~\ref{Rem:PolyTechnical}. Thus, if $\SolSpaceT{f_{r-1}}{\HH[t]_{r-1}}=\SolSpaceT{f_{r-1}}{\HH(x_1)\dots(x_l)(y_1)\dots(y_k)[t]_{r-1}}$ and $\SolSpaceT{\tilde{f}_r}{\HH(x_1)\dots(x_l)}=\SolSpaceT{\tilde{f}_r}{\HH}$, then $\VV=\VV'$ by Remark~\ref{Rem:PolyTechnical}. Given $B_{r-1}$, $\tilde{B}_{r}$ from above one gets a basis of $\VV'$.\\
{\bf(2)}~If $\sigma(t)=\alpha\,t$, then
$\SolSpaceaV{\alpha^r,-1}{\tilde{f}_r}{\HH(x_1)\dots(x_l)(y_1)\dots(y_k)}=\SolSpaceaV{\alpha^r,-1}{\tilde{f}_r}{\HH}$
by Lemma~\ref{Lemma:NoExtForPi}. Hence, if
$\SolSpaceT{f_{r-1}}{\HH(x_1)\dots(x_l)(y_1)\dots(y_k)[t]_{r-1}}=\SolSpaceT{f_{r-1}}{\HH[t]_{r-1}}$,
$\VV=\VV'$ by Remark~\ref{Rem:PolyTechnical}. Given the bases $B_{r-1}$, $\tilde{B}_{r}$ as stated above, one gets a basis of $\VV'$.
\end{proof}

\section{Constructing \pisiDE-extensions}\label{Sec:ConstructDeltaExt}

Subsequently, we prove the following theorem which will establish Result~\ref{Res:DepthConstruct} in Section~\ref{Sec:MainProofs}.

\begin{thm}\label{Thm:ExtendForTeleOrdered}
Let $\dfield{\FF}{\sigma}$ be an ordered \pisiDE-extension of
$\dfield{\GG}{\sigma}$ and $f\in\FF$. Then there is a
\sigmaSE-extension $\dfield{\EE}{\sigma}$ of $\dfield{\FF}{\sigma}$ such that $\dfield{\EE}{\sigma}$ can be brought to an ordered \pisiDE-extension of $\dfield{\GG}{\sigma}$ and such that there is a $g\in\EE$ as in~\eqref{Equ:Telef}. If
$\dfield{\GG}{\sigma}$ is $\sigma$-computable, such an
$\dfield{\EE}{\sigma}$ and $g$ can be given explicitly.
\end{thm}

\noindent In order to accomplish this task, we consider the
following more general situation.

\begin{defn}\label{Def:CompleteA}
Let $\dfield{\FF}{\sigma}$ be a \pisiSE-extension of
$\dfield{\GG}{\sigma}$ and $\vect{f}\in\FF^n$. Then
$\dfield{\FF}{\sigma}$ is \notion{$(\vect{f},\dd)$--complete}, if for
any \pisiSE-extension\footnote{
Note that for later applications we could restrict to the case that all $x_i$ are \sigmaSE-extensions.}
$\dfield{\FF(x_1)\dots(x_u)}{\sigma}$ of $\dfield{\FF}{\sigma}$ with
extension depth $\leq\dd$ we have
$\SolSpaceT{f}{\FF(x_1)\dots(x_u)}=\SolSpaceT{f}{\FF}$.
\end{defn}

\noindent That is to say, we show the following theorem.

\begin{thm}\label{Thm:AlgTheoremA}
Let $\dfield{\FF}{\sigma}$ be an ordered \pisiDE-extension of
$\dfield{\GG}{\sigma}$, $\dd\geq0$ and $\vect{f}\in\FF^n$. Then
there is a \sigmaSE-extension $\dfield{\SA}{\sigma}$ of
$\dfield{\FF}{\sigma}$ which is $(\vect{f},\dd)$--complete and which
can be brought to an ordered \pisiDE-extension of $\dfield{\GG}{\sigma}$. If
$\dfield{\GG}{\sigma}$ is $\sigma$-computable, then such an $\dfield{\SA}{\sigma}$ and a basis of $\SolSpaceT{f}{\SA}$ can be given explicitly.
\end{thm}

\noindent Then Theorem~\ref{Thm:ExtendForTeleOrdered} is implied by the
following lemma.

\begin{lem}\label{Lemma:BasicComplete}
Let $\dfield{\FF}{\sigma}$ be a \pisiSE-extension of
$\dfield{\GG}{\sigma}$ and $\vect{f}\in\FF^n$. If
$\dfield{\FF}{\sigma}$ is $(\vect{f},\depth{\vect{f}}+1)$--complete,
then $\dim\SolSpaceT{f}{\FF}=n+1$.
\end{lem}

\begin{proof}
Suppose $\dim\VV<n+1$, i.e., there is a $\vect{c}\in\KK^n$ such that there is no $g\in\FF$ with $\sigma(g)-g=\vect{c}\vect{f}=:f$. Thus
there is the \sigmaSE-extension $\dfield{\FF(s)}{\sigma}$ of
$\dfield{\FF}{\sigma}$ with $\sigma(s)=s+f$ and
$\depth{s}\leq\depth{\vect{f}}+1$. Hence $\dfield{\FF}{\sigma}$ is
not $(\vect{f},\depth{\vect{f}}+1)$--complete.
\end{proof}

\vspace*{-0.1cm}

\noindent Namely, we conclude by
Theorem~\ref{Thm:AlgTheoremA} that there is a \sigmaSE-extension
$\dfield{\SA}{\sigma}$ of $\dfield{\FF}{\sigma}$ which is
$(\vecT{f},\depth{\FF}+1)$--complete and which can be brought to an ordered \pisiDE-extension of $\dfield{\GG}{\sigma}$. Hence by
Lemma~\ref{Lemma:BasicComplete} there is a $g\in\SA$
such that~\eqref{Equ:Telef}.

In most applications one works with a \pisiSE-field over $\GG$, with $\sigma(k)=k+1$ for some $k\in\FF$. In this case, the following shortcut can be applied; the proof is similar to Prop.~\ref{Prop:TwoNestedExt}.

\vspace*{-0.1cm}

\begin{lem}\label{Lemma:ShortCut2}
Let $\dfield{\FF}{\sigma}$ be a \pisiSE-field over $\GG$ and $\vect{f}\in\FF^n$. If
$\sigma(g)-g\in\const{\GG}{\sigma}^*$ for some
$g\in\FF$, then $\dfield{\FF}{\sigma}$ is $(\vect{f},1)$--complete.
\end{lem}

\vspace*{-0.3cm}

\subsection{A constructive proof}\label{Sec:FirstAlg}

\vspace*{-0.1cm}

The proof of Theorem~\ref{Thm:AlgTheoremA} will be obtained by
refining the reduction of Section~\ref{Sec:Reduction}. Namely, let $\dd>0$, let $\dfield{\EE}{\sigma}$ with $\EE=\FF(t_1)\dots(t_e)$ be an ordered
\pisiDE-extension of $\dfield{\GG}{\sigma}$ where $e=0$ or $\depth{t_1}\geq\depth{\FF}$, and let
$\vect{f}\in\EE^n$. Then loosely speaking, we will obtain an
$(\vect{f},\dd)$--complete extension $\dfield{\SA}{\sigma}$ of
$\dfield{\EE}{\sigma}$ by constructing step by
step a tower of extensions, say
$\FF=\FF_0\leq\FF_1\leq\FF_2\leq\dots\leq\FF_l$, where
$\dfield{\FF_i}{\sigma}$ is a \sigmaDE-extension of
$\dfield{\FF_{i-1}}{\sigma}$ for $1\leq i\leq l$. Within this
construction problem~\CP\ in Figure~\ref{Fig:PolyRed} will be refined to the following subproblem: we are given a vector
$\vect{f'}$ with entries from $\FF_{i-1}$, and we have to enrich $\FF_{i-1}$
by \sigmaDE-extensions to $\FF_{i}$ such that $\FF_{i}$ becomes
$(\vect{f'},\dd-1)$-compete. Note that during this extension process it is crucial that
$\dfield{\FF_i(t_1)\dots(t_e)}{\sigma}$ forms a \pisiDE-extension
of $\dfield{\FF_i}{\sigma}$ for each $1\leq i\leq l$.
In order to get a grip on this situation, we introduce
the following definition, which reduces to
Definition~\ref{Def:CompleteB} when $e=0$.

\vspace*{-0.15cm}

\begin{defn}\label{Def:CompleteB}
Let $\dfield{\EE}{\sigma}$ be a \pisiSE-extension of
$\dfield{\FF}{\sigma}$ with $\EE=\FF(t_1)\dots(t_e)$ and
$\vect{f}\in\FF^n$. Then $\dfield{\EE}{\sigma}$ is
\notion{$(\vect{f},\dd,\FF)$--complete}, if for any \pisiSE-extension
$\dfield{\EE(x_1)\dots(x_u)}{\sigma}$ of $\dfield{\EE}{\sigma}$ over
$\FF$ with extension depth $\leq\dd$ we have
$\SolSpaceT{f}{\FF(x_1)\dots(x_u)}=\SolSpaceT{f}{\FF}$.
\end{defn}

\vspace*{-0.15cm}

\noindent Subsequently, we prove the following theorem which implies Theorems~\ref{Thm:AlgTheoremA} and~\ref{Thm:ExtendForTeleOrdered}.

\vspace*{-0.15cm}

\begin{thm}\label{Thm:AlgTheoremB}
Let $\dd\geq0$ and let $\dfield{\FF(t_1)\dots(t_e)}{\sigma}$ be an ordered
\pisiDE-extension of $\dfield{\GG}{\sigma}$; if $e>0$, $\depth{t_1}\geq\dd$. Let
$\vect{f}\in\FF^n$. Then there is a \sigmaSE-extension\footnote{For further remarks on this construction see page~\pageref{Page:Ordering}.}
$\dfield{\FF'(t_1)\dots(t_e)}{\sigma}$ of
$\dfield{\FF(t_1)\dots(t_e)}{\sigma}$ with extension depth $\leq\dd$
such that $\dfield{\FF'(t_1)\dots(t_e)}{\sigma}$ is $(\vect{f},\dd,\FF')$--complete and such that
$\dfield{\FF'(t_1)\dots(t_e)}{\sigma}$ is an ordered
\pisiDE-extension of $\dfield{\GG}{\sigma}$. If
$\dfield{\GG}{\sigma}$ is $\sigma$-computable, such an $\dfield{\FF'(t_1)\dots(t_e)}{\sigma}$ and a basis of
$\SolSpaceT{f}{\FF'(t_1)\dots(t_e)}$ can be given explicitly.
\end{thm}

\vspace*{-0.15cm}

We will show Theorem~\ref{Thm:AlgTheoremB} by induction on the depth
$\dd$. The base case $\dd=0$ is covered by Lemma~\ref{Lemma:ShortCut}.1; the proof of Lemma~\ref{Lemma:ShortCut} is immediate with Lemma~\ref{Lemma:BasicComplete}.

\vspace*{-0.15cm}

\begin{lem}\label{Lemma:ShortCut}
Let $\dfield{\EE}{\sigma}$ be a \pisiSE-extension of
$\dfield{\GG}{\sigma}$, let
$\vect{f}\in\FF^n$ and set $\VV:=\SolSpaceT{f}{\FF}$. Then the following holds.
\begin{description}
\item[(1)] $\dfield{\EE}{\sigma}$ is $(\vect{f},0,\FF)$--complete.
\item[(2)] If $\dim\VV=n+1$ (or $\dim\SolSpaceT{f}{\EE}=n+1)$, $\dfield{\EE}{\sigma}$ is $(\vect{f},i,\FF)$--complete for all $i\geq0$.
\item[(3)] $\dfield{\FF}{\sigma}$ is
$(\vect{f},\depth{\vect{f}}+1,\FF)$--complete iff $\dim\SolSpaceT{f}{\FF}=n+1$.
\end{description}
\end{lem}

In the following let $\dd>0$ and let $\dfield{\FF(t_1)\dots(t_e)}{\sigma}$ be an ordered \pisiDE-extension of $\dfield{\GG}{\sigma}$; if $e>0$,  then $\depth{t_1}\geq\dd$.

\medskip

\noindent{\it Simplification I.} Note that it suffices to restrict to the case that
$\depth{t_i}=\dd$  for $1\leq i\leq e$. Otherwise, let $r\geq0$ be maximal such that
$\depth{t_r}=\dd$. Then we show that there exists such a
\sigmaSE-extension $\dfield{\FF'(t_1)\dots(t_r)}{\sigma}$ of
$\dfield{\FF(t_1)\dots(t_r)}{\sigma}$ as required. Finally, by
Proposition~\ref{Prop:ExtensionOnTop} we get the desired
\pisiDE-extension $\dfield{\FF'(t_1)\dots(t_r)\dots(t_e)}{\sigma}$ of
$\dfield{\GG}{\sigma}$.

\smallskip

\noindent The induction step uses another induction on the number of extensions in $\FF$ with depth~$\dd$. The base case and the induction step of this ``internal induction'' are considered in Sections~\ref{Sec:CompletePhase} and~\ref{Sec:ReductionPhase}, respectively.

\subsubsection{The completion phase}\label{Sec:CompletePhase}

\medskip

The case $\delta(\FF)<\dd$ (including $\FF=\GG$) is covered by the following consideration.

\medskip

\noindent{\it Simplification II.}
We can assume that $\dd=\depth{\FF}+1$ by
Lemma~\ref{Lemma:ShortCut}: If we find such a \sigmaSE-extension
$\FF'(t_1)\dots(t_e)$ of $\dfield{\FF(t_1)\dots(t_e)}{\sigma}$ which is
$(\vect{f},\depth{\vect{\FF}}+1)$--complete, then
$\dim\SolSpaceT{f}{\FF'(t_1)\dots(t_e)}=n+1$, and thus
$\FF'(t_1)\dots(t_e)$ is $(\vect{f},i)$--complete for any $i\geq0$. With this preparation the following lemma gives the key idea.

\vspace*{-0.2cm}

\begin{lem}\label{Lemma:DepthOptbySubProb}
Let $\dd>0$ and let $\dfield{\EE}{\sigma}$ with $\EE:=\FF(t_1)\dots(t_e)$
be an ordered \pisiDE-extension of $\dfield{\FF}{\sigma}$ with
$\depth{\FF}<\dd$ and $\depth{t_i}=\dd$ for $1\leq i\leq e$. Let
$\vect{f}=\vecT{f_1,\dots,f_n}\in\FF^n$ and suppose that
$\dfield{\EE}{\sigma}$ is $(\vect{f},\dd-1)$--complete. Then any
\sigmaSE-extension $\dfield{\EE(s_1)\dots(s_r)}{\sigma}$ of
$\dfield{\EE}{\sigma}$ with $\sigma(s_i)-s_i\in\{f_1,\dots,f_n\}$ for $1\leq
i\leq r$ is depth-optimal; in particular, $\depth{s_i}=\dd$.
\end{lem}

\vspace*{-0.2cm}

\begin{proof}
Let $\dfield{\EE(s_1)\dots(s_r)}{\sigma}$ be such a
\sigmaSE-extension of $\dfield{\EE}{\sigma}$ with constant field $\KK$. First note that
$\depth{s_i}=\dd$ for all $1\leq i\leq r$: If there is an $s_v$ with $\depth{s_v}<\dd$ and $\sigma(s_v)=s_v+f_j$, then by reordering we get the \sigmaSE-extension $\dfield{\EE(s_v)}{\sigma}$ of $\dfield{\EE}{\sigma}$ with extension depth $<\dd$; a contradiction that $\dfield{\EE}{\sigma}$ is $(\vect{f},\dd-1)$--complete. Now suppose that $s_u$, $1\leq u\leq r$, is not depth-optimal with
$\sigma(s_{u})=s_u+f_j$; set $\HH:=\EE(s_1)\dots(s_{u-1})$. Then there is a \sigmaSE-extension
$\dfield{\HH(x_1)\dots(x_v)}{\sigma}$ of
$\dfield{\HH}{\sigma}$ with
$\delta(x_i)\leq\delta(f_j)<\dd$ for $1\leq i\leq v$ such that there is a
$g\in\HH(x_1)\dots(x_v)\setminus\HH$
with $\sigma(g)-g=f_j$. The \sigmaSE-extension
$\dfield{\EE(x_1)\dots(x_v)(s_1)\dots(s_{u-1})}{\sigma}$ of
$\dfield{\EE}{\sigma}$ is obtained by reordering
(note that $\depth{x_i}<\dd,\depth{s_i}=\dd$). By
Prop.~\ref{Prop:PropertiesInPiSi}.2, $g=\sum_{i=1}^{u-1}c_is_i+g'$ where $c_i\in\KK$, $g'\in\EE(x_1)\dots(x_v)\setminus\EE$. Hence
$\sigma(g')-g'=\vect{c}\vect{f}$ for some $\vect{c}\in\KK^n$, i.e.,
$\dfield{\EE}{\sigma}$ is not $(\vect{f},\dd-1)$--complete, a contradiction.
\end{proof}

Namely, by our induction assumption we apply Theorem~\ref{Thm:AlgTheoremB} and take a \sigmaSE-extension  $\dfield{\HH(t_1)\dots(t_e)}{\sigma}$
of $\dfield{\FF(t_1)\dots(t_e)}{\sigma}$ with extension depth $<\dd$ which is an ordered \pisiDE-extension $\dfield{\GG}{\sigma}$ and which is $(\vect{f},\dd-1)$--complete.
Then we adjoin step by step \sigmaSE-extensions such that for all $1\leq j\leq n$ there is a $g$ with $\sigma(g)-g=f_j$ as follows:

\small
\noindent{\tiny1}\hspace*{0.2cm} $i:=0$. FOR $1\leq j\leq n$ DO\\
{\tiny2}\hspace*{0.5cm} IF $\depth{f_j}=\dd-1$ and $\nexists g\in\HH(s_1)\dots(s_{i})$ s.t.~$\sigma(g)=g+f_j$ THEN
adjoin the \sigmaSE-extension\\
\hspace*{1cm} $\dfield{\HH(s_1)\dots(s_{i+1})}{\sigma}$ of $\dfield{\HH(s_1)\dots(s_{i})}{\sigma}$ with $\sigma(s_{i+1})=s_{i+1}+f_j$; i:=i+1. FI\\
{\tiny3}\hspace*{0.2cm} OD
\normalsize

\smallskip

\noindent Finally, we get a \sigmaSE-extension, say
$\dfield{\HH(t_1)\dots(t_e)(s_1)\dots(s_r)}{\sigma}$ of
$\dfield{\FF(t_1)\dots(t_e)}{\sigma}$; note that this extension process can be constructed explicitly if $\dfield{\GG}{\sigma}$ is $\sigma$-computable.  We complete the base case (of the internal induction) by the following arguments.

\begin{description}
\item[$\bullet$]By Lemma~\ref{Lemma:DepthOptbySubProb} $\dfield{\HH(t_1)\dots(t_e)(s_1)\dots(s_r)}{\sigma}$ is an ordered \pisiDE-extension of $\dfield{\GG}{\sigma}$. By
Prop.~\ref{Prop:ReorderEqualDepth} we get the \sigmaDE-extension
$\dfield{\HH'(t_1)\dots(t_e)}{\sigma}$ of $\dfield{\HH}{\sigma}$
with $\HH':=\HH(s_1)\dots(s_r)$. Hence $\dfield{\HH'(t_1)\dots(t_e)}{\sigma}$ is an ordered \pisiDE-extension of $\dfield{\GG}{\sigma}$.
\item[$\bullet$] Since
$\dim\SolSpaceT{f}{\HH'(t_1)\dots(t_e)}=n+1$, $\dfield{\HH'(t_1)\dots(t_e)}{\sigma}$
is $(\vect{f},\dd,\HH')$--complete by Lemma~\ref{Lemma:ShortCut}; if $\dfield{\GG}{\sigma}$ is $\sigma$-computable, a basis of $\SolSpaceT{f}{\HH'}$ can be given~explicitly.
\end{description}

\begin{exmp}\label{Exp:Truncatedh}
Take the ordered \pisiDE-field $\dfield{\FF(b)}{\sigma}$ over $\KK$ with $\FF=\KK(k)(q)$ and~\eqref{Equ:TrunDF}; let $\vect{f}=\vecT{\frac{1}{1+k+m}}$. By Lemma~\ref{Lemma:ShortCut2} $\dfield{\FF(b)}{\sigma}$ is $(\vect{f},1)$--complete. Since there is no $g\in\FF(b)$ with $\sigma(g)-g=\frac{1}{1+k+m}$, we get the \sigmaDE-extension $\dfield{\FF(b)(h)}{\sigma}$ of $\dfield{\FF(b)}{\sigma}$ with $\sigma(h)=h+\frac{1}{1+k+m}$ by Lemma~\ref{Lemma:DepthOptbySubProb}. By Proposition~\ref{Prop:ReorderEqualDepth} we obtain the ordered \pisiDE-field $\dfield{\FF(h)(b)}{\sigma}$. By construction, $\VV:=\SolSpaceT{f}{\FF(h)(b)}=\{\vecT{1,h},\vecT{0,1}\}$, i.e., $\dim\VV=2$. Therefore $\dfield{\FF(h)(b)}{\sigma}$ is $(\vect{f},2,\FF(h))$--complete. Note: Since $\dfield{\FF(b)(s)}{\sigma}$ with $\sigma(s)=s+q\,b$ is an ordered \pisiDE-field with $\depth{s}>\depth{h}$, we get the ordered \pisiDE-field $\dfield{\FF(h)(b)(s)}{\sigma}$ by Proposition~\ref{Prop:ExtensionOnTop}.
\end{exmp}

\begin{exmp}\label{Exp:TruncatedH}
Take the ordered \pisiDE-field $\dfield{\FF(s)}{\sigma}$ with $\FF=\KK(k)(q)(h)(b)$ from Example~\ref{Exp:Truncatedh}; let $\vect{f}=\vecT{-bqh,bq}$. First note that $\dfield{\FF(s)}{\sigma}$ is $(\vect{f},2)$--complete and that there is no $g\in\FF(s)$ with $\sigma(g)-g=-bqh$; see Example~\ref{Exp:TruncatedHSubCom}. Hence we can construct the \sigmaDE-extension $\dfield{\FF(s)(H)}{\sigma}$ of $\dfield{\FF(s)}{\sigma}$ with $\sigma(H)=s-bqH$; by reordering we get the ordered \pisiDE-field $\dfield{\FF(H)(s)}{\sigma}$. A basis of $\SolSpaceT{f}{\FF(H)(s)}$ is $\{\vecT{1,0,H},\vecT{0,1,s},\vecT{0,0,1}\}$.
Clearly, $\dfield{\FF(H)(s)}{\sigma}$ is $(\vect{f},3,\FF(H))$--complete.
\end{exmp}

\subsubsection{The reduction phase }\label{Sec:ReductionPhase}
We suppose that $\depth{\FF}\geq\dd>0$, i.e., $\FF=\HH(t)$ where
$\dfield{\HH(t)}{\sigma}$ is a \pisiDE-extension of
$\dfield{\HH}{\sigma}$ with $\depth{t}\geq\dd$ and
$\depth{t}\geq\depth{\HH}\geq\depth{t}-1$. As above, $\dfield{\HH(t)(t_1)\dots(t_e)}{\sigma}$ is a \pisiSE-extension of $\dfield{\HH(t)}{\sigma}$; in particular with the Simplification~I: if $e>0$, then
\begin{equation}\label{Equ:tiEqual}
\depth{t}=\depth{t_1}=\dots=\depth{t_e}.
\end{equation}
\noindent With the following definition and Corollary~\ref{Cor:PolyRatExt} we obtain Corollary~\ref{Cor:CompleteRatCrit}.

\begin{defn}
Let $\dfield{\EE}{\sigma}$ be a \pisiSE-extension of
$\dfield{\HH}{\sigma}$ with $\EE=\HH(t)(t_1)\dots(t_e)$ and
$\vect{f}\in\HH[t]^n_r$. Then $\dfield{\EE}{\sigma}$ is
\notion{$(\vect{f},\dd,\HH[t]_r)$--complete}, if for any
\pisiSE-extension $\dfield{\EE(x_1)\dots(x_u)}{\sigma}$ of
$\dfield{\EE}{\sigma}$ over $\FF$ with extension depth $\leq\dd$ we have
$\SolSpaceT{f}{\HH(x_1)\dots(x_u)[t]_r}=\SolSpaceT{f}{\HH[t]_r}$.
\end{defn}

\begin{cor}\label{Cor:CompleteRatCrit}
Let $\dd>0$ and let $\dfield{\HH(t)(t_1)\dots(t_e)}{\sigma}$ be a
\pisiSE-extension of $\dfield{\HH}{\sigma}$ with $\delta(t)\geq\dd$; if $e>0$, then~\eqref{Equ:tiEqual}.
Let $\vect{f}\in\HH(t)^n$, and define $\vect{h}$ and $\vect{p}$
by~\eqref{Equ:Splitf}. Let $R$ be a basis of
$\SolSpaceT{h}{\fracF{\HH(t)}}$ and set $\VV:=\SolSpaceT{f}{\HH(t)}$.
\begin{description}
\item[(1)] If $R=\{\}$, $\dfield{\HH(t)(t_1)\dots(t_e)}{\sigma}$ is $(\vect{f},\HH(t),\dd)$--complete;  $\VV=\{0\}^n\times\KK$. Otherwise:

\item[(2)] Take $\vect{p'}\in\HH[t]^m,b$ by~\eqref{Equ:PolyF},~\eqref{Equ:DegBound}. If
$\dfield{\HH'(t)(t_1)\dots(t_e)}{\sigma}$ is a \pisiSE-extension of
$\dfield{\HH(t_1)\dots(t_e)}{\sigma}$ that is
$(\vect{p'},\HH'[t]_b,\dd)$-complete,
$\dfield{\HH'(t)(t_1)\dots(t_e)}{\sigma}$ is
$(\vect{f},\HH'(t),\dd)$-complete. If $P$ is a basis of
$\SolSpaceT{f}{\HH'[t]_b}$, a basis of
$\VV$ can be constructed by Remark~\ref{Rem:RatDetails}.
\end{description}
\end{cor}

\begin{exmp}\label{Exp:TruncatedFullRat}
Consider the ordered \pisiDE-field $\dfield{\HH(s)}{\sigma}$ with $\HH:=\KK(k)(q)(b)$ and~\eqref{Equ:TrunDF}; let $\vect{f}=\vecT{\frac{bq+s}{1+k+m}}$. As in  Example~\ref{Exp:TruncatedNaiveRat} we get $\vect{h}=\vecT{0}$, $\vect{p}=\vect{f}$, $R=\{\vecT{1,0},\vecT{0,1}\}$, $\vect{p'}=\vect{f}$ and $r=2$. In Examples~\ref{Exp:TruncatedFullPoly} and~\ref{Exp:TruncatedFullPoly2} we will construct the \sigmaSE-extension $\dfield{\HH'(s)}{\sigma}$ of
$\dfield{\HH(s)}{\sigma}$ with $\HH':=\KK(k)(q)(h)(b)(H)$ and~\eqref{Eqn:TruncatedDoubleExt} where $\dfield{\HH'(s)}{\sigma}$ is an ordered \pisiDE-field which is $(\vect{f},3,\HH'[s]_2)$--complete; we obtain the basis $P=\{\vecT{1,sh+H},\vecT{0,1}\}$ of $\SolSpaceT{p'}{\HH'[s]_2)}$. Hence $\dfield{\HH'(s)}{\sigma}$ is $(\vect{f},3,\HH')$-com\-plete by Corollary~\ref{Cor:CompleteRatCrit}; a basis of $\SolSpaceT{f}{\HH'(s)}$ is $P$.
\end{exmp}

\noindent Let $r:=b>0$ and set $\HH_r:=\HH$ and
$\vect{f_r}:=\vect{p'}\in\HH[t]^m_r$. Define $\vect{\tilde{f}_r}$ as
in~\eqref{Equ:Defineftilde}. If $t$ is a \piE-extension, we can apply
Corollary~\ref{Cor:MakeComplete} (see below) and obtain the reduction $r\to r-1$.

Otherwise, if $t$ is a \sigmaSE-extension, the following
preprocessing step is necessary. By the induction assumption we can apply Theorem~\ref{Thm:AlgTheoremB} and get a \sigmaSE-extension of
$\dfield{\HH_{r}}{\sigma}$ with extension depth $<\dd$ which can be brought to an ordered \pisiDE-extension
$\dfield{\HH_{r-1}}{\sigma}$ of $\dfield{\GG}{\sigma}$ and which is $(\vect{\tilde{f}_r},\dd-1,\HH_{r-1})$--complete. By
Proposition~\ref{Prop:ExtensionOnTop} we can adjoin the extensions $t_i$ on top and get the ordered \pisiDE-extension $\dfield{\HH_{r-1}(t_1)\dots(t_e)}{\sigma}$ of $\dfield{\GG}{\sigma}$. Now we are ready apply
Corollary~\ref{Cor:MakeComplete} ($\HH_{r-1}$ is replaced by $\HH$) and proceed with the reduction $r\to r-1$.

\begin{cor}\label{Cor:MakeComplete}
Let $\dd>0$ and let $\dfield{\HH(t)(t_1)\dots(t_e)}{\sigma}$ be a \pisiSE-extension of $\dfield{\GG}{\sigma}$ with $\delta(t)\geq\dd$; if $e>0$, then~\eqref{Equ:tiEqual}.
Let $\dfield{\HH'(t)(t_1)\dots(t_e)}{\sigma}$ be a \pisiSE-extension of $\dfield{\HH(t)(t_1)\dots(t_e)}{\sigma}$ with extension depth $\leq\dd$.
Let $r>0$ and
$\vect{f_r}\in\HH[t]^m_r$, and define $\vect{\tilde{f}_r}$ and
$\vect{f_{r-1}}$ as in~\eqref{Equ:Defineftilde}
and~\eqref{Equ:DefinefdeltaM}.
\begin{description}
\item[(1) $\sigma(t)-t\in\FF$:] If $\dfield{\HH}{\sigma}$ is
$(\vect{\tilde{f}_r},\dd-1)$--complete and in addition $\dfield{\HH'(t)(t_1)\dots(t_e)}{\sigma}$ is
$(\vect{f_{r-1}},\HH'[t]_{r-1},\dd)$--complete, then
$\dfield{\HH'(t)(t_1)\dots(t_e)}{\sigma}$ is
$(\vect{f_r},\HH[t]_r,\dd)$--complete.\\
If $\tilde{B}_r$ and $B_{r-1}$ are bases of $\SolSpaceT{\tilde{f}_r}{\HH}$ and
$\SolSpaceT{f_{r-1}}{\HH'[t]_{r-1}}$, respectively, we get
a basis of $\SolSpaceT{f_r}{\HH'[t]_r}$ following
Remark~\ref{Rem:PolyTechnical}.
\item[(2) $\alpha:=\tfrac{\sigma(t)}{t}\in\FF$:]
If $\dfield{\HH'(t)(t_1)\dots(t_e)}{\sigma}$ is
$(\vect{f_{r-1}},\HH'[t]_{r-1},\dd)$-complete, then it follows that
$\dfield{\HH'(t)(t_1)\dots(t_e)}{\sigma}$ is
$(\vect{f_r},\HH'[t]_r,\dd)$--complete.
If $\tilde{B}_r$ and $B_{r-1}$ are bases of the solution spaces $\SolSpaceaV{\alpha^r,-1}{\tilde{f}_r}{\HH}$ and
$\SolSpaceT{f_{r-1}}{\HH'[t]_{r-1}}$, respectively, we get
a basis of $\SolSpaceT{f_r}{\HH'[t]_r}$ following
Remark~\ref{Rem:PolyTechnical}.
\end{description}
\end{cor}

\begin{proof}
Let $\dfield{\EE}{\sigma}$ be a \pisiSE-extension of $\dfield{\HH'}{\sigma}$ with extension depth $\dd$. Reorder it to $\dfield{\HH'(x_1)\dots(x_l)(y_1)\dots(y_k)(t)(t_1)\dots(t_e)}{\sigma}$ with $\depth{x_i}<\dd$ for $1\leq i\leq l$ and $\depth{y_i}=\dd$ for $1\leq i\leq k$. Then by Corollary~\ref{Cor:IncremRedRef} $\SolSpaceT{f}{\EE}=\SolSpaceT{f}{\FF(t_1)\dots(t_e)}$.
\end{proof}

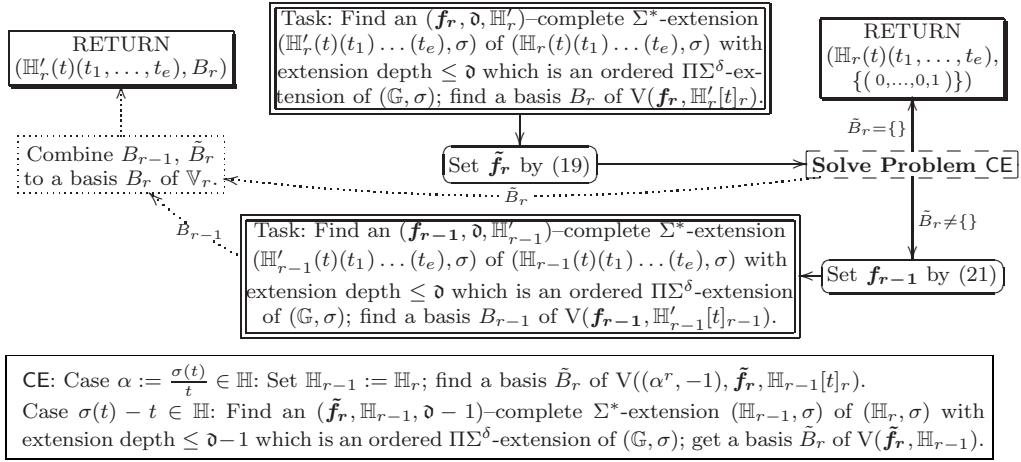
\begin{figure}[h]
\scriptsize
$$\xymatrix@R=0.3cm@C=0.1cm{
*+[F-,]\txt{RETURN\\
$(\HH'_{r}(t)(t_1,\dots,t_e),B_r)$}&*+[F=]{\txt{Task: Find an
$(\vect{f_r},\dd,\HH'_{r})$--complete \sigmaSE-extension\\
$\dfield{\HH'_r(t)(t_1)\dots(t_e)}{\sigma}$ of $\dfield{\HH_{r}(t)(t_1)\dots(t_e)}{\sigma}$ with\\
extension depth $\leq\dd$ which is an ordered
\pisiDE-ex-\\ tension of $\dfield{\GG}{\sigma}$; find a basis
$B_r$ of $\SolSpaceT{f_r}{\HH'_r[t]_r}$.}}\ar[d]&*+[F-,]\txt{RETURN\\
$(\HH_{r}(t)(t_1,\dots,t_e)$,\\
$\{\vecTSS{0,\dots,0,1}\})$}\\
*+[F.]\txt{Combine $B_{r-1}$, $\tilde{B}_r$\\
to a basis $B_r$ of $\VV_{r}$.}\ar@{.>}[u]&*+[F-:<3pt>]\txt{Set
$\vect{\tilde{f}_r}$
by~\eqref{Equ:Defineftilde}}\ar[r]&*+[F--]\txt{{\bf Solve Problem~\CE}}
\ar@{.>}@/^1.pc/[ll]|{\tilde{B}_r}\ar[u]^<(0.26){\tilde{B}_r=\{\}}\ar[d]^{\tilde{B}_r\neq\{\}}\\
&*+[F=]{\txt{Task: Find an
$(\vect{f_{r-1}},\dd,\HH'_{r-1})$--complete \sigmaSE-extension\\
$\dfield{\HH'_{r-1}(t)(t_1)\dots(t_e)}{\sigma}$ of $\dfield{\HH_{r-1}(t)(t_1)\dots(t_e)}{\sigma}$ with\\
extension depth $\leq\dd$ which is an ordered \pisiDE-extension\\
of $\dfield{\GG}{\sigma}$; find a basis $B_{r-1}$ of
$\SolSpaceT{f_{r-1}}{\HH'_{r-1}[t]_{r-1}}$.}}\ar@{.>}@/^2pc/[lu]|<(.33){B_{r-1}}
&*+[F-:<3pt>]\txt{Set $\vect{f_{r-1}}$
by~\eqref{Equ:DefinefdeltaM}}\ar[l] }$$

\MyFrame{\CE: Case $\alpha:=\frac{\sigma(t)}{t}\in\HH$: Set $\HH_{r-1}:=\HH_r$; find a
basis $\tilde{B}_r$ of
$\SolSpaceaV{\alpha^r,-1}{\tilde{f}_r}{\HH_{r-1}[t]_r}$.\\
Case $\sigma(t)-t\in\HH$: Find an
$(\vect{\tilde{f}_r},\HH_{r-1},\dd-1)$--complete \sigmaSE-extension
$\dfield{\HH_{r-1}}{\sigma}$ of $\dfield{\HH_r}{\sigma}$ with extension depth
$\leq\dd-1$ which is an ordered
\pisiDE-extension of $\dfield{\GG}{\sigma}$;
get a basis $\tilde{B}_r$ of
$\SolSpaceT{\tilde{f}_r}{\HH_{r-1}}$.}

\normalsize \caption{The refined polynomial
reduction.}\label{Fig:PolyDepthRed}
\end{figure}

\noindent Summarizing, we obtain a reduction for $r=b,...,1$, which can be
illustrated in Figure~\ref{Fig:PolyDepthRed}.

\vspace*{0.2cm}

\begin{exmp}\label{Exp:TruncatedFullPoly}
We continue the reduction from Example~\ref{Exp:TruncatedFullRat} with $r=2$.
\begin{description}
\item[$r=2$:] Set $\vect{f_2}:=\vect{f}=\vecT{\frac{bq+s}{1+k+m}}$.
By~\eqref{Equ:Defineftilde} we get
$\vect{\tilde{f}_2}=\vecT{0}$. Clearly, $\dfield{\HH}{\sigma}$ is $(\vect{\tilde{f}_2},2)$--complete with the basis $\tilde{B}_2=\{\vecT{1,0},\vecT{0,1}\}$ of $\SolSpaceT{\tilde{f}_2}{\HH}$.

\item[$r=1$:]
We get $\vect{f_{1}}=\vecT{\frac{bq+s}{1+k+m},-b^2q^2-2bqs}$ by~\eqref{Equ:DefinefdeltaM} and $\vect{\tilde{f}_1}=\vecT{\frac{1}{1+k+m},-2bq}$ by~\eqref{Equ:Defineftilde}. We can construct the \sigmaSE-extension $h$ with $\depth{h}=2$ which gives the ordered \pisiDE-field $\dfield{\KK(k)(q)(h)(b)}{\sigma}$ and which is $(\vect{\tilde{f}_1},2)$--complete; a basis of $\SolSpaceT{\tilde{f}_1}{\KK(k)(q)(h)(b)}$ is $\tilde{B}_1=\{\vecT{1,0,h},\vecT{0,0,-1}\}$; see Example~\ref{Exp:TruncatedSub}. This gives $\vect{f_0}=\vecT{-bqh,qp}$.

\end{description}

\end{exmp}

\noindent If $r=0$, we need a \sigmaSE-extension
$\dfield{\HH'_0(t)(t_1)\dots(t_e)}{\sigma}$ of
$\dfield{\HH_0(t_1)\dots(t_e)}{\sigma}$ with extension depth
$\leq\dd$ which is an ordered \pisiDE-extension of $\dfield{\GG}{\sigma}$ and which is
$(\vect{f_0},\dd,\HH'_0)$--complete.

\begin{exmp}\label{Exp:TruncatedFullPoly2}
We continue Example~\ref{Exp:TruncatedFullPoly} for the case $r=0$. By Ex.~\ref{Exp:TruncatedH} we get the ordered \pisiDE-field $\dfield{\HH'(s)}{\sigma}$ with $\HH'=\KK(k)(q)(h)(b)(H)$ which is $(\vect{f_0},3,\HH)$--complete; a basis of  $\SolSpaceT{\tilde{f}_0}{\HH'}$ is $\tilde{B}_0=\{\vecT{1,0,H},\vecT{0,0,1}\}$.
Hence we obtain the bases $B_1=\{\vecT{1,0,sh+H},\vecT{0,0,1}\}$ of $\SolSpaceT{f_1}{\HH'(s)}$ and $B_2=\{\vecT{1,sh+H},\vecT{0,1}\}$ of $\SolSpaceT{f_2}{\HH'(s)}$.
\end{exmp}

\noindent Note that
$\vect{f_0}$ is a vector in $\HH_0$ which is a smaller field in the following sense: $\dfield{\HH_0}{\sigma}$ is a \sigmaSE-extension of $\dfield{\HH}{\sigma}$
with extension depth $<\dd$, but the extension $t$ with
$\depth{t}=\dd$ is eliminated (it pops up in the tower
$(t)(t_1)\dots(t_e)$ above). Eventually, all  extensions with depth $\dd$ are eliminated, and we get a difference field with depth $\dd-1$; see Section~\ref{Sec:CompletePhase}.

\begin{exmp}\label{Exp:TruncatedSub}
We are given the ordered \pisiDE-field $\dfield{\HH(b)}{\sigma}$ over $\KK$ with $\HH=\KK(k)(q)$ and $\vect{f}=\vecT{\frac{1}{1+k+m},-2bq}$. Following Figure~\ref{Fig:RatRed} we get $\vect{p'}=\vect{f}$. The degree bound is~$1$ by~\eqref{Equ:DegBound}. We start the reduction of Figure~\ref{Fig:PolyDepthRed} with $r=1$, set $\vect{f_1}:=\vect{p'}=\vect{f}$, and get $\vect{\tilde{f}_0}=\vecT{0,-2q}$ by~\eqref{Equ:Defineftilde}. A basis of $\SolSpaceaV{\frac{1+m+K}{1+k},-1}{\tilde{f}_0}{\HH}$ is $\{\vecT{1,0,0}\}$. Hence $\vect{f_0}=\vecT{\frac{1}{1+k+m}}$ by~\eqref{Equ:DefinefdeltaM}. Now we need an ordered \pisiDE-field $\dfield{\HH'(b)}{\sigma}$ which is a \sigmaSE-extension of $\dfield{\HH(b)}{\sigma}$ with extension depth $<2$ and which is $(\vect{f_0},2,\HH')$--complete; note that $b$ is eliminated. By Example~\ref{Exp:Truncatedh} we get the \pisiDE-field $\dfield{\HH'(b)}{\sigma}$ with $\HH'=\KK(k)(q)(h)$ and the basis $B_0=\{\vecT{1,h},\vecT{0,1}\}$ of $\SolSpaceT{f_0}{\HH'}$. Completing the reduction, we get the basis $\{\vecT{1,0,h},\vecT{0,0,1}\}$ of $\SolSpaceT{f}{\HH'(b)}$. By construction, $\dfield{\HH'(b)}{\sigma}$ is $(\vect{f},2)$--complete.
\end{exmp}

\subsection{Some refinements for \piE-extensions and polynomial extensions}

\vspace*{-0.2cm}

We sum up the construction from above: The derived \sigmaDE-extensions are defined by entries of some vectors $\vect{f'}$ which occur within the reduction process; see
Section~\ref{Sec:CompletePhase}. Internally, those vectors $\vect{f'}$
are determined by the reduction presented in
Figure~\ref{Fig:RatRed} (the rational reduction)
and Figure~\ref{Fig:PolyDepthRed} (the polynomial reduction).
Exploiting additional properties in difference fields, we can predict how the $\vect{f'}$ and therefore the derived \sigmaDE-extensions look like. The first result is needed in Lemma~\ref{Lemma:SwapXY}.

\begin{cor}\label{Cor:NoPiExt}
Let $\dfield{\FF(y)}{\sigma}$ be a \pisiSE-extension of $\dfield{\GG}{\sigma}$ with $\sigma(y)/y\in\FF$ such that $\dfield{\FF(y)}{\sigma}$ can be brought to an ordered \pisiDE-extension of $\dfield{\GG}{\sigma}$; let $\vect{f}\in\FF^n$ and $\dd\geq0$. Then there is a \sigmaSE-extension of $\dfield{\FF(y)}{\sigma}$ over $\FF$ with extension depth $\leq\dd$ which can be brought to an ordered \pisiDE-extension of $\dfield{\GG}{\sigma}$ and which is $(\vect{f},\dd)$--complete.
\end{cor}

\begin{proof}
If $\depth{y}\geq\dd$, the corollary follows by Theorem~\ref{Thm:AlgTheoremB} and Proposition~\ref{Prop:PropertiesInPiSi}.2. Let $\depth{y}<\dd$. We refine the inductive proof in Section~\ref{Sec:ConstructDeltaExt}. Suppose that the reduction holds for $\dd-1$.
As in Section~\ref{Sec:ConstructDeltaExt} we assume that $\dfield{\FF(t_1)\dots(t_e)}{\sigma}$ is a \pisiSE-extension of $\dfield{\GG}{\sigma}$; if $e>0$, then~\eqref{Equ:tiEqual}.
Now reorder $\dfield{\FF(y)}{\sigma}$ to an ordered \pisiDE-extension $\dfield{\FF'}{\sigma}$ of $\dfield{\GG}{\sigma}$.
If $\depth{\FF'}<\dd$, we construct a \sigmaSE-extension over $\FF$ as required; see Section~\ref{Sec:CompletePhase}. If $\depth{\FF'}\geq\dd$, set $\FF=\HH(t)$ with $\depth{t}\geq\dd$ as in Section~\ref{Sec:ReductionPhase} with $\sigma(t)=\alpha\,t+\beta$; note that $t\neq y$, since $\depth{y}<\dd$. Then by Corollary~\ref{Cor:CompleteRatCrit} $\vect{p'}\in\HH[t]^m$. Define $b$ by~\eqref{Equ:DegBound} and set $r:=b$, $\HH_r:=\HH$. Now we apply the reduction as given in Figure~\ref{Fig:PolyDepthRed}. Since $\vect{\tilde{f}_r}$ is free of $y$, we can apply the induction assumption: we can take --as required-- a \sigmaSE-extension $\dfield{\HH_{r-1}}{\sigma}$ of $\dfield{\HH_r}{\sigma}$ where the new \sigmaSE-extensions do not depend on $y$. Note that $\SolSpaceaV{\alpha^r,-1}{\tilde{f}_r}{\HH_r}$ is free of $y$ by Proposition~\ref{Prop:PropertiesInPiSi}.2 (if $t$ is a \sigmaSE-extension) or Lemma~\ref{Lemma:NoExtForPi} (if $t$ is a \piE-extension). Hence $\vect{f_{r-1}}$ is free of $y$. Suppose we reach the base case (after at most $r$ steps) with $\vect{f_0}\in\HH_0$, free of $y$, where $\dfield{\HH_0(t)(t_1)\dots(t_e)}{\sigma}$ is a \sigmaSE-extension of $\dfield{\HH(t)(t_1)\dots(t_e)}{\sigma}$ which is free of $y$. By the reduction of the extensions with depth~$\depth{t}$, the corollary follows.
\end{proof}

\noindent Corollary~\ref{Cor:PolyExt} can be shown completely analogously by using the following Lemma~\ref{Lemma:PolyClosure}.

\begin{lem}[(\cite{Schneider:08d},Thm.~2.7)]\label{Lemma:PolyClosure}
Let $\dfield{\HH(t_1)\dots(t_e)}{\sigma}$ be a polynomial \pisiSE-extension
of $\dfield{\HH}{\sigma}$. For all $g\in\HH[t_1,\dots,t_e]$,
$\sigma(g)-g\in\HH[t_1,\dots,t_e]$ iff
$g\in\HH[t_1,\dots,t_e]$.
\end{lem}

\begin{cor}\label{Cor:PolyExt}
Let $\dfield{\HH(t_1)\dots(t_e)}{\sigma}$ be an ordered \pisiDE-extension of
$\dfield{\GG}{\sigma}$ with $\depth{\HH}=d-1$ and  $\depth{t_1}\geq
d$ such that the \sigmaDE-extension
$\dfield{\HH(t_1)\dots(t_e)}{\sigma}$ of $\dfield{\HH}{\sigma}$ is polynomial;\\ let $\vect{f}\in\HH[t_1,\dots,t_e]^n$ and $\dd\geq0$.
\begin{description}
\item[(1)] Then there is \sigmaSE-extension
$\dfield{\HH'(s_1)\dots(s_r)(t_1)\dots(t_e)}{\sigma}$ of
$\dfield{\HH(t_1)\dots(t_e)}{\sigma}$ with $\depth{s_i}\geq d$ for $1\leq i\leq r$ and
$\depth{\HH'}=d-1$ which can be brought to an ordered \pisiDE-extension of $\dfield{\GG}{\sigma}$ and which is $(\vect{f},\dd)$--complete.
\item[(2)] In
particular, the extension $\dfield{\HH'(s_1)\dots(s_r)(t_1)\dots(t_e)}{\sigma}$
of $\dfield{\HH'}{\sigma}$ is polynomial.
\item[(3)] We have $\SolSpaceT{f}{\HH'(s_1)\dots(s_r)(t_1)\dots(t_e)}\subseteq\KK^n\times\HH'[s_1,\dots,s_r][t_1,\dots,t_e]$.
\item[(4)] It can be constructed explicitly, if $\dfield{\GG}{\sigma}$ is $\sigma$-computable.
\end{description}
\end{cor}

\begin{exmp}
Consider the polynomial \pisiDE-extension $\dfield{\KK(k)(q)(b)(s)}{\sigma}$ of $\dfield{\KK(k)}{\sigma}$ with~\eqref{Equ:TrunDF} and let $f\in\KK(k)[q,b,s]$. By Corollary~\ref{Cor:PolyExt} our construction will always yield a \sigmaSE-extension $\dfield{\EE}{\sigma}$ of $\dfield{\KK(k)(q)(b)(s)}{\sigma}$ with $\EE=\KK(k)(q)(b)(s)(s_1)\dots(s_r)$ where $\dfield{\EE}{\sigma}$ is a polynomial \pisiSE-extension of $\dfield{\KK(k)}{\sigma}$; in particular, any solution $g$ of~\eqref{Equ:Telef} is in $\KK(k)[q,b,s][s_1,\dots,s_r]$.
\end{exmp}

\subsection{Algorithmic considerations: An optimal algorithm}\label{Sec:DepthSpeedUp}

The building blocks from above can be summarized to Algorithm~\ref{Alg:MultipleSolveSolutionSpace}.

\medskip

\begin{algor}{\label{Alg:MultipleSolveSolutionSpace}}
{Solving problem~\SSU.}
{\texttt{FindDepthCompleteExt}($\vect{f},\dd,\FF,\FF(t_1)\dots(t_e)$)}
{An ordered \pisiDE-extension $\dfield{\FF(t_1)\dots(t_e)}{\sigma}$
of a $\sigma$-computable $\dfield{\GG}{\sigma}$, $\vect{f}\in\FF^n$, $\dd\geq0$.}
{A \sigmaSE-extension
$\dfield{\FF'(t_1)\dots(t_e)}{\sigma}$ of
$\dfield{\FF(t_1)\dots(t_e)}{\sigma}$ s.t.\ $\dfield{\FF'(t_1)\dots(t_e)}{\sigma}$ is an ordered\\[-0.1cm] \hspace*{0.78cm}\pisiDE-extension of $\dfield{\GG}{\sigma}$ and
 is $(\vect{f},\dd,\FF')$--complete; a basis of $\SolSpaceT{f}{\FF'}$.}
\item\label{Alg:BaseCase}
IF $\dd=0$ or\footnote{We are in the base case or we apply  Lemma~\ref{Lemma:ShortCut2}.} ($\dd=1$ and $\const{\GG}{\sigma}=\GG$ and $\sigma(k)=k+1$ for some $k\in\FF$) THEN
\item\shiftS Compute a basis $B$ of $\SolSpaceT{f}{\FF}$; RETURN $(B,\FF(t_1)\dots(t_e))$. FI
\item IF $e\geq1$ and $\depth{t_e}>\dd$ THEN\hspace{1cm}\textsf{(*Simplification I*)}

\item\shiftS Let $r\geq0$ be minimal such that $\depth{t_r}=\dd$.
\item\shiftS Execute
$(B,\dfield{\FF'(t_1)\dots(t_r)}{\sigma})$=\texttt{FindDepthCompleteExt}($\vect{f},\dd,\FF,\FF(t_1)\dots(t_r)$).
\item\shiftS RETURN $(B,\FF'(t_1)\dots(t_e))$ FI


\item IF $\depth{\FF}<\dd$ THEN $\dd:=\depth{\FF}+1$\hspace{1cm}\textsf{(*Simplification~II*)}

\item\shiftS\label{Alg:SubExt} Let\footnote{For an operative improvement towards an optimal algorithm see Lemma~\ref{Lemma:BasisMani}.} $(B,\HH(t_1)\dots(t_e))$=\texttt{FindDepthCompleteExt}($\vect{f},\dd-1,\FF(t_1)\dots(t_e),\FF(t_1)\dots(t_e)$)

\item\label{Alg:ExtCommand}\shiftS
Construct a \sigmaSE-extension $\dfield{\HH'(t_1)\dots(t_e)}{\sigma}$
of $\dfield{\HH(t_1)\dots(t_e)}{\sigma}$ and a basis $B'$ of\\
\shiftS $\SolSpaceT{f}{\HH'}$ as in Sec.~\ref{Sec:CompletePhase}. Return $(B', \HH'(t_1)\dots(t_e))$ FI

\item[] \textsf{Reduction phase:} Let $\FF:=\HH(t)$ with $\delta(t)\geq\depth{\HH}\geq\dd$; if $e>0$, then~\eqref{Equ:tiEqual}.

\item\label{Alg:RedPhaseStart} Follow the rational reduction as in Figure~\ref{Fig:RatRed}. Let $R$ be
a basis of $\SolSpaceT{r}{\fracF{\HH(t)}}$.

\item IF $R=\{\}$, THEN RETURN
$(\{\vecT{0,\dots,0,1}\},\FF(t_1)\dots(t_e))$ FI

\item\label{Alg:SubExt2} Take $\vect{p'}\in\HH[t]^m$, $b$ by~\eqref{Equ:PolyF} and~\eqref{Equ:DegBound}.
Apply the polynomial reduction\footnote{If $\sigma(t)-t\in\HH$ and $r>0$, \CE\ is solved by $(B_r,\HH_{r-1})$:=\texttt{FindDepthCompleteExt}($\vect{f}_r,\dd-1,\HH_r,\HH_r)$.} from Figure~\ref{Fig:PolyDepthRed}
for $r=b,\dots,1$.
If the reduction stops earlier, return the corresponding result.
Otherwise, take $\vect{f_0}$ with the
computed ordered \pisiDE-extension
$\dfield{\HH_0(t)(t_1)\dots(t_e)}{\sigma}$ of
$\dfield{\GG}{\sigma}$.

\item\label{Alg:KeepDepth} Execute
$(B_0,\HH'(t)(t_1)\dots(t_e)$:=\texttt{FindDepthCompleteExt}($\vect{f_0},\dd,\HH_0,\HH_0(t)(t_1)\dots(t_e)$).

\item\label{Alg:RedPhaseEnd} Finish the reductions from Figures~\ref{Fig:PolyDepthRed} and~\ref{Fig:RatRed}; let  $B'$ be the basis of $\SolSpaceT{f}{\HH'(t)}$.

\item RETURN $(B',\HH'(t)(t_1)\dots(t_e))$.
\end{algor}

\medskip

Now suppose that we remove lines~\ref{Alg:SubExt}
and~\ref{Alg:ExtCommand} and return instead $(B,\dfield{\FF(t_1)\dots(t_e)}{\sigma})$ where $B$ is
a basis of $\SolSpaceT{f}{\FF}$. Then this modified version of
Algorithm~\ref{Alg:MultipleSolveSolutionSpace} boils down to the recursive reduction presented in Section~\ref{Sec:Reduction}; see Remark~\ref{Rem:RatPolyRed}.
In other words, the execution of
lines~\ref{Alg:SubExt} and~\ref{Alg:ExtCommand} is the heart of our new algorithm.
In the sequel, we will optimize this part further. For this task we will refine the reduction phase (lines~\ref{Alg:RedPhaseStart}--\ref{Alg:RedPhaseEnd}) as follows.

Let $\dfield{\FF(t_1)\dots(t_e)}{\sigma}$ be a \pisiSE-extension of
$\dfield{\FF}{\sigma}$ over $\FF$ with $\dd=\depth{t_i}$ for $1\leq i\leq e$ and $\depth{\FF}<\dd$; moreover let $V=\{1\leq i\leq e|\sigma(t_i)-t_i\in\FF\}$ and let $\vect{f'}\in\FF(t_1,\dots,t_e)^n$. Note: by executing \texttt{FindDepthCompleteExt}($\vect{f'},\dd,\FF,\FF(t_1)\dots(t_e))$)
it calls itself with depth $\dd$ in line~\ref{Alg:KeepDepth}; for all other recursive calls (in line~\ref{Alg:SubExt2}, see footnote) we use depth $<\dd$. Finally, if we enter the completion phase (lines~\ref{Alg:SubExt} and~\ref{Alg:ExtCommand}) with depth $\dd$ we can assume that $\vect{f}$ contains all the elements $\sigma(t_i)-t_i$ for $i\in V$. This follows by Lemmata~\ref{Lemma:EntriesInGroundField}
and~\ref{Lemma:EntriesInPolCase}. Given such a refined reduction, we can simplify the completion phase as follows.

\begin{lem}\label{Lemma:BasisMani}
Let $\dfield{\EE}{\sigma}$ be a \pisiSE-extension of
$\dfield{\FF}{\sigma}$ over $\FF$ with $\EE=\FF(t_1)\dots(t_e)$ such that $\dd=\depth{t_i}$ for $1\leq i\leq e$ and $\depth{\FF}<\dd$; let $\KK=\const{\FF}{\sigma}$ and
$V=\{1\leq i\leq e|\sigma(t_i)-t_i\in\FF\}$. Let $\vect{f}\in\FF^n$ where the
the last $|V|$ entries are $\sigma(t_i)-t_i\in\FF$ for $i\in V$. Then:
\begin{description}
\item[(1)] If $\dfield{\FF}{\sigma}$ is $(\vect{f},\dd-1)$--complete, then $\dfield{\EE}{\sigma}$ is $(\vect{f},\dd-1)$--complete.

\item[(2)] If a basis of $\SolSpaceT{f}{\FF}$ is given,
by row-operations in $\KK^n\times\FF$ over $\KK$ one can construct a \sigmaSE-extension $\dfield{\EE(s_1)\dots(s_r)}{\sigma}$ of
$\dfield{\EE}{\sigma}$ s.t.\  $\dfield{\EE(s_1)\dots(s_r)}{\sigma}$ is
$(\vect{f},\FF(s_1)\dots(s_r),\dd)$--com\-plete; a basis of
$\SolSpaceT{f}{\FF(s_1)\dots(s_r)}$ can be extracted with no extra cost.
\end{description}
\end{lem}

\begin{proof}
{\bf(1)} Suppose that $\dfield{\EE}{\sigma}$ is not $(\vect{f},\dd-1)$--complete. Then there is a \sigmaSE-extension $\dfield{\EE(x_1)\dots(x_r)}{\sigma}$ of $\dfield{\FF}{\sigma}$ with $g\in\EE(x_1)\dots(x_r)\setminus\EE$ and $\vect{c}\in\KK^n$ such that $\sigma(g)-g=\vect{c}\vect{f}$. By
Proposition~\ref{Prop:PropertiesInPiSi},
$g=\sum_{i\in V}d_{i}t_i+w$ for some $d_i\in\KK$ and $w\in\FF(x_1)\dots(x_r)\setminus\FF$. Thus there is $\vect{e}\in\KK^n$ such that $\sigma(w)-w=\vect{e}\vect{f}$ and therefore $\dfield{\FF}{\sigma}$ is not $(\vect{f},\dd-1)$--complete.\\
{\bf(2)} Suppose that the entries $\sigma(t_{i})-t_{i}$ with $i\in V$ occur in the last $u=|V|$ entries of $\vect{f}$; in particular suppose that they are sorted in the order as the corresponding extensions $t_i$ occur in $\EE$. Take a basis of
$\SolSpaceT{f}{\FF}$ and apply row operations such that one gets a basis $B=\{\vecT{c_{i1},\dots,c_{in},g_i},1\leq i\leq m\}\stackrel{.}{\cup}\{\vecT{0,\dots,0,1}\}$ where $\vect{C}=\vecT{c_{ij}}$ is in reduced form.
If $\vect{C}$ is the identity matrix, $\dfield{\EE}{\sigma}$ is $(\vect{f},\FF,\dd)$--complete by Lemma~\ref{Lemma:ShortCut}.2 and we are done.\\
\indent Otherwise, let $T\neq\{\}$ be the set of all integers $1\leq k\leq n-u$ such that the $k$th column does not have a corner element in $\vect{C}$. Suppose that $T=\{j_1,\dots,j_r\}$ with $n-u\geq j_1>j_2>\dots> j_r\geq 1$. Now consider the difference field
$\dfield{\EE(s_1)\dots(s_r)}{\sigma}$ extension of
$\dfield{\EE}{\sigma}$ where $\EE(s_1)\dots(s_r)$ is a rational
function field and for all $1\leq k\leq r$ we have
$\sigma(s_k)-s_k=f_{j_k}$. We prove that this is a
\sigmaSE-extension which is
$(\vect{f},\FF,\dd)$--complete.\\
\indent Let $0\leq k\leq r$ be minimal such that $s_k$ is not a \sigmaSE-extension.
Then there is a $g\in\EE(s_1)\dots(s_{k-1})$ with
$\sigma(g)-g=f_{j_k}$. Hence, by
Proposition~\ref{Prop:PropertiesInPiSi},
$g=\sum_{i=1}^{k-1}c_{i}s_{i}+\sum_{i\in V}d_it_{i}+w$ with
$w\in\FF$,$c_i,d_i\in\KK$. Hence
$f_{j_k}-\sum_{i=1}^{k-1}c_{i}f_{j_i}-\sum_{i=n-u}^nd'_if_i=
\sigma(w)-w$ for some $d'_i\in\KK$; thus
$\vect{b}=\vecT{0,\dots,0,1,\dots}\in\SolSpaceT{f}{\FF}$ where 1 is at the $j_k$th position. Since $\vect{C}$ is in row reduced form and the $j_k$th position has no corner entry, we cannot generate $\vect{b}$; a contradiction that $B$ is a basis. Hence $\dfield{\EE(s_1)\dots(s_r)}{\sigma}$ is a \sigmaSE-extension of $\dfield{\EE}{\sigma}$.\\
We get besides
$B$ the solutions
$B'=\{\vecT{0,\dots,1,\dots,0,s_k}|1\leq k\leq r\}$ where the $1$ is at the $j_k$th position and
$B''=\{\vecT{0,\dots,1,\dots,0,t_k}|1\leq k\leq e\}$  where the $1$ is at the $(n-e+k)$th position. Since the $n+1$ elements in
$B\stackrel{.}{\cup}B'\stackrel{.}{\cup}B''$ are lin.\
independent, $\dfield{\EE(s_1)\dots(s_r)}{\sigma}$ is
$(\vect{f},\FF,\dd)$--complete by
Lemma~\ref{Lemma:ShortCut}. Clearly, $B\stackrel{.}{\cup}B'$ is a
basis of $\SolSpaceT{f}{\FF(s_1)\dots(s_r)}$.
\end{proof}

\noindent\textbf{Crucial improvements.} The first consequence is that
we can replace line~\ref{Alg:SubExt} by executing the function call
$\text{$(B,\HH)$=\texttt{FindDepthCompleteExt}($\vect{f},\dd-1,\FF,\FF$)}.$
Then by Prop.~\ref{Prop:ExtensionOnTop} we get an ordered \pisiDE-extension $\dfield{\HH(t_1)\dots(t_e)}{\sigma}$ of $\dfield{\GG}{\sigma}$. Moreover, by Lemma~\ref{Lemma:BasisMani}.1 $\dfield{\HH(t_1)\dots(t_e)}{\sigma}$ is \hbox{$(\vect{f},\dd-1)$}--complete, as needed in Lemma~\ref{Lemma:DepthOptbySubProb}.

Finally, by Lemma~\ref{Lemma:BasisMani}.2 we simplify the construction of step~\ref{Alg:ExtCommand} as follows: By
analyzing the basis $B$ of $\SolSpaceT{f}{\HH}$, that has been
computed already in step~\ref{Alg:SubExt}, we get a \sigmaSE-extension of $\dfield{\HH(t_1)\dots(t_r)}{\sigma}$ of $\dfield{\GG}{\sigma}$ which can be brought to an ordered \pisiDE-extension $\dfield{\HH'(t_1)\dots(t_r)}{\sigma}$ and which is  $(\vect{f},\HH',\dd)$--complete; we extract a basis of $\SolSpaceT{f}{\HH'}$.

\begin{exmp}\label{Exp:TruncatedHSubCom}
In Example~\ref{Exp:TruncatedH} we claim that $\dfield{\FF(s)}{\sigma}$ with $\FF=\KK(k)(q)(h)(b)$ is $(\vect{f},2)$--complete where $\vect{f}=\vecT{-bqh,bq}$. Note that $\sigma(s)-s=bq$. Hence by Lemma~\ref{Lemma:BasisMani}.1 it suffices to show that $\dfield{\FF}{\sigma}$ is $(\vect{f},2)$--complete. With our algorithm this can be easily checked; during this check we get the basis $\{\vecT{0,0,1}\}$ of $\SolSpaceT{f}{\FF}$. Following the proof of Lemma~\ref{Lemma:BasisMani}.2,  $\dfield{\FF(s)(H)}{\sigma}$ is a \sigmaSE-extension of $\dfield{\FF(s)}{\sigma}$ with $\sigma(H)=H-bqh$
and we get the basis $\{\vecT{0,0,1},\vecT{1,0,H}\}$ of $\SolSpaceT{f}{\FF(s)(H)}$. By Lemma~\ref{Lemma:DepthOptbySubProb} $H$ is depth-optimal.
\end{exmp}

We emphasize that the modified algorithm differs from the reduction presented in Section~\ref{Sec:Reduction} (which is similar to Karr's algorithm) by just analyzing the sub-results and by inserting extensions if necessary.

\medskip

\textit{In a nutshell, running our new algorithm which computes an appropriate \pisiDE-exten\-sion and which outputs the corresponding solution to problem~\PT\ is not more expensive than choosing such a \pisiDE-extension manually and solving problem~\PT\ with the recursive algorithm from Section~\ref{Sec:Reduction} (or Karr's algorithm). On the contrary, adjoining the extensions only when it is required during the reduction keeps the computations as simple and therefore as cheap as possible.}

\section{Proving the main results (from Section~\ref{Sec:MainResults})}\label{Sec:MainProofs}
We need the following preparation to prove Result~\ref{Res:Reordering}.

\begin{lem}\label{Lemma:PiNotNeeded}
Let $\dfield{\FF(y)}{\sigma}$ be a \piE-extension of $\dfield{\FF}{\sigma}$ which can be brought to an ordered \pisiDE-extension of $\dfield{\GG}{\sigma}$. Let $f\in\FF$ and let $\dfield{\EE}{\sigma}$ be a \sigmaSE-extension of $\dfield{\FF}{\sigma}$ with extension depth $\dd$ and $g\in\EE$ such that~\eqref{Equ:Telef}. Then there is a \sigmaSE-extension $\dfield{\SA}{\sigma}$ of $\dfield{\FF}{\sigma}$ with extension depth $\leq\dd$ and $g'\in\SA$ such that $\sigma(g')-g'=f$ and $\depth{g'}\leq\depth{g}$.
\end{lem}

\begin{proof}
Write $\EE=\FF(y)(s_1)\dots(s_e)$ with $\dd=\max_i\depth{s_i}$.
Since we can bring $\dfield{\FF(y)}{\sigma}$ to an ordered \pisiDE-extension of
$\dfield{\GG}{\sigma}$, we can apply
Corollary~\ref{Cor:NoPiExt}: There is a
\sigmaSE-extension $\dfield{\SA}{\sigma}$ of $\dfield{\FF(y)}{\sigma}$ over $\FF$ with $\SA=\FF(y)(x_1)\dots(x_r)$
which can be brought to an ordered \pisiDE-extension of $\dfield{\GG}{\sigma}$ and in which we have $g'\in\SA$ s.t.\ $\sigma(g')-g'=f$; by Proposition~\ref{Prop:PropertiesInPiSi}.2, $g'\in\FF(x_1)\dots(x_r)$.
By Theorem~\ref{Thm:ConstructSigmaExtOnTopOrdered} we can take a \sigmaSE-extension $\dfield{\SA'}{\sigma}$ of $\dfield{\SA}{\sigma}$ and an $\FF(y)$-monomorphism $\fct{\tau}{\EE}{\SA'}$ s.t.~\eqref{Equ:DepthIneq} for all $a\in\EE$. Note that $\sigma(\tau(g))-\tau(g)=f=\sigma(g')-g'$. Since $\tau(g),g'\in\SA'$, $\tau(g)=g'+c$ for some $c\in\const{\GG}{\sigma}$. Therefore, $\depth{g}\geq\depth{\tau(g)}=\depth{g'}$.
Since $\depth{\tau(s_i)}\leq\depth{s_i}\leq\dd$ for $1\leq i\leq r$, $g'\in\FF(x_i|\depth{x_i}\leq\dd)=:\SA''$. By construction, $\dfield{\SA''}{\sigma}$ is a \sigmaSE-extension of $\dfield{\FF}{\sigma}$ with extension depth $\leq\dd$.
\end{proof}

\begin{lem}\label{Lemma:SwapXY}
Let $\dfield{\FF(x)(y)}{\sigma}$ be a \pisiDE-extension of
$\dfield{\FF}{\sigma}$ and suppose that  $\dfield{\FF(x)}{\sigma}$ and $\dfield{\FF(y)}{\sigma}$
can be brought to ordered \pisiDE-extensions of
$\dfield{\GG}{\sigma}$. Then the \pisiSE-extension
$\dfield{\FF(y)(x)}{\sigma}$ of~$\dfield{\FF}{\sigma}$ is depth-optimal.
\end{lem}

\begin{proof}
First we show that $\dfield{\FF(y)}{\sigma}$ is a \pisiDE-extension of
$\dfield{\FF}{\sigma}$. If $y$ is a
\piE-extension, we are done. Otherwise, let $y$
be a \sigmaSE-extension with $\sigma(y)=y+f$ which is not
depth-optimal. Hence, we can take a \sigmaSE-extension
$\dfield{\FF(s_1)\dots(s_e)}{\sigma}$ of $\dfield{\FF}{\sigma}$ with
extension depth $\leq\depth{f}$ and $g\in\FF(s_1)\dots(s_e)$ such that~\eqref{Equ:Telef}. There are
two cases.\\
{\bf Case~1a:} $x$ is a \piE-extension. By
Corollary~\ref{Cor:ExtendPiOverSigma}
$\dfield{\FF(s_1)\dots(s_e)(x)}{\sigma}$ is a \piE-extension of
$\dfield{\FF(s_1)\dots(s_e)}{\sigma}$. By reordering,
$\dfield{\FF(x)(s_1)\dots(x_e)}{\sigma}$ is a \sigmaSE-extension of $\dfield{\FF(x)}{\sigma}$. Consequently,
$\dfield{\FF(x)(y)}{\sigma}$ is not
a \sigmaDE-extension of $\dfield{\FF(x)}{\sigma}$,
a contradiction.\\
{\bf Case~1b:} $x$ is a \sigmaSE-extension. Bring
$\dfield{\FF(x)}{\sigma}$ to an ordered \pisiDE-extension
$\dfield{\SA}{\sigma}$ of $\dfield{\GG}{\sigma}$. Hence, by
Thm.~\ref{Thm:ConstructSigmaExtOnTopOrdered} there is a
\sigmaSE-extension $\dfield{\EE}{\sigma}$ of $\dfield{\SA}{\sigma}$
with extension depth $\leq\depth{f}$ and an $\FF$-monomorphism
$\fct{\tau}{\FF(s_1)\dots(s_e)}{\EE}$ with
$\sigma(\tau(g))-\tau(g)=f$. Since $\SA=\FF(x)$ (as fields), $\dfield{\FF(x)(y)}{\sigma}$
is not a \sigmaDE-extension of $\dfield{\FF(x)}{\sigma}$; a
contradiction.

Second, we show that
$\dfield{\FF(y)(x)}{\sigma}$ is a \pisiDE-extension of $\dfield{\FF(y)}{\sigma}$. If $x$ is a \piE-extension, we are done. Otherwise, let $x$ be a \sigmaSE-extension. If $y$ is a \sigmaSE-extension and $\depth{y}\geq\depth{x}$, the statement follows by Lemma~\ref{Lemma:ShiftSumToLeft} and by Proposition~\ref{Prop:ReorderEqualDepth}. What remains to consider are the cases that $y$ is a \piE-extension or that $y$ is a \sigmaSE-extension with $\depth{y}<\depth{x}$. Now suppose that $\dfield{\FF(y)(x)}{\sigma}$ is a \sigmaSE-extension of $\dfield{\FF(y)}{\sigma}$ with $\sigma(x)=x+f$ which is not depth-optimal. Hence, we can take a \sigmaSE-extension $\dfield{\FF(y)(s_1)\dots(s_e)}{\sigma}$ of $\dfield{\FF(y)}{\sigma}$ with extension depth $\leq\depth{f}$ and $g\in\FF(y)(s_1)\dots(s_e)$ such that~\eqref{Equ:Telef}.\\
{\bf Case~2a:} $y$ is a \piE-extension. By Lemma~\ref{Lemma:PiNotNeeded}, there is a \sigmaSE-extension $\dfield{\SA}{\sigma}$ of $\dfield{\FF}{\sigma}$ with extension depth $\leq\depth{f}$ and $g'\in\SA$ such that $\sigma(g')-g'=f$; hence $\dfield{\FF(x)}{\sigma}$ is not a \sigmaDE-extension of $\dfield{\FF}{\sigma}$, a contradiction.\\
{\bf Case~2b:} $y$ is a \sigmaSE-extension with $\depth{y}>\depth{x}$. Reorder
$\dfield{\FF(y)}{\sigma}$ to a \pisiDE-extension of
$\dfield{\GG}{\sigma}$. By Theorem~\ref{Thm:DepthStableOrdered}
$\delta(g)\leq\delta(f)+1$. Since $\delta(f)+1=\delta(x)<\delta(y)$,
$g$ is free of $y$. Hence, $\dfield{\FF(s_1)\dots(s_e)}{\sigma}$ is a \sigmaSE-extension of $\dfield{\FF}{\sigma}$ with $g\in\FF(s_1)\dots(s_e)$
such that~\eqref{Equ:Telef}, and therefore $\dfield{\FF(x)}{\sigma}$ is not a
\sigmaDE-extension of $\dfield{\FF}{\sigma}$; a
contradiction.
\end{proof}

\noindent$\bullet$~{\bf Result~\ref{Res:Reordering}.} If $e=0,1$ nothing has to be shown. Let
$\dfield{\GG(t_1)\dots(t_e)(x)}{\sigma}$ be a \pisiDE-extension of $\dfield{\GG}{\sigma}$ with $e\geq1$ and suppose the theorem holds for $e\geq1$ extension. Choose any possible reordering. If $x$ stays on top, by the induction assumption all extensions below are depth-optimal. $x$ remains depth-optimal, since the field below has not changed. This shows this case. Otherwise, suppose that $t_i$ for some $1\leq i\leq e$ is on top. Then we can reorder our field to the \pisiSE-extension $\dfield{\HH(t_i)(x)}{\sigma}$
of $\dfield{\GG}{\sigma}$ with $\HH:=\GG(t_1)\dots(t_{i-1})(t_{i+1})\dots(t_{e})$. By the induction assumption we can bring $\dfield{\HH(t_i)}{\sigma}$
and $\dfield{\HH(x)}{\sigma}$
to ordered \pisiDE-extensions of $\dfield{\GG}{\sigma}$. Thus, we can apply Lemma~\ref{Lemma:SwapXY} and get the \pisiDE-extension
$\dfield{\HH(x)(t_i)}{\sigma}$
of $\dfield{\GG}{\sigma}$. By the induction assumption we can bring the extensions in $\HH$ to the desired order without changing the \pisiDE-property.

\smallskip

\noindent$\bullet$~{\bf  Result~\ref{Res:DepthConstruct}.} This
follows by Theorem~\ref{Thm:ExtendForTeleOrdered},
Corollary~\ref{Cor:PolyExt} and Result~\ref{Res:Reordering}. In particular, $\dfield{\EE}{\sigma}$ and $g\in\EE$ can be computed as follows.

\smallskip
\small
\noindent{\tiny1}\hspace*{0.4cm}Reorder $\dfield{\FF}{\sigma}$ to an ordered \pisiSE-extension of $\dfield{\GG}{\sigma}$.\\ \noindent{\tiny2}\hspace*{0.4cm}Execute $(B,\EE)$:=\texttt{FindDepthCompleteExt}($\vecT{f},\depth{f}+1,\FF,\FF$) and extract $g$ from $B$ s.t.~\eqref{Equ:Telef}.
\normalsize

\medskip

\noindent$\bullet$~{\bf Result~\ref{Res:DepthStable}.} This is a
direct consequence of Theorem~\ref{Thm:DepthStableOrdered} and
Result~\ref{Res:Reordering}.

\smallskip

\noindent$\bullet$~{\bf  Result~\ref{Res:ConstructSigmaExtOnTop}.}
This follows by Theorem~\ref{Thm:ConstructSigmaExtOnTopOrdered} and
Result~\ref{Res:Reordering}.

\smallskip

\noindent$\bullet$~{\bf Result~\ref{Res:ExtensionDepthStable}.} This is implied by the following more general statement: there is a
\sigmaSE-extension $\dfield{\SA}{\sigma}$ of $\dfield{\EE}{\sigma}$
and a \pisiDE-extension $\dfield{\DD}{\sigma}$ of
$\dfield{\FF}{\sigma}$ with an $\FF$-isomorphism
$\fct{\tau}{\SA}{\DD}$ as in~\eqref{Equ:DepthIneq} for all
$a\in\EE$; we can assume that $\EE$ is ordered.\\
We prove this result by induction on the number of extensions in
$\EE$. For $\EE=\FF$, take $\DD:=\FF$ and $\SA=:\FF$ with
$\tau=\text{id}_{\FF}$. Now suppose we have shown the result for
$\EE=\FF(t_1)\dots(t_{e-1})$ with $e\geq1$. I.e., we are given a
\pisiDE-extension $\dfield{\DD}{\sigma}$ of $\dfield{\FF}{\sigma}$,
a \sigmaSE-extension $\dfield{\SA}{\sigma}$ of
$\dfield{\EE}{\sigma}$ with $\SA=\EE(s_1)\dots(s_u)$ and an
$\FF$-isomorphism $\fct{\tau}{\SA}{\DD}$ as in~\eqref{Equ:DepthIneq} for all $a\in\EE$. Let
$\dfield{\EE(t)}{\sigma}$ be a \pisiSE-extension of
$\dfield{\EE}{\sigma}$ with $\delta(t)\geq\delta(\EE)$.\\
{\bf Case 1:} Suppose that $t$ is a \piE-extension with
$\sigma(t)=\alpha t$. Then by Corollary~\ref{Cor:ExtendPiOverSigma}
we can construct the \piE-extension $\dfield{\SA(t)}{\sigma}$ of
$\dfield{\SA}{\sigma}$. Moreover, by
Proposition~\ref{Prop:HomProperties}.2 we can construct the
\piE-extension $\dfield{\DD(x)}{\sigma}$ of $\dfield{\DD}{\sigma}$
with $\sigma(x)=\tau(\alpha)x$ and can extend the $\FF$-isomorphism
$\tau$ to $\fct{\tau}{\SA(t)}{\DD(x)}$ with $\tau(t)=x$. By
reordering, we get the \sigmaSE-extension
$\dfield{\EE(t)(s_1)\dots(s_u)}{\sigma}$ of
$\dfield{\EE(t)}{\sigma}$ with the $\FF$-isomorphism
$\fct{\tau}{\EE(t)(s_1)\dots(s_u)}{\DD(x)}$. As
$\delta(\tau(\alpha))\leq\delta(\alpha)$, it follows that
$\delta(\tau(t))\leq\delta(t)$. Hence~\eqref{Equ:DepthIneq}
for all $a\in\EE(t)$.\\
{\bf Case 2:} Suppose that $t$ is a \sigmaSE-extension with
$\sigma(t)=t+\beta$. We consider two subcases\\
{\bf Case 2a:} If there is a $g\in\SA$ with $\sigma(g)-g=\beta$, let $j\geq1$ be
minimal such that $g\notin\EE(s_1)\dots(s_{j-1})$. Then by Theorem~\ref{Thm:PiSigma}.1 there is the
\sigmaSE-extension $\dfield{\EE(s_1)\dots(s_{j-1})(t)}{\sigma}$ of
$\dfield{\EE(s_1)\dots(s_{j-1})}{\sigma}$ with $\sigma(t)=t+\beta$. Furthermore,
there is an $\EE(s_1)\dots(s_{j-1})$-isomorphism
$\fct{\rho}{\EE(s_1)\dots(s_{j-1})(t)}{\EE(s_1)\dots(s_{j-1})(s_j)}$
with $\rho(t)=g$ by Prop.~\ref{Prop:HomProperties}.1.
By reordering we get the \sigmaSE-extension
$\dfield{\EE(t)(s_1)\dots(s_{j-1})}{\sigma}$ of
$\dfield{\EE(t)}{\sigma}$. Now we can construct a
\sigmaSE-extension $\dfield{\SA'}{\sigma}$ of
$\dfield{\EE(t)(s_1)\dots(s_{j-1})}{\sigma}$ with an
$\EE(t)(s_1)\dots(s_{j-1})$-iso\-morphism $\fct{\rho}{\SA'}{\SA}$ by
Prop.~\ref{Prop:HomProperties}.3. Hence we arrive at an
$\FF$-isomorphism $\fct{\tau'}{\SA'}{\DD}$ with
$\tau':=\tau\circ\rho$. Finally, observe that for all $a\in\EE$ we have
$\tau'(a)=\tau(\rho(a))=\tau(a)$ and
$\tau'(t)=\tau(\rho(t))=\tau(g)$. Since
$\sigma(\tau(g))-\tau(g)=\tau(\beta)$,
$\delta(\tau(g))\leq\delta(\tau(\beta))+1$ by
Result~\ref{Res:DepthStable}. With
$\delta(\tau(\beta))+1\leq\delta(\beta)+1=\delta(t)$ it follows that
$\delta(\tau'(t))\leq\delta(t)$. Since
$\delta(\tau'(a))=\depth{\tau(a)}\leq\delta(a)$ for all $a\in\EE$, we get
$\delta(\tau'(a))\leq\delta(a)$ for all $a\in\EE(t)$.\\
{\bf Case 2b:} Suppose that there is no $g\in\SA$ with
$\sigma(g)-g=\beta$. Then there is no $g\in\DD$ with
$\sigma(g)-g=\tau(\beta)$. By Result~\ref{Res:DepthConstruct} there
is a \sigmaDE-extension $\dfield{\DD(y_1)\dots(y_v)}{\sigma}$ of
$\dfield{\DD}{\sigma}$ such that $\sigma(g)-g=\tau(\beta)$ for some
$g\in\DD(y_1)\dots(y_v)\setminus\DD(y_1)\dots(y_{v-1})$. Moreover,
by Proposition~\ref{Prop:HomProperties}.3 it follows that there is a
\sigmaSE-extension $\dfield{\SA(x_1)\dots(x_{v-1})}{\sigma}$ of
$\dfield{\SA}{\sigma}$ and an $\FF$-isomorphism
$\fct{\tau'}{\SA(x_1)\dots(x_{v-1})}{\DD(y_1)\dots(y_{v-1})}$ where
$\tau'(a)=\tau(a)$ for all $a\in\EE$. Furthermore, we can construct
the \sigmaSE-extension $\dfield{\SA(x_1)\dots(x_{v-1})(t)}{\sigma}$
of $\dfield{\SA(x_1)\dots(x_{v-1})}{\sigma}$ with
$\sigma(t)=t+\beta$ by Proposition~\ref{Prop:PropertiesInPiSi}.1.
Finally, we can construct the $\FF$-isomorphism
$\fct{\tau''}{\SA(x_1)\dots(x_{v-1})(t)}{\DD(y_1)\dots(y_{v-1})(y_v)}$
with $\tau''(a)=\tau'(a)$ for all $a\in\SA(x_1)\dots(x_{v-1})$ and
$\tau''(t)=g$ by Proposition~\ref{Prop:HomProperties}.1. By reordering
of $\SA(x_1)\dots(x_{v-1})(t)$ we obtain the \sigmaSE-extension
$\dfield{\SA'}{\sigma}$ of $\dfield{\EE(t)}{\sigma}$ with $\SA'=\EE(t)(s_1)\dots(s_u)(x_1)\dots(x_{v-1})$. As above,
$\delta(\tau''(t))=\delta(g)\leq\delta(\tau(\beta))+1\leq
\delta(\beta)+1=\delta(t)$. Since
$\tau''(a)=\tau'(a)=\tau(a)$ for all $a\in\EE$,
$\delta(\tau''(a))\leq\delta(a)$ for all $a\in\EE$. Thus $\delta(\tau''(a))\leq\delta(a)$ for all $a\in\EE(t)$.\\
Note that this construction can be given explicitly, if $\dfield{\GG}{\sigma}$ is $\sigma$-computable.

\smallskip

\noindent$\bullet$~{\bf Result~\ref{Res:ProdFree}.} The induction base $e=0$ is obvious. Suppose Result~\ref{Res:ProdFree} holds for $e\geq0$ extensions, and consider a \pisiSE-extension
$\dfield{\FF(t_1)\dots(t_{e+1})}{\sigma}$ of $\dfield{\FF}{\sigma}$ with $g\in\FF(t_1)\dots(t_{e+1})$ s.t.~\eqref{Equ:Telef}. Then by assumption there is a \sigmaDE-extension
$\dfield{\FF(t_1)(s_1)\dots(s_r)}{\sigma}$ of $\dfield{\FF(t_1)}{\sigma}$ with $g'\in\FF(t_1)(s_1)\dots(s_r)$ such that $\delta(g')\leq\delta(g)$ and $\sigma(g')-g'=f$. Now we apply Result~\ref{Res:ExtensionDepthStable}
(if $t_1$ is a \sigmaSE-extension) and Lemma~\ref{Lemma:PiNotNeeded} together with Result~\ref{Res:ExtensionDepthStable} (if $t_1$ is a \piE-extension): It follows that there is a
\sigmaDE-extension $\dfield{\SA}{\sigma}$ of $\dfield{\FF}{\sigma}$ with $g''\in\SA$ s.t.~$\delta(g'')\leq\delta(g')$ and $\sigma(g'')-g''=f$. Since $\depth{g''}\leq\depth{g}$, we are done.

\smallskip

\noindent$\bullet$~{\bf Result~\ref{Res:ProdSumDef}.} This is a direct consequence of Results~\ref{Res:DepthConstruct} and~\ref{Res:ProdFree}.

\smallskip

\noindent$\bullet$~{\bf Result~\ref{Res:DOT}.} Let $\dfield{\EE}{\sigma}$ be such a \sigmaDE-extension of $\dfield{\FF}{\sigma}$ with $\KK=\const{\FF}{\sigma}$; take $g\in\EE$ as in~\eqref{Equ:Telef}. {\bf(1)}~Let
$\dfield{\HH}{\sigma}$ be a \pisiSE-extension of
$\dfield{\FF}{\sigma}$ with $g'\in\HH$ s.t.~$\sigma(g')-g'=f$.
By Result~\ref{Res:ProdFree} there is a
\sigmaDE-extension $\dfield{\SA}{\sigma}$ of $\dfield{\FF}{\sigma}$ with $\SA=\FF(x_1)\dots(x_r)$
and $h\in\SA$ such that $\sigma(h)-h=f$ and $\delta(h)\leq\delta(g')$.
By Result~\ref{Res:ConstructSigmaExtOnTop} we get a
\sigmaDE-extension $\dfield{\EE'}{\sigma}$ of $\dfield{\EE}{\sigma}$
and an $\FF$-monomorphism $\fct{\tau}{\SA}{\EE'}$ as in~\eqref{Equ:DepthIneq} for all $a\in\SA$. Hence
$\delta(\tau(h))\leq\delta(h)\leq\delta(g')$. Since $\tau(h),g\in\EE'$ and
$\sigma(\tau(h))-\tau(h)=f$, $\tau(h)=g+c$ for some
$c\in\KK$. Hence $\delta(\tau(h))=\delta(g)$.\\
{\bf(2)}~Suppose in addition that $\depth{s_e}=\dd$ and  $g\in\EE\setminus\FF(s_1,\dots,x_{e-1})$. By the above considerations,  $\depth{\tau(x_i)}\leq\depth{x_i}$ for $1\leq i\leq r$ and $\tau(h)=g+c$ for some $c\in\KK$. Hence there is an $i$ with $1\leq i\leq r$ s.t.\ $s_e$ occurs in $\tau(x_i)$. Hence $\dd=\depth{s_e}\leq\depth{\tau(x_i)}\leq\depth{x_i}$.

\smallskip

\noindent$\bullet$~{\bf Result~\ref{Res:ParaTele}.} By
Theorem~\ref{Thm:AlgTheoremA} we can take a \sigmaDE-extension
$\dfield{\EE}{\sigma}$ of $\dfield{\FF}{\sigma}$ which is
$(\vect{f},\dd)$--complete. Now let
$\dfield{\HH}{\sigma}$ be any \pisiSE-extension of $\dfield{\FF}{\sigma}$ with extension depth $\leq\dd$ and $g\in\HH$, $\vect{c}\in\KK^n$ s.t.~\eqref{Equ:ParaTele}. Then by
Results~\ref{Res:DepthConstruct} and~\ref{Res:DOT}.1 we take a \sigmaDE-extension
$\dfield{\SA}{\sigma}$ of $\dfield{\FF}{\sigma}$ with $g'\in\SA$ such that
$\sigma(g')-g'=\vect{c}\vect{f}=:f$ and $\delta(g')\leq\delta(g)$.
Moreover, by Result~\ref{Res:ConstructSigmaExtOnTop} we take a \sigmaSE-extension $\dfield{\EE'}{\sigma}$ of $\dfield{\EE}{\sigma}$ with extension depth $\leq\dd$ and an $\FF$-monomorphism $\fct{\tau}{\SA}{\EE'}$ s.t.\  $\depth{h}\leq\depth{g'}$ for  $h:=\tau(g')$. Since $\sigma(h)-h=f$ and $\dfield{\EE}{\sigma}$ is $(\vect{f},\dd)$--complete, $h\in\EE$; in particular, $\depth{h}\leq\depth{g}$.
$\dfield{\EE}{\sigma}$ can be constructed explicitly:

\smallskip
\small
\noindent{\tiny1}\hspace*{0.4cm}Reorder $\dfield{\FF}{\sigma}$ to an ordered \pisiSE-extension of $\dfield{\GG}{\sigma}$.\\
\noindent{\tiny2}\hspace*{0.4cm}$(B,\EE)$:=\texttt{FindDepthCompleteExt}($\vect{f},\depth{\vect{f}},\FF,\FF$).
\normalsize

\section{Applications from Particle Physics}\label{Sec:Application}

We conclude our article by non-trivial applications from particle physics~\cite{Schneider:07h,Schneider:07i}.
For the computations we used the summation package \SigmaP~\cite{Schneider:07a} which contains in its inner core our new difference field theory.

\subsection{Finding recurrence relations with smaller order}

In massive higher
order calculations of Feynman diagrams~\cite{Schneider:07h} the sum
$$A(N)=\sum_{i=1}^{\infty}
\frac{B(N,i)}{i+N+2}S_1(i)S_1(N+i),$$
where $B(N,i)=\tfrac{\Gamma(N)\Gamma(i)}{\Gamma(N+i)}$ denotes the beta function~\cite[p.~5]{AAR}, arose. It turns out that
our refined creative telescoping method produces --analogous to Example~\ref{Equ:CreaSol}-- a recurrence with minimal order:
\begin{align*}
(N+2)&A(N)-(N+3)A(N+1)=2\tfrac{N^5+5N^4+21N^3+38N^2+28N+8}{N^4(N+1)^2(N+2)^2}\\
&+
      2\tfrac{(-1)^N}{N(N+2)}\Big(
                   -\tfrac{(3N+4)(\zeta_2+2S_{-2}(N))}
                         {(N+1)(N+2)}
                   -2\zeta_3-2S_{-3}(N)-2\zeta_2S_1(N)-4S_{1,-2}(N)
                \Big)\\
       &+\tfrac{1}{N+1}(S_2(N)-\zeta_2)
         +\tfrac{N^6+8N^5+31N^4+66N^3+88N^2+64N+16}{N^3(N+1)^2(N+2)^3}S_1(N);
\end{align*}
note that standard creative telescoping produces a recurrence of order $4$ only; see~\cite[p.~6]{Schneider:07h}. Given this optimal recurrence of order $1$, the closed form\footnote{$\zeta_k$ denotes the Riemann zeta function at $k$; e.g., $\zeta_2=\pi^2/6$.}
\small
\begin{equation*}
\begin{split}
     A(N)=&\tfrac{2(-1)^N}{N(N+1)(N+2)}\Biggl[
                                        2S_{-2,1}(N)
                                       -3S_{-3}(N)
                                       -2S_{-2}(N)S_1(N)
                                       -\zeta_2S_1(N)
                                     -\zeta_3
                                       -\tfrac{2S_{-2}(N)+\zeta_2}{N+1}
                                   \Biggr]\\
&        -2\tfrac{S_3(N)-\zeta_3}{N+2}
        -\tfrac{S_2(N)-\zeta_2}{N+2}S_1(N)
      +\tfrac{2+7N+7N^2+5N^3+N^4}
              {N^3(N+1)^3(N+2)}S_1(N)
          +2\tfrac{2+7N+9N^2+4N^3+N^4}
              {N^4(N+1)^3(N+2)}
\end{split}
\end{equation*}
\normalsize
can be read off immediately.\\
We remark that in this example the algebraic object $(-1)^N$ occurs which cannot be handled in a direct fashion in \pisiSE-fields. As it turns out, our algorithmic framework can be slightly extended such that it works also in this case; the technical details are omitted here.

Similar examples for our refined creative telescoping method can be found, e.g., in~\cite{Schneider:03,Schneider:T05b,Schneider:06e,Schneider:08c}.

\vspace*{-0.2cm}

\subsection{Simplification of d'Alembertian solution}

\vspace*{-0.2cm}

As worked out in~\cite{Schneider:07i} \SigmaP\ could reproduce the evaluation\footnote{For the original computation~\cite{Vermaseren:2005qc} the package~\textsf{Summer}~\cite{Vermaseren:99} based on Form was used which is specialized to manipulate huge expressions in terms of harmonic sums.} of a Feynman diagram that occurred in~\cite{Vermaseren:2005qc} during the computation of the third-order QCD corrections to deep-inelastic scattering by photon exchange. More precisely, in Mellin space the related Feynman diagram could be expressed in terms of the recurrence

\vspace*{-0.5cm}

\begin{multline*}
-N(N+1)^2 (N+2)(3N+7)F(N)+(N+1)(N+2)^2(N+3)(3N+4)F(N+1)\\
+N(N+1)(N+2)(N+3)(3N+7)F(N+2)\\
-(N+1)(N+2)(N+3)(N+4)(3 N+4)F(N+3)=f(N)
\end{multline*}

\vspace*{-0.2cm}

\noindent with inhomogeneous part
\begin{align*}
f(N&)=\Big(1-(-1)^N\Big)\Big(\tfrac{6 N^4+38
   N^3+81 N^2+66 N+14}{N (N+1) (N+2)}\big(24 \zeta_3+16S_{-3}(N)\big)\\
&+\tfrac{16 (N+2) (N+3) (3 N+4)}{(N+1)^4}+\tfrac{16
   \left(6 N^7+56 N^6+213 N^5+429 N^4+496 N^3+339 N^2+138 N+28\right)}{N^2 (N+1)^2 (N+2)^2}S_{-2}(N)\Big)\\
&-\tfrac{8 \left(12 N^7+115 N^6+462 N^5+1026
   N^4+1383 N^3+1152 N^2+552 N+112\right)}{N^2 (N+1)^2
   (N+2)^2}S_{-2}(N)\\
&-\tfrac{16\left(9 N^4+61 N^3+144 N^2+138 N+42\right)\zeta_3}{N
   (N+1) (N+2)}-\tfrac{8 \left(12 N^4+81 N^3+189 N^2+180 N+56\right)}{N (N+1) (N+2)}S_{-3}(N)\\
&+\tfrac{8 \left(3 N^4+18 N^3+30 N^2+7 N-12\right)}{(N+1)^2 (N+2)^2}S_2(N)+\tfrac{8 \left(N^2+9 N+12\right) }{(N+1)
   (N+2)}S_3(N)
\end{align*}
\normalsize
and initial values in terms of $\zeta$-values (which are not printed here). \SigmaP\ easily computes the general d'Alembertian solution

\vspace*{-.5cm}

\begin{multline}\label{Equ:dAlembertSum}
F(N)=c_1\frac{1}{N+1}+c_2\frac{(-1)^N}{N+1}+c_3\frac{(-1)^N N S_{-1}(N)-2}{N(N+1)}\\[-0.7cm]
   \underbrace{-\frac{1}{N+1}\sum_{k=4}^N(-1)^k\underbrace{\sum_{j=4}^k\frac{(-1)^j (3 j-2)}{(j-2) (j-1) j}\overbrace{\sum_{i=4}^j\frac{f(i-3)}{(3 i-5) (3 i-2)}}^{=A(j)}}_{=B(k)}}_{=C(N)}$$
\end{multline}

\vspace*{-0.3cm}

\noindent for constants $c_1,c_2,c_3$. Checking initial values shows that
$c_1=\frac{41 \zeta_3}{7}$, $c_2=\frac{1}{7}(53\zeta_3-70\zeta_5)$ and $c_3=-\frac{12 \zeta_3}{7}$. Now the main task is to simplify~\eqref{Equ:dAlembertSum} further.
With, e.g., Karr's algorithm~\cite{Karr:81} the inner sum $A(j)$ can be eliminated and one gets a rather big expression for $A(j)$ in terms of single nested harmonic sums $S_i(j)$. In other words, we obtain an expression for~\eqref{Equ:dAlembertSum} where the depth is reduced by one. To get a representation with optimal nested depth, we execute our refined algorithm; the result is an expression for $C(N)$ in terms of two nested sum expressions only:
$$C(N)=-\frac{(-1)^N B(N)}{2(N+1)}-\frac{1}{4 (N+1)}\sum_{k=4}^N\frac{f(k-3)}{(k-1)(k-2)}+\frac{(3 N-2)
   (3 N+1)}{4(N-1) N (N+1)}A(N).$$
Note that the depth optimality of the sum representation is justified by results from~\cite{Schneider:T07f}.

\noindent Finally, splitting these sums by partial fraction decomposition, we get the solution~\cite{Schneider:07i}:
\begin{align*}
F&(N)=\frac{2 \Big(6 \zeta_3 +
  (-1)^N(6 \zeta_3 -5 N^2
  \zeta_5)\Big)}{N^2 (N+1)}
-\frac{8(1+(-1)^N)
  S_{-5}(N)}{N+1}-\frac{4 \Big(1+(-1)^N\Big)
  S_5(N)}{N+1}\\[-0.1cm]
&+S_2(N) \Big(\frac{4 S_3(N)}{N+1}-\frac{4
  \zeta_3 }{N+1}\Big)+S_{-3}(N) \Big(\frac{8(1+(-1)^N)}{N^2
  (N+1)}-\frac{4(2+(-1)^N) S_{-2}(N)}{N+1}\\
&-\frac{4
  S_2(N)}{N+1}\Big)+S_{-2}(N) \Big(\frac{-12 \zeta_3  N^3+(-1)^N
  (8-8 N^3 \zeta_3)+8}{N^3 (N+1)}-\frac{(4+8
  (-1)^N) S_3(N)}{N+1}\Big)\\[-0.1cm]
&+\frac{8 S_{-3,-2}(N)}{N+1}+
  \frac{(4-4(-1)^N)S_{-3,2}(N)}{N+1}+\frac{(4+12
(-1)^N) S_{-2,3}(N)}{N+1}-\frac{8S_{2,3}(N)}{N+1}.
\end{align*}
\normalsize
For further examples how one can simplify d'Alembertian solutions~\cite{Abramov:94} with our algorithms see, e.g., \cite{Schneider:03,Schneider:T05b,Schneider:06e,Schneider:08c}.
We note that in the derived result no algebraic relations between the harmonic sums occur. In the next section we show how \SigmaP\ eliminates, or equivalently, finds such algebraic relations explicitly and efficiently.

\subsection{Finding algebraic relations of nested sums}\label{Sec:AlgRel}

During the calculation of Feynman integrals harmonic sums arise frequently; see, for instance, \cite{Bluemlein:99,Vermaseren:99,Vermaseren:2005qc,Schneider:07h,Schneider:07i} for further literature. In order to derive compact representations of such computations, one can use, e.g., results from~\cite{Bluemlein:04} where all relations of harmonic sums are classified in general and tabulated up to nested depth $6$. Alternatively, we illustrate how this task can be handled efficiently in the general \pisiDE-field setting.
Consider, e.g., the sums
$S_{4,2}(N)$, $S_{2,4}(N)$, $S_{2,1,1,1,1}(N)$, $S_{1,2,
   1,1,1}(N)$, $S_{1,1,2,1,1}(N)$, $S_{1,1,1,2,1}(N)$, $S_{1,1,1,1,2}(N)$ which are algebraically independent -- except the last one: here the relation
\begin{equation}\label{Equ:S5Rel}
\begin{split}
S&_{1,1,1,1,2}(N)=\frac{1}{8} \Big(2 S_{1}(N)^6+7 S_2(N)
   S_1(N)^4+4 S_2(N)^2 S_1(N)^2\\
&+8S_{1,1,1,2}(N) S_1(N)+8 S_{1,1,2,1}(N)S_1(N)+8 S_{1,2,1,1}(N) S_1(N)\\
&+8S_{2,1,1,1}(N) S_1(N)+S_2(N)^3+24S_{1,1,1}(N)^2+8 S_{2,4}(N)+8S_{4,2}(N)\\
&+\big(-4 S_1(N)^2-2
   S_2(N)\big) S_4(N)+\big(-16 S_1(N)^3-24 S_2(N)S_1(N)\big)S_{1,1,1}(N)\\
&-8S_{1,1,1,2,1}(N)-8 S_{1,1,2,1,1}(N)-8S_{1,2,1,1,1}(N)-8 S_{2,1,1,1,1}(N)\Big)
\end{split}
\end{equation}
pops up.
With the naive reduction from Section~\ref{Sec:Reduction}, \SigmaP\ finds~\eqref{Equ:S5Rel} by representing the sums in a \pisiSE-field $\dfield{\QQ(t_1)\dots(t_{18})}{\sigma}$ where the depths of the \sigmaSE-extensions $t_1,\dots,t_{18}$ are:
$1,2,2,2,3,3,3,3,3,4,4,4,4,5,5,5,5,6,$
respectively.
With a standard notebook (2.16 GHz) we needed $772$ seconds to construct this \pisiSE-field in order to get the relation~\eqref{Equ:S5Rel}.

Applying our new algorithms, we can represent the harmonic sums in a \pisiDE-field with again $18$ extensions, but this time the depths are
$1,2,2,2,2,2,2,3,3,3,3,3,3,3,3,3,3,3,$
respectively. E.g., the extensions correspond to the sum representations~\eqref{Equ:S24S42} and

\vspace*{-0.5cm}

\begin{align*}
S_{2,1,1,1,1}(N)=&\frac{1}{24}\sum_{k=1}^N\frac{S_1(k)^4+6 S_2(k) S_1(k)^2+8 S_3(k)
   S_1(k)+3 S_2(k)^2+6 S_4(k)}{k^2},\\
S_{1,2,1,1,1}(N)=&
\frac{1}{6}\Big(\sum_{k=1}^N\frac{-(k S_1(k)-1) \left(S_1(k)^3+3
   S_2(k) S_1(k)+2 S_3(k)\right)}{k^3}\\
&+S_1(N)\sum_{k=1}^N\frac{S_1(k)^3+3 S_2(k) S_1(k)+2S_3(k)}{k^2}\Big), \;\text{etc};
\end{align*}
\normalsize

\vspace*{-0.2cm}

\noindent note that the representation of $S_{2,4}(N)$ and $S_{4,2}(N)$ in the corresponding \pisiDE-field has been carried out in details in Example~\ref{Exp:S12S21}.
In total we needed 37 seconds (instead of 772 seconds) to construct the underlying \pisiDE-field.
Based on this optimal \pisiDE-field representation, by backwards transformation the relation~\eqref{Equ:S5Rel} can be found automatically.

\medskip

\noindent\textbf{Acknowledgement.} I would like to thank Peter Paule for very enjoyable and helpful discussions.

\bibliographystyle{abbrv}

\begin{thebibliography}{10}
\bibitem{Abramov:94}
S.~Abramov and M.~Petkov{\v s}ek.
\newblock D'{A}lembertian solutions of linear differential and difference
  equations.
\newblock In J.~von~zur Gathen, editor, {\em Proc. ISSAC'94}, pages 169--174.
  ACM Press, 1994.

\bibitem{AAR}
G.~Andrews, R.~Askey, and R.~Roy.
\newblock {\em Special Functions}.
\newblock Cambridge University Press, 2006.

\bibitem{Bauer:99}
A.~Bauer and M.~Petkov{\v{s}}ek.
\newblock Multibasic and mixed hypergeometric {Gosper}-type algorithms.
\newblock {\em J.~Symbolic Comput.}, 28(4--5):711--736, 1999.

\bibitem{Schneider:07h}
I.~Bierenbaum, J.~Bl{\"u}mlein, S.~Klein, and C.~Schneider.
\newblock Difference equations in massive higher order calculations.
\newblock In {\em Proc. ACAT 2007}, volume PoS(ACAT)082, 2007.

\bibitem{Bluemlein:04}
J.~Bl{\"u}mlein.
\newblock Algebraic relations between harmonic sums and associated quantities.
\newblock {\em Comput. Phys. Commun.}, 159(4):19--54, 2004.

\bibitem{Bluemlein:99}
J.~Bl\"umlein and S.~Kurth.
\newblock Harmonic sums and {M}ellin transforms up to two-loop order.
\newblock {\em Phys. Rev.}, D60, 1999.

\bibitem{Bron:00}
M.~Bronstein.
\newblock On solutions of linear ordinary difference equations in their
  coefficient field.
\newblock {\em J.~Symbolic Comput.}, 29(6):841--877, 2000.

\bibitem{Chyzak:00}
F.~Chyzak.
\newblock An extension of {Z}eilberger's fast algorithm to general holonomic
  functions.
\newblock {\em Discrete Math.}, 217:115--134, 2000.

\bibitem{Schneider:06c}
K.~Driver, H.~Prodinger, C.~Schneider, and J.~Weideman.
\newblock Pad\'e approximations to the logarithm~{III}: Alternative methods and
  additional results.
\newblock {\em Ramanujan~J.}, 12(3):299--314, 2006.

\bibitem{Gosper:78}
R.~Gosper.
\newblock Decision procedures for indefinite hypergeometric summation.
\newblock {\em Proc. Nat. Acad. Sci. U.S.A.}, 75:40--42, 1978.

\bibitem{Singer:99}
P.~Hendriks and M.~Singer.
\newblock Solving difference equations in finite terms.
\newblock {\em J.~Symbolic Comput.}, 27(3):239--259, 1999.

\bibitem{Karr:81}
M.~Karr.
\newblock Summation in finite terms.
\newblock {\em J.~ACM}, 28:305--350, 1981.

\bibitem{Karr:85}
M.~Karr.
\newblock Theory of summation in finite terms.
\newblock {\em J.~Symbolic Comput.}, 1:303--315, 1985.

\bibitem{Schneider:06e}
M.~Kauers and C.~Schneider.
\newblock Application of unspecified sequences in symbolic summation.
\newblock In J.~Dumas, editor, {\em Proc. ISSAC'06.}, pages 177--183. ACM
  Press, 2006.

\bibitem{Schneider:06d}
M.~Kauers and C.~Schneider.
\newblock Indefinite summation with unspecified summands.
\newblock {\em Discrete Math.}, 306(17):2021--2140, 2006.

\bibitem{Schneider:07f}
M.~Kauers and C.~Schneider.
\newblock Symbolic summation with radical expressions.
\newblock In C.~Brown, editor, {\em {Proc. ISSAC'07}}, pages 219--226, 2007.

\bibitem{Schneider:T05b}
M.~Kuba, H.~Prodinger, and C.~Schneider.
\newblock Generalized reciprocity laws for sums of harmonic numbers.
\newblock SFB-Report 2005-17, J. Kepler University Linz, 2005.

\bibitem{Schneider:07i}
S.~Moch and C.~Schneider.
\newblock Feynman integrals and difference equations.
\newblock In {\em Proc. ACAT 2007}, volume PoS(ACAT)083, 2007.

\bibitem{Schneider:08c}
R.~Osburn and C.~Schneider.
\newblock Gaussian hypergeometric series and extensions of supercongruences.
\newblock {\em Math. Comp.}, 2008.
\newblock To appear.

\bibitem{Paule:95}
P.~Paule.
\newblock Greatest factorial factorization and symbolic summation.
\newblock {\em J.~Symbolic Comput.}, 20(3):235--268, 1995.

\bibitem{PauleRiese:97}
P.~Paule and A.~Riese.
\newblock A {M}athematica q-analogue of {Z}eil\-ber\-ger's algorithm based on
  an algebraically motivated aproach to $q$-hypergeometric telescoping.
\newblock In M.~Ismail and M.~Rahman, editors, {\em Special Functions, q-Series
  and Related Topics}, volume~14, pages 179--210. Fields Institute Toronto,
  AMS, 1997.

\bibitem{Schneider:03}
P.~Paule and C.~Schneider.
\newblock Computer proofs of a new family of harmonic number identities.
\newblock {\em Adv. in Appl. Math.}, 31(2):359--378, 2003.

\bibitem{Schneider:07g}
P.~Paule and C.~Schneider.
\newblock Truncating binomial series with symbolic summation.
\newblock {\em INTEGERS. Electronic Journal of Combinatorial Number Theory},
  7:1--9, 2007.
\newblock \#A22.

\bibitem{Schneider:07c}
R.~Pemantle and C.~Schneider.
\newblock {When is 0.999... equal to 1?}
\newblock {\em Amer. Math. Monthly}, 114(4):344--350, 2007.

\bibitem{Risch:69}
R.~Risch.
\newblock The problem of integration in finite terms.
\newblock {\em Trans. Amer. Math. Soc.}, 139:167--189, 1969.

\bibitem{Risch:70}
R.~Risch.
\newblock The solution to the problem of integration in finite terms.
\newblock {\em Bull. Amer. Math. Soc.}, 76:605--608, 1970.

\bibitem{Schneider:T01}
C.~Schneider.
\newblock Symbolic summation in difference fields.
\newblock Technical Report 01-17, RISC-Linz, J.~Kepler University, November
  2001.
\newblock PhD Thesis.

\bibitem{Schneider:04d}
C.~Schneider.
\newblock The summation package {S}igma: Underlying principles and a rhombus
  tiling application.
\newblock {\em Discrete Math. Theor. Comput. Sci.}, 6(2):365--386, 2004.

\bibitem{Schneider:04a}
C.~Schneider.
\newblock Symbolic summation with single-nested sum extensions.
\newblock In J.~Gutierrez, editor, {\em Proc. ISSAC'04}, pages 282--289. ACM
  Press, 2004.

\bibitem{Schneider:05f}
C.~Schneider.
\newblock Finding telescopers with minimal depth for indefinite nested sum and
  product expressions.
\newblock In M.~Kauers, editor, {\em Proc. ISSAC'05}, pages 285--292. ACM,
  2005.

\bibitem{Schneider:05d}
C.~Schneider.
\newblock A new {S}igma approach to multi-summation.
\newblock {\em Adv. in Appl. Math. Special Issue Dedicated to Dr. David P.
  Robbins}, 34(4):740--767, 2005.

\bibitem{Schneider:05c}
C.~Schneider.
\newblock Product representations in ${\Pi}{\Sigma}$-fields.
\newblock {\em Ann. Comb.}, 9(1):75--99, 2005.

\bibitem{Schneider:05a}
C.~Schneider.
\newblock Solving parameterized linear difference equations in terms of
  indefinite nested sums and products.
\newblock {\em J. Differ. Equations Appl.}, 11(9):799--821, 2005.

\bibitem{Schneider:07d}
C.~Schneider.
\newblock {Simplifying Sums in $\Pi\Sigma$-Extensions}.
\newblock {\em J. Algebra Appl.}, 6(3):415--441, 2007.

\bibitem{Schneider:07a}
C.~Schneider.
\newblock Symbolic summation assists combinatorics.
\newblock {\em S\'em.~Lothar. Combin.}, 56:1--36, 2007.
\newblock Article B56b.

\bibitem{Schneider:T07f}
C.~Schneider.
\newblock Symbolic summation finds optimal nested sum representations.
\newblock SFB-Report 2007-26, SFB F013, J. Kepler University Linz, 2007.

\bibitem{Schneider:08d}
C.~Schneider.
\newblock Parameterized telescoping proves algebraic independence of sums.
\newblock {\em Ann. Comb.}, 2008.
\newblock To appear.

\bibitem{Singer:85}
M.~Singer, B.~Saunders, and B.~Caviness.
\newblock An extension of {L}iouville's theorem on integration in finite terms.
\newblock {\em SIAM J. Comput.}, 14(4):966--990, 1985.

\bibitem{Vermaseren:99}
J.~Vermaseren.
\newblock Harmonic sums, {M}ellin transforms and integrals.
\newblock {\em Int. J.~Mod. Phys.}, A14:2037--2976, 1999.

\bibitem{Vermaseren:2005qc}
J.~Vermaseren, A.~Vogt, and S.~Moch.
\newblock The third-order qcd corrections to deep-inelastic scattering by
  photon exchange.
\newblock {\em Nucl. Phys.}, B724:3--182, 2005.

\bibitem{Zeilberger:91}
D.~Zeilberger.
\newblock The method of creative telescoping.
\newblock {\em J.~Symbolic Comput.}, 11:195--204, 1991.

\end{thebibliography}

\end{document}